\documentclass[preprint,3p,12pt,times]{elsarticle}

\usepackage{graphicx}
\usepackage{epsfig}
\usepackage{epstopdf}
\usepackage{amssymb}
\usepackage{amsthm}
\usepackage{bm}

\usepackage[colorlinks=true,citecolor=blue,linkcolor=black]{hyperref}
\usepackage{color}
\usepackage{amsmath}
\usepackage{multirow}
\usepackage[justification=centering]{caption}
\usepackage{subcaption}
\usepackage{color}

\usepackage{booktabs} 

\journal{International Journal of Mechanical Sciences}

\begin{document}

\begin{frontmatter}



\title{A Total Lagrangian SPH Method for Modelling Damage and Failure in Solids}

\author[essteyr]{Md Rushdie Ibne Islam}
\author[essteyr,baku]{Chong Peng\corref{essteyr1}}
\ead{chong.peng@essteyr.com}

\cortext[essteyr1]{Corresponding Author}
\address[essteyr]{ESS Engineering Software Steyr GmbH, Berggasse 35, 4400 Steyr, Austria}
\address[baku]{Institut f{\"{u}}r Geotechnik, Universit{\"{a}}t f{\"{u}}r Bodenkultur, Feistmantelstrasse 4, 1180 Vienna, Austria} 

\begin{abstract}
An algorithm is proposed to model crack initiation and propagation within the total Lagrangian Smoothed Particle Hydrodynamics (TLSPH) framework. TLSPH avoids the two main deficiencies of conventional SPH, i.e., tensile instability and inconsistency, by making use of the Lagrangian kernel and gradient correction, respectively. In the present approach, the support domain of a particle is modified, where it only interacts with its immediately neighbouring particles. A virtual link defines the level of interaction between each particle pair. The state of the virtual link is determined by damage law or cracking criterion. The virtual link approach allows easy and natural modelling of cracking surfaces without explicit cracking treatments such as particle splitting, field enrichment or visibility criterion. The performance of the proposed approach is demonstrated via a few numerical examples of both brittle and ductile failure under impact loading.     
\end{abstract}


\end{frontmatter}


\section{Introduction}\label{sec-1}
Initiation and propagation cracks, propagation and their interaction are often encountered in solid mechanics problems. Materials under different loading conditions may undergo various stages before complete failure. Therefore, one of the primary aims of computational mechanics is to simulate the cracks and their propagation accurately. Earlier attempts are mostly based on grid/mesh-based methods such as the Finite Element Method (FEM). Despite being popular and successful methods for modelling of solids, FEM and other grid-based methods suffer from significant drawbacks in modelling discrete cracks due to the presence of grids or meshes \cite{belytschko1996difficulty}. Many attempts are made to model the cracks and their propagation through special treatments such as mesh refinement and field enrichments etc. \citep{martha1993arbitrary, melenk1996partition, de1999elastoplastic, fries2010extended}. However, it is still quite challenging to model multiple crack paths, their propagation and interaction. Furthermore, if the material undergoes finite deformation, the grid/mesh suffers from mesh entanglement, mesh distortion etc., which affect the accuracy of the solution significantly. Moreover, the grid-based methods require good quality of mesh for accurate computation. This presents a challenge when the geometry of the problem domain is complex. 

Meshfree or particle-based methods \cite{li2002meshfree} presents a way out of these problems. Due to the absence of mesh, the particle-based techniques do not suffer from mesh distortion or entanglement and can model large deformations of materials. Among the meshfree methods, Smoothed Particle Hydrodynamics (SPH) \cite{gingold1977smoothed, lucy1977numerical, liu2010smoothed} is a well-established and completely meshfree method which has a long application history. SPH was first developed for astronomical studies \cite{gingold1977smoothed, monaghan1985refined} and later was used in fluid mechanics problems \cite{monaghan1988introduction, monaghan1989problem, monaghan1994simulating}. The early application of SPH in solid mechanics is the high-speed impact where material undergoes large deformation \cite{libersky1991smooth, libersky1993high}. In SPH, the entire computational domain is discretised into discrete particles. These particles represent a certain volume of the material in the computational domain. The field variables are approximated by the SPH interpolation, where the values at a point are approximated by a smoothing function defined on a support domain with finite radius, which is known as the kernel function. The kernel functions are constructed in such a way that for any particle the influence of other particles on it decreases with the increase in the particle distance. The derivatives of the field variables are also computed based on the discrete values of the particles. This ensures that the computation is entirely free from the grid or mesh.

Despite being an early success, SPH suffers from some severe deficiencies: (a) tensile instability \cite{swegle1995smoothed}, i.e. the particle tends to form a cluster locally which represents a numerical fracture of the material, (b) an hourglass mode caused by the collocation scheme where interpolation and numerical integration are performed using the same set of particles \cite{belytschko2000unified, randles2000normalized} and (c) inconsistency \cite{randles1996smoothed} of the SPH approximation, i.e. the SPH formulation does not provide zeroth or first order consistency in the approximation. Different approaches are suggested to overcome these problems such as - (a) artificial pressure/stress correction \citep{monaghan2000sph, gray2001sph} to remove tensile instability, (b) stress point method \cite{dyka1995approach, belytschko2000unified} to overcome hourglass mode and (c) gradient correction \cite{chen1999corrective} to improve the consistency. The artificial pressure/stress correction can alleviate tensile instability and can give rise to satisfactory results provided that the parameter is tuned suitably. However, this parameter calibration needs to be performed for every problem, which is not only inconvenient but also introduces an undesirable arbitrariness into the simulation. Belytschko et al. \cite{belytschko2000unified} investigated the cause of the tensile instability and observed that the use of the updated particle position for the computation of kernel function is the reason and the Lagrangian kernel does not suffer from this issue \citep{belytschko2002stability, rabczuk2004stable}. In this regard, the total-Lagrangian Smoothed Particle Hydrodynamics (TLSPH) was developed \citep{vignjevic2006sph, belytschko2000unified, bonet2002alternative}. In TLSPH, the kernel functions are always calculated on the reference configuration. Hence, it does not suffer from any tensile instability.

Although SPH can model large plastic deformation in its original form, it is incapable of modelling discrete cracks as it is a continuum-based method. In this regard, several treatments such as visibility criterion, particle splitting etc. \citep{rabczuk2004cracking, rabczuk2007three, rabczuk2010simple, organ1996continuous, ren2010meshfree} can be used to model the material damage and explicit crack surfaces in particle methods. However, these techniques are computationally intensive. Chakraborty and Shaw \cite{chakraborty2013pseudo} developed a pseudo-spring approach for the crack treatment. In this method, the particles are connected through a set of springs and these springs define the level of interaction between particles. Once a spring fails, it is assumed that the crack passes through the spring. This treatment is quite simple as it does not require any visibility criterion or particle splitting algorithms. This method is successfully applied to capture the brittle failure, adiabatic shear failure, ceramic-metal composite failure etc. \citep{shaw2015beyond, chakraborty2014crack, chakraborty2015prognosis, chakraborty2017computational, islam2017computational}. Zhou and his group \citep{zhou2015novel, bi2016numerical, bi2017numerical, yin2018numerical} used a similar virtual bond based approach to model cracks in rocks. However, as these approaches use the Eulerian kernel function, they suffer from the tensile instability problem and need to tune the parameters.

In the present work, an algorithm within the TLSPH framework is proposed. As mentioned before, in TLSPH the kernel is always calculated in the reference configuration. The bell-shaped kernel functions ensure that the contribution of the immediate neighbour particles in the approximations is the largest. Therefore, in the present approach, the definition of a neighbourhood for a particle is modified. The interaction of any particle is restricted to its immediate neighbours only. However, as the neighbourhood of particles is restricted to its immediate neighbour, this might cause the under-integration of the field variables. To overcome this, the kernel gradient correction \cite{chen1999corrective} is applied in the simulations. This restores the consistency in the particle approximation as well as helps in overcoming the particle deficiency problem at particles with incomplete support domain. Each particle is connected with its immediate neighbours with a set of virtual links. These links do not provide any extra stiffness to the system. However, they define the extent of interaction between particles. The damage state of these virtual links is evaluated based on the material damage laws. An unbroken link defines the complete interaction between particles. The interaction between particles goes through a continuous change as the damage state of the connecting links changes. Once a connection is completely broken, the interaction between connecting particles is stopped, and it is assumed that the crack path passes through the connecting link or the particle pair. The present approach does not require any crack tracking algorithm. The crack path and its propagation can be tracked automatically through the broken links in the domain. The performance of the present approach is demonstrated via a few numerical examples of brittle and ductile failure under impact loadings. 

The paper is organised as follows. In the next section, the basic formulation of TLSPH is briefly revisited. The treatment for material damage and fracture is proposed in section \ref{sec_pl}. The pin-ball contact algorithm between two bodies is discussed briefly in section \ref{con_al}. The performance of the present algorithm is demonstrated via the rigid-plastic analysis of perfect beam, crack propagation in a notched beam, the Kalthoff's experiment, and the mode I and mixed mode crack propagation in a deep beam in section \ref{num}. Some conclusions are drawn in section \ref{conclu}.

\section{Total-Lagrangian Smoothed Particle Hydrodynamics (TLSPH) formulation}\label{sec-sph} 
Smoothed Particle Hydrodynamics (SPH) is a Lagrangian particle method where the particles interact through a kernel function. In standard SPH, the kernel function is defined based on the current configuration, which means that particles can enter or exit each other's support domains as the material deforms. As a result, the standard SPH kernel function is called Eulerian kernel even though that SPH is a Lagrangian particle method. In SPH the conservation equations of mass, momentum and energy as shown in equations \ref{eq1}, \ref{eq2} and \ref{eq3} are solved at each time step in the current configuration or frame. The extensive information on SPH can be found in \citep{liu2010smoothed, liu2006restoring} and their references therein. 

\begin{equation}\label{eq1}
    \dfrac{\mathrm{d}\rho}{\mathrm{d}t}=-\rho\nabla\cdot\bm v
\end{equation}
\begin{equation}\label{eq2}
    \dfrac{\mathrm{d}\bm v}{\mathrm{d}t}=\dfrac{1}{\rho}\nabla \bm\sigma
\end{equation}
\begin{equation}\label{eq3}
    \dfrac{\mathrm{d}e}{\mathrm{d}t}=\dfrac{1}{\rho} \bm \sigma :(\nabla\otimes\bm v)
\end{equation}
where $\rho$ denotes the material density, $\bm v$ and $\bm \sigma$ are respectively the velocity vector and Cauchy stress tensor in the current configuration, $e$ is the specific internal energy, $\mathrm{d}/\mathrm{d}t$ is time derivative taken in a moving Lagrangian frame. $\nabla$ denotes the divergence or gradient operator, and $\otimes$ is the outer product between two vectors.

In this section, the TLSPH formulation \citep{vignjevic2006sph, de2013total, ganzenmuller2015hourglass, leroch2016smooth, rausch2017modeling} of particle approximation and conservation equations are discussed briefly. In the total Lagrangian description of SPH, the conservation and the constitutive equations are solved in the undeformed or reference configuration $\bm{X}$ only. The changes in the field variables are used to compute the current deformed configuration $\bm{x}$. The current description $\bm{x}$ is related to the reference description $\bm{X}$ through a mapping $\bm{\phi}$ as

\begin{equation}
   \bm{x} = \bm{\phi} \left(\bm{X},t\right)
\end{equation} 

The displacement $\bm{u}$ is computed as

\begin{equation}
   \bm{u} = \bm{x} - \bm{X}
\end{equation}

The mass, momentum and energy conservation equation \ref{eq1}, \ref{eq2} and \ref{eq3} \cite{de2013total} can be expressed in terms of the reference configurations as follows

\begin{equation}\label{ref1}
  \rho = J^{-1} \rho_0
\end{equation}

\begin{equation}\label{ref2}
  \dfrac{\mathrm{d}v}{\mathrm{d}t} = \frac{1}{\rho_0} \nabla_0 \cdot \bm{P}
\end{equation}

\begin{equation}\label{ref3}
   \dfrac{\mathrm{d}e}{\mathrm{d}t} = \dfrac{1}{\rho_0} \bm{P}:\dot{\bm{F}}
\end{equation}
where, $0$ indicates that the values are computed at the undeformed reference configuration $\bm{X}$, $\bm{P}$ denotes the first Piola Kirchhoff stress, $J$ is the Jacobian computed as $J = \mathrm{det}(\bm{F})$, where $\bm{F}$ denotes the deformation gradient matrix calculated as

\begin{equation}
   \bm{F} = \dfrac{\mathrm{d}\bm{x}}{\mathrm{d}{\bm{X}}} = \dfrac{\mathrm{d}\bm{u}}{\mathrm{d}{\bm{X}}} + \bm{I}
\end{equation}

The deformation gradient matrix $\bm{F}$ relates the current and reference configuration, i.e. it defines the deformation of a line element in the current configuration based on the reference configuration. $\bm I$ is an identity tensor.

\subsection{Particle approximation in TLSPH}
In TLSPH, any field variable $f(\bm X_i)$ is approximated based on the reference configuration as follows

\begin{equation}
   f(\bm X_i) = \sum_j f(\bm X_j) W(\bm X_{ij}) \dfrac{m_j}{\rho_{0j}}
\end{equation}
where $f(\bm X_j)$ is the field variable value at $j$-th particle, $\bm X_{ij} = \bm X_i - \bm X_j$ is the vector from particle $j$ to particle $i$, $m_j/\rho_{0j}$ represents the volume of $j$-th particle in the reference configuration and $W(\bm X_{ij})$ is the kernel function defined in the undeformed reference configuration. In this work, the following cubic B spline function is used for the approximation.

\begin{equation}\label{kernel}
    W(q, h)=\alpha_d 
\begin{cases}
    1-\dfrac{3}{2} q^2 +\dfrac{3}{4} q^3,& \text{if } 0\le q\le 1\\
    \dfrac{1}{4}(2-q)^3,              & \text{if } 1\le q\le 2\\
    0                                 & \text{otherwise}
\end{cases}
\end{equation}
where, $\alpha_d=10/7\pi h^2$ in 2D and $q=|\bm{X}_i-\bm{X}_j|/h$ is the normalised distance associated with a particle pair with smoothing length $h$.

The derivative of the function $f(\bm X_i)$ is approximated as

\begin{equation}
   \nabla f(\bm X_i) = \sum_j f(\bm X_j) \nabla_i W(\bm X_{ij}) \dfrac{m_j}{\rho_{0j}}
\end{equation}
where $\nabla_i W(\bm X_{ij})$ is the kernel derivative computed at $\bm X_i$ based on the reference configuration. Monaghan \cite{monaghan1992smoothed} suggested symmetric approximation form for the zeroth order completeness in the derivative approximation with the Eulerian kernel. Similarly, the symmetric estimate for derivative approximation in TLSPH writes as follows 

\begin{equation}
   \nabla f(\bm X_i) = - \sum_j \left[f(\bm X_i) - f(\bm X_j) \right] \nabla_i W(\bm X_{ij}) \dfrac{m_j}{\rho_{0j}}
\end{equation}

Consequently, the particle approximation form for the deformation gradient and its rate are 

\begin{equation}\label{deformation}
   \bm{F}_i = - \sum_j \left(\bm u_i - \bm u_j \right) \otimes \nabla_i W(\bm X_{ij}) \dfrac{m_j}{\rho_{0j}} + \bm I \\
            = - \sum_j \left(\bm x_i - \bm x_j \right) \otimes \nabla_i W(\bm X_{ij}) \dfrac{m_j}{\rho_{0j}}
\end{equation}

\begin{equation}\label{rate_deformation}
   \bm{\dot{F}}_i = - \sum_j \left(\bm v_i - \bm v_j \right) \otimes \nabla_i W(\bm X_{ij}) \dfrac{m_j}{\rho_{0j}}
\end{equation}

The conservation equations \ref{ref2} and \ref{ref3} can be expressed in particle form as

\begin{equation}\label{con1}
   \dfrac{\mathrm{d} \bm v_i}{\mathrm{d}t} = \sum_j m_j \left( \dfrac{\bm P_i}{\rho_{0i}^2} + \frac{\bm P_j}{\rho_{0j}^2} - \bm{\Pi}_{ij} \right)  \nabla_i W(\bm X_{ij})
\end{equation}

\begin{equation}\label{con2}
   \frac{\mathrm{d}e_i}{\mathrm{d}t} = \dfrac{\bm P_i}{\rho_{0i}} : \sum_j \left(\bm v_i - \bm v_j \right) \otimes \nabla_i W(\bm X_{ij}) \dfrac{m_j}{\rho_{0j}}
\end{equation}
where, the first Piola Korchhoff stress $\bm{P}$ is calculated as $\bm{P} = J \bm{F}^{-1} \bm{\sigma}$. $\bm{\Pi}_{ij}$ is the artificial viscosity computed as $\bm{\Pi}_{ij} = J \bm{F}^{-1} \pi_{ij}$. In the presence of jump in field variables or shock, the artificial viscosity is used to for stabilised SPH computation. In the present work, the artificial viscosity \cite{monaghan1983shock} is used in the following form. 

\begin{equation}\label{artificial}
    \pi_{ij}= 
\begin{cases}
    \dfrac{-\beta_1 \bar{C}_{ij}\mu_{ij} + \beta_2 \mu^2_{ij}}{\bar{\rho}_{ij}},& \text{if } \bm v_{ij}\cdot\bm x_{ij}\le 0\\
    0,              & \text{otherwise}
\end{cases}
\end{equation}
where, $\beta_1$, $\beta_2$ are the controlling parameters for artificial viscosity, $\mu_{ij}=h(\bm{v}_{ij}\cdot\bm{x}_{ij})/(\bm{x}^2_{ij}+\epsilon h^2)$; $\epsilon$ is a small number used to avoid singularity (here $\epsilon=0.01$), sound speed is computed as $C=\sqrt{E/\rho}$ in the materials, $E$ is the Young's modulus, $\bm{x}_{ij}=\bm{x}_i-\bm{x}_j$, $\bm{v}_{ij}=\bm{v}_{i}-\bm{v}_{j}$, $\bar{C}_{ij}=(C_i+C_j)/2$ and $\bar{\rho_{ij}}=(\rho_i+\rho_j)/2$. The controlling parameters of artificial viscosity $\beta_1$ and $\beta_2$ are obtained through numerical experiments so that the fluctuations due to the shock or jump present in the field variables are removed without over damping the computation.  

\subsection{Constitutive model}
The Cauchy stress $\bm{\sigma}$ is computed based on the hydrostatic pressure $p$ and the deviatoric stress $\bm{s}$ as $\bm{\sigma}=\bm{s}-p\bm I$. Material frame invariant Jaumann stress rate is used to compute the deviatoric stress components as 

\begin{equation}\label{eq11}
    \dot{\bm s}=2\mu \left(\dot{\bm \epsilon}-\frac{1}{3}\dot{\varepsilon}\bm I\right)+ \bm s \bm \omega - \bm \omega \bm s    
\end{equation}
where, $\mu$ is the shear modulus of the material; $\dot{\bm \epsilon}$ and $\bm \omega$ are the strain rate tensor and spin tensor, $\dot\varepsilon$ is the sum of the diagonal components of the strain rate tensor. The strain rate and spin tensors are obtained from the velocity gradient tensor $\bm{l} = \bm{\dot{F} F^{-1}}$ as

\begin{equation}\label{eq12}
    \dot{\bm \epsilon}=\dfrac{1}{2} \left( \bm l + \bm l^{\mathrm{T}} \right)
\end{equation} 

\begin{equation}\label{eq13}
    \bm \omega=\dfrac{1}{2} \left( \bm l - \bm l^{\mathrm{T}}\right)
\end{equation} 

In the present work, the linear function of compressibility \cite{eliezer1986introduction} is used to compute the pressure $p$ as, 
\begin{equation}\label{eos}
    p=K\left(\frac{\rho}{\rho_0}-1\right)
\end{equation}
where $K=E/(3-6 \nu)$ is the bulk modulus, $\nu$ is the poison's ratio of the material. 

The material plasticity is incorporated by the pressure-independent Von-Mises yield criterion, in which the yield function is defined as $y_f=\sqrt{J_2}-\sigma_y/\sqrt{3}$, where $\sigma_y$ is the yield stress of material and $J_2=\bm s : \bm s /2$ is the second invariant of the deviatoric stress tensor. Return mapping algorithm to bring back the deviatoric stress to the yield surface is implemented using the Wilkins criterion as $\bm s_n=c_f \bm s$, where $c_f=\min \left(\sigma_y/\sqrt{3J_2},1\right)$ and $\bm s_n$ is the corrected deviatoric stress tensor. The following equations compute the increment of plastic strain, the increment of effective plastic strain and the accumulated plastic work density

\begin{equation}
   \Delta \bm \epsilon_{pl} = \dfrac{1-c_f}{2\mu} \bm s    
\end{equation}
\begin{equation}
   \Delta \epsilon_{pl} = \sqrt{\dfrac{2}{3}  \Delta \bm \epsilon_{pl} : \Delta \bm \epsilon_{pl}} = \dfrac{1-c_f}{3\mu} \sqrt{\frac{3}{2} \bm s : \bm s} 
\end{equation}
\begin{equation}
   \Delta w_p =  \Delta \bm \epsilon_{pl} : \bm s_{n}          
\end{equation}   

\section{Treatment for material damage and fracture}\label{sec_pl}
When solid materials undergo large deformation, the adjacent material points will always remain as neighbour points unless the body has crack surfaces formed and loses continuity. Translating this condition into an SPH point of view, a particle should always interact with the same set of neighbouring particles. However, this requirement is not satisfied in standard SPH. On the other hand, in TLSPH the particle interactions are computed based on the initial configuration, where the deforming property of solids is naturally modelled. However, the original form of TLSPH cannot handle problems with explicit material separation such as cracking and fragmentation, as the material is always treated as a continuous body because the particle connectivity is fixed. Therefore, the original form of TLSPH is incapable of modelling material damage. Additional treatment is needed to model discontinuities such as cracking surface.

A pseudo-spring analogy for modelling cracks and damage in solid materials in standard SPH with immediate neighbour interaction is proposed in \cite{chakraborty2013pseudo}. The algorithm is extended in modelling of crack paths and their interaction for brittle and ductile materials in \citep{chakraborty2014crack, chakraborty2015prognosis, chakraborty2017computational, islam2017computational}. However, the major disadvantage of this approach is that it suffers from tensile instability. The artificial/Monaghan pressure is used in that computational framework to overcome tensile instability. However, the artificial/Monaghan pressure coefficient needs to be determined through numerical calibrations for each problem. As TLSPH avoids tensile instability, TLSPH is an appropriate method for cracking modelling in solids.

\begin{figure}[hbtp!]
\centering
\includegraphics[width=0.8\textwidth]{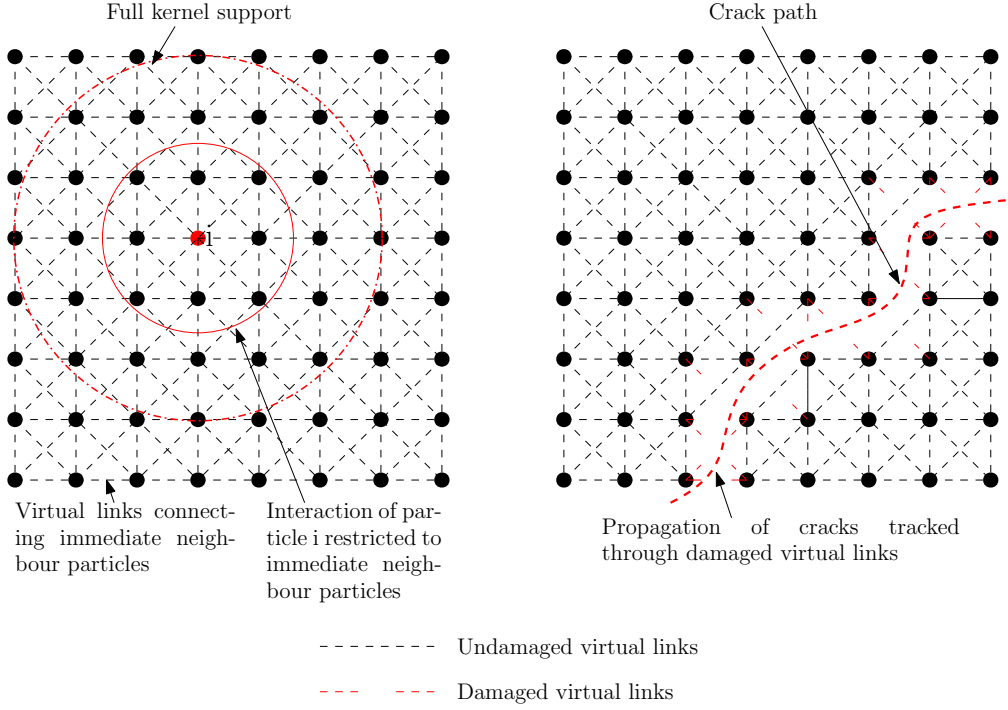}
\caption{Virtual links on immediate neighbours and crack propagation through damaged links}\label{imnb}
\end{figure}

In SPH, the bell-shaped kernel functions ensure that the influence of particles inversely varies with the distance from the particle of interest, i.e. the influence of the immediate neighbour particles is at maximum whereas the influence of the particles near the kernel boundary is minimum. This property of the kernel function inspires the assumptions of the present cracking treatment method. In the present method, the neighbour definition of particles is modified, where a particle only interacts with its immediate neighbours as shown in Figure \ref{imnb}.

Moreover, the particles are connected through a set of virtual links. These links are termed ``virtual" because they do not provide any extra stiffness to the system but only defines the level of interaction between particles. The material constitutive properties guide the damage evolution of these links and consequently define the interaction. In the initial stage, as there is no damage in the links, the particles interact normally without any reduction. As the material deforms, the damage index increases in these links, resulting in reduced interaction between connecting particles. The interaction between particles ceases to exist once the link is completely damaged and it is assumed that the crack path propagates through that damaged link as shown in Figure \ref{imnb}. With the help of the virtual link, the complete crack path can be tracked automatically through these damaged links (Figure \ref{imnb}). The presented method avoids the use of particle splitting or visibility criterion; thus, it simplifies the formulation and implementation and reduces the computational cost.

\subsection{Consistency correction}
As the neighbour is redefined in the present algorithm, but the smoothing length is kept unchanged, the approximation might not preserve the zeroth- or first-order consistency. Hence, a gradient correction as proposed in \cite{chen1999corrective} is employed to restore the consistency in the approximation as follows.

\begin{equation}\label{csph_app_1}
    \nabla_i \hat{W}(\bm X_{ij}) = \bm{K}_i  \nabla_i W(\bm X_{ij})
\end{equation}
where, 
\begin{equation}\label{csph_app_2}
   \bm K_i=\bm B_i^{-1},~~~\mathrm{where}~~~\bm{B}_i=-\sum_j \bm{X}_{ij} \otimes \nabla_i W(\bm X_{ij}) \dfrac{m_j}{\rho_{0j}}
\end{equation}

This correction also removes the truncation error caused by the incomplete support domain.

\subsection{Modified conservation equations}
An interaction coefficient $f_{ij}$ is introduced to define the state of interaction between the particle pair $i$ and $j$. The computation of $f_{ij}$ is purely based on the damage index $D_{ij}$ of the particle pair as

\begin{equation}\label{int}
   f_{ij} = 1 - D_{ij}
\end{equation} 
Initially, the material is assumed to be undamaged i.e. $D_{ij} = 0$. This means complete interaction between the particle pair or $f_{ij}=1$. Once failure starts, the damage index $D_{ij}$ increases, thus $f_{ij}$ becomes less than $1$, implying partial or reduced interaction between particles ($0 < D_{ij} < 1$ i.e. $0 < f_{ij} < 1$). When the damage index $D_{ij}$ reaches $1$, $f_{ij}$ becomes $0$ (Equation \ref{int}). This implies a broken link, i.e., the interaction between the connecting particles through the broken link ceases to exist completely. Therefore, a discontinuous crack surface is implicitly modelled through the broken link. 

\begin{figure}[hbtp!]
\centering
\begin{subfigure}[t]{0.3\textwidth}    
\includegraphics[width=\textwidth,trim={10 10 620 10}, clip]{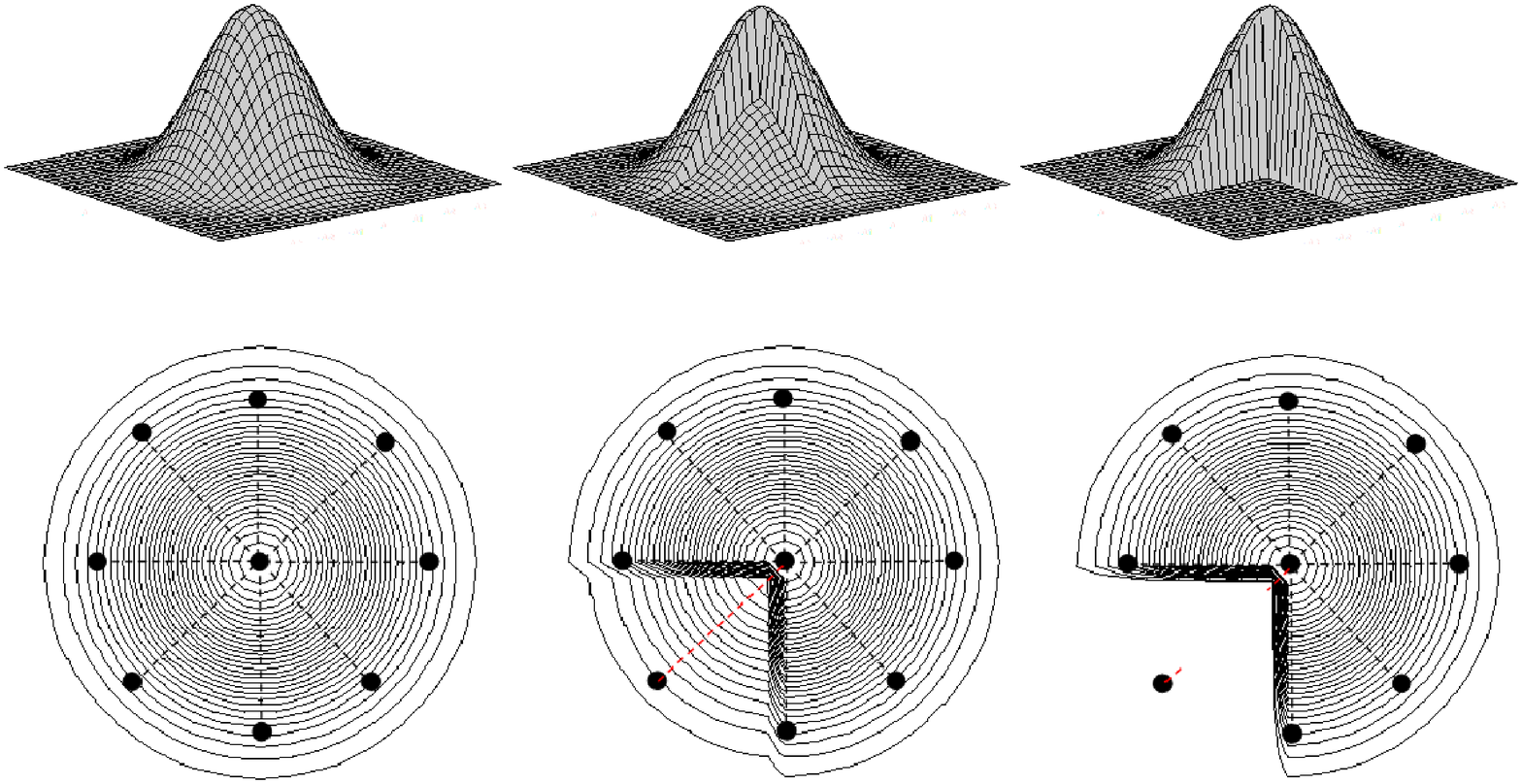}
\caption{Kernel: $f_{ij} = 1$} 
\end{subfigure}
\begin{subfigure}[t]{0.26\textwidth}
\includegraphics[width=\textwidth,trim={340 10 340 10}, clip]{kernel_function.eps}
\caption{Kernel: $0<f_{ij}<1$} 
\end{subfigure}
\begin{subfigure}[t]{0.3\textwidth}    
\includegraphics[width=\textwidth,trim={620 10 10 10}, clip]{kernel_function.eps}
\caption{Kernel: $f_{ij} = 0$} 
\end{subfigure}
\begin{subfigure}[t]{0.3\textwidth}
\includegraphics[width=\textwidth,trim={10 10 630 10}, clip]{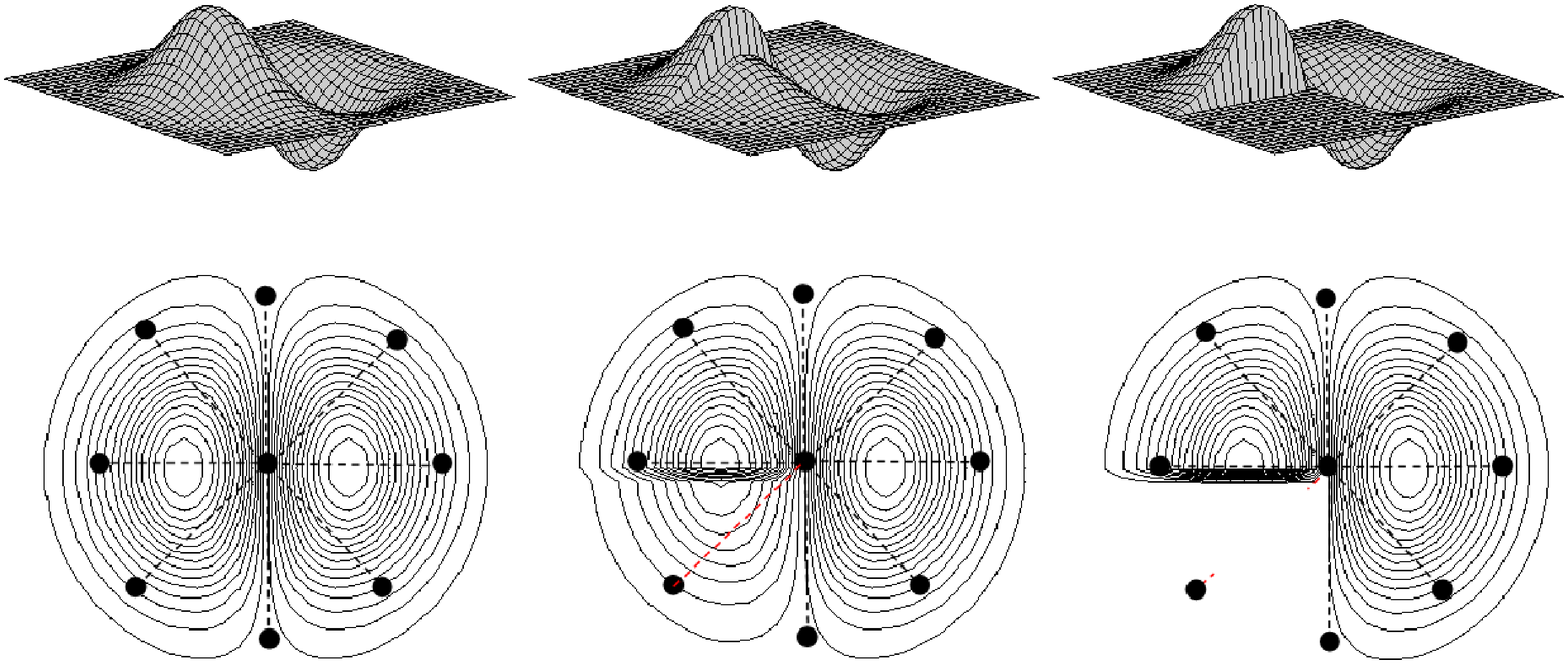}
\caption{Derivative: $f_{ij} = 1$} 
\end{subfigure}
\begin{subfigure}[t]{0.28\textwidth}    
\includegraphics[width=\textwidth,trim={330 10 330 10}, clip]{kernel_derivative.eps}
\caption{Derivative: $0<f_{ij}<1$} 
\end{subfigure}
\begin{subfigure}[t]{0.3\textwidth}
\includegraphics[width=\textwidth,trim={630 10 10 10}, clip]{kernel_derivative.eps}
\caption{Derivative: $f_{ij} = 0$} 
\end{subfigure}
\caption{Modification of kernel in 2D based on damage state of virtual links}\label{fig_kernel2D}
\end{figure}

The interaction function $f_{ij}$ is multiplied with the corrected kernel $ \nabla_i \bar{W}(\bm X_{ij}) = f_{ij} \nabla_i \hat{W}(\bm X_{ij})$. The shapes of the modified kernel function $\bar{W}(\bm X_{ij})$ and its derivative $\nabla_i \bar{W}(\bm X_{ij})$ for different interaction coefficients $f_{ij}$ are shown in Figure \ref{fig_kernel2D}. Let $N^i$ be the set of particles in the modified neighbourhood of particle $i$ in the reference configuration. $N^i_U$ and $N^i_D$ are the sets of neighbouring particles connected to the particle $i$ through the undamaged and damaged virtual links, respectively. The damage index $D_{ij}$ is zero and the interaction coefficient $f_{ij}$ is one for the particles in the set $N^i_U$. Similarly, for the particles in the set $N^i_D$, $0 < D_{ij} \leq 1 $ and  $1 > f_{ij} \geq 0$ hold. In short, $f_{ij} = 1$ means no damage, $ 0 < f_{ij} < 1$ means partial damage and $f_{ij} = 0$ means fully damaged virtual link, or generation of new crack opening. With the modified kernel gradients, the conservation equatons \ref{con1} and \ref{con2} accordingly have the following form 

\begin{equation}\label{con1m}
   \dfrac{\mathrm{d} \bm v_i}{\mathrm{d} t} = \sum_{j \in N^i_U} m_j \left( \dfrac{\bm P_i}{\rho_{0i}^2} + \dfrac{\bm P_j}{\rho_{0j}^2} - \bm{\Pi}_{ij} \right)  \nabla_i \hat{W}(\bm X_{ij}) + \sum_{j \in N^i_D} m_j \left( \dfrac{\bm P_i}{\rho_{0i}^2} + \dfrac{\bm P_j}{\rho_{0j}^2} - \bm{\Pi}_{ij} \right)  \left( f_{ij} \nabla_i \hat{W}(\bm X_{ij}) \right)
\end{equation}

\begin{equation}\label{con2m}
   \frac{\mathrm{d} e_i}{\mathrm{d} t} = \dfrac{\bm P_i}{\rho_{0i}} : \sum_{j \in N^i_U} \left(\bm v_i - \bm v_j \right) \nabla_i \hat{W}(\bm X_{ij}) \dfrac{m_j}{\rho_{0j}} + \dfrac{\bm P_i}{\rho_{0i}} : \sum_{j \in N^i_D} \left(\bm v_i - \bm v_j \right) \left( f_{ij} \nabla_i \hat{W}(\bm X_{ij}) \right) \frac{m_j}{\rho_{0j}}
\end{equation}

\section{Contact froce}\label{con_al}
\begin{figure}[hbtp!]
\centering
\includegraphics[width=0.7\textwidth]{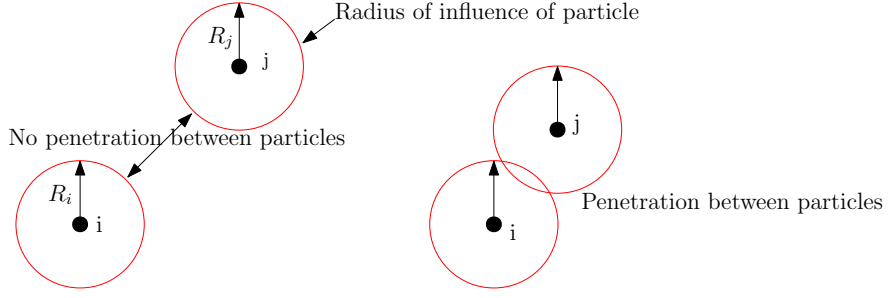}
\caption{Penetration between particles for contact force}\label{contact}
\end{figure}

An explicit contact algorithm is necessary for TLSPH to model the interaction between different bodies due to the use of the reference configuration. In the present work, the pin-ball contact algorithm as proposed in \cite{campbell2000contact} is used to model the multi-body contact in TLSPH. In this approach, it is assumed that each SPH particle is a virtual contact body, which is a circle in 2D and a sphere in 3D, having a radius of $kh$, where $k$ is a constant factor. Once the virtual contact bodies of the particles belonging to different objects overlap each other, the contact is activated as shown in Figure \ref{contact}. The material properties and the relative motions of the objects in contact determine the contact force.

For a particle pair $i$-$j$, the magnitude of the overlapping is determined as

\begin{equation}\label{eq9contact}
   p_d=(R_i+R_j)-|\bm{x}_i-\bm{x}_j|
\end{equation}
where, $R_i$ and $R_j$ are the radii of the virtual contact bodies of $i$-th and $j$-th particle, respectively. $p_d>0$ means there is overlap between the two bodies. As a result, the contact force between the particle pair is evaluated \cite{belytschko1993splitting} as

\begin{equation}
    F_{ij} = K_p~\min(F^1_{ij}, F^2_{ij}) 
\end{equation}
where,
\begin{eqnarray}   
    F^1_{ij}=\begin{cases}
    \dfrac{\rho_i \rho_j R^3_i R^3_j}{\rho_i R^3_i + \rho_j R^3_j} \dfrac{\dot{p_d}}{\Delta t},          & \dot{p_d} > 0\\
    0,              &  \dot{p_d} < 0
\end{cases}\\
    F^2_{ij}=\left(\dfrac{\mu_i \mu_j}{\mu_i + \mu_j}\sqrt{\dfrac{R_i R_j}{R_i + R_j}}\right)p_d^{1.5}
\end{eqnarray}
where $\Delta t$ is the time step, $\dot{p_d}= |\bm v_i- \bm v_j|$ is the rate of penetration and $K_p$ is a scale factor generally chosen (through numerical experiment) based on particle spacing, impact velocity etc. The modified momentum equation \ref{con1m} is modified by including the contact force  

\begin{equation}\label{contact1}  
\dfrac{\mathrm{d} \bm v_i}{\mathrm{d} t}=\left(\dfrac{\mathrm{d} \bm v_i}{\mathrm{d} t}\right)_{\mathrm{Eq.}~\ref{con1m}}+\dfrac{\bm x_i -\bm x_j}{|\bm{x}_i-\bm{x}_j|}\frac{F_{ij}}{m_i}\\
\end{equation}

\section{Numerical simulation}\label{num}
In this section, the performance of the proposed TLSPH method with the virtual link is investigated. The numerical results are compared with analytical, numerical, and experimental results available in the literature. Four cases, i.e., the deflection of a beam under impact, the crack propagation in a notched beam, the Kalthoff-Winkler experiment, and different modes of failure in a deep beam are modelled. Overall, the results are found to be in good agreement with the reference results. The present approach does not introduce unphysical numerical fracture or failure in the computation.

\subsection{Rigid-Plastic analysis for perfect beam}\label{sec_fullbeam}
Unconventional particle connectivity is used in the presented method because only the immediately neighbouring particles are considered in the simulation. To assess the influence of this, firstly, the midpoint deflection of an Aluminium beam is modelled. The beam is of 142.24 mm length; the cross-section is 6.35 mm $\times$ 6.35 mm. In the experiment, it is deflected under the impact of a cylindrical projectile of 50 mm length and 14.74 mm diameter \citep{chen2004experimental}. The case is idealised into a 2D problem with unit width. The mass of the projectile is scaled \cite{chakraborty2015prognosis} to keep the mass ratio between the projectile and target the same and to keep the transmitted impulse constant. The material parameters for the beam and projectile are shown in Table \ref{rp_p1_t1}. The stiffness and yield stress of the projectile is much higher than those of the beam. Consequently, the deformation of the projectile is negligible. It behaves almost like a rigid body in this simulation. Therefore, the analytical solution of the deflection at the midpoint of the beam in \cite{liu1987experimental} can be used as a reference solution, which reads

\begin{table}[h!]
\caption{Material parameters for the steel projectile and the aluminium beam.}\label{rp_p1_t1}
\centering
\begin{tabular}{ccccc}
\toprule
~  & $\rho$ (kg/m$^3$) & $E$ (GPa) & $\nu$ & $\sigma_y$ (MPa) \\
\cmidrule{1-5}
Steel projectile & 7850 & 200 & 0.3 & 600 \\
Aluminum beam & 2680 & 68.95 & 0.33 & 277.8 \\
\bottomrule                                                          
\end{tabular}
\end{table}

\begin{figure}[hbtp!]
\centering
\includegraphics[width=0.6\textwidth]{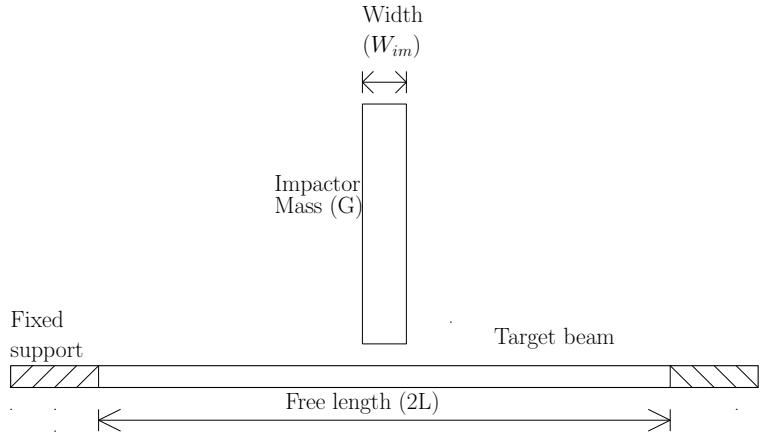}
\caption{Clamped aluminium beam struck by a steel projectile at the mid-span.}\label{bp}
\end{figure}

\begin{equation}\label{wf}
\dfrac{w_f}{H} = \dfrac{1}{2} \left(-1 + \sqrt{1 + \dfrac{G V^2 L/(M_p H)}{1+ L/(2L-L)}}\right)
\end{equation}
where $G$ is the mass of the projectile; $L$ is the distance between the impact location to the support with 2$L$ the length of the free part of the beam; $H$ is the thickness of the beam; $M_p=0.25\sigma_y BH^2$ is the plastic moment of the beam, where $B$ is the width and $\sigma_y$ is the yield stress; and $v_0$ is the initial velocity of the projectile.

\begin{figure}[hbtp!]
\centering
\includegraphics[width=0.5\textwidth]{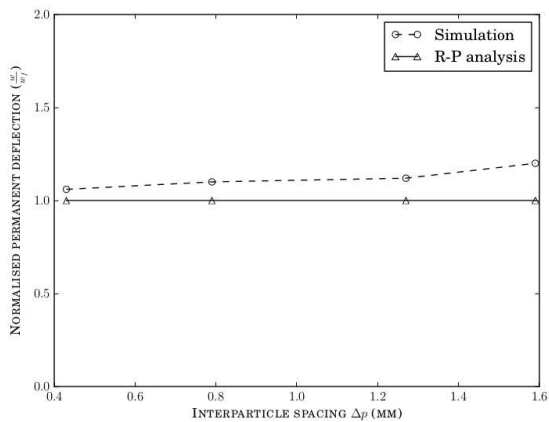}
\caption{Normalised permanent displacement for different discretisation with present formulation ($v_0=20$ m/s). }\label{convergence}
\end{figure}

\begin{figure}[hbtp!]
\centering
\includegraphics[width=0.5\textwidth]{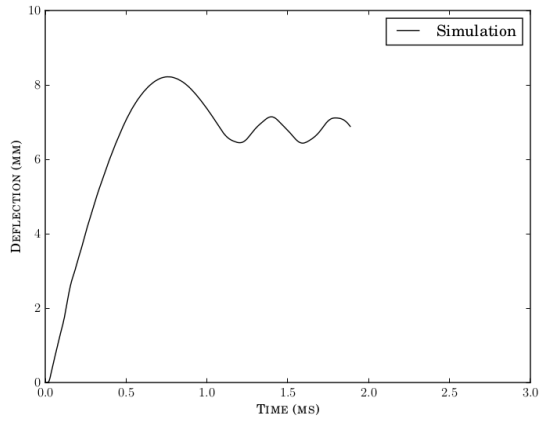}
\caption{Deflection for the midpoint of the aluminium beam ($\Delta p =0.423$ mm, $v_0=20$ m/s)}\label{beam4}
\end{figure}

\begin{figure}[hbtp!]
\centering
\begin{subfigure}[t]{0.4\textwidth}    
\includegraphics[width=\textwidth, trim={10 50 10 300}, clip]{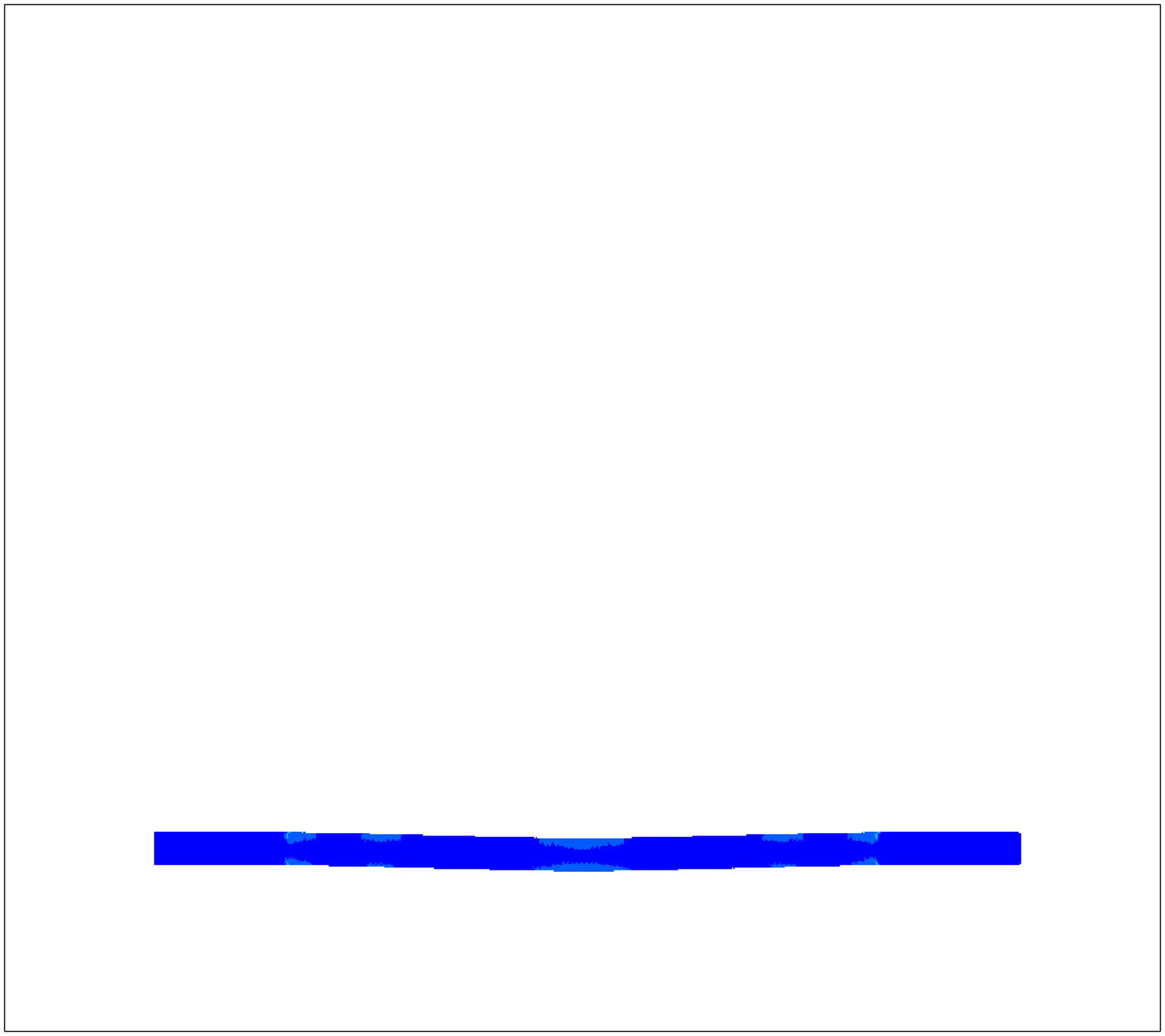}
\caption{Time = $0.1$ ms} 
\end{subfigure}
\begin{subfigure}[t]{0.4\textwidth}
\includegraphics[width=\textwidth, trim={10 50 10 300}, clip]{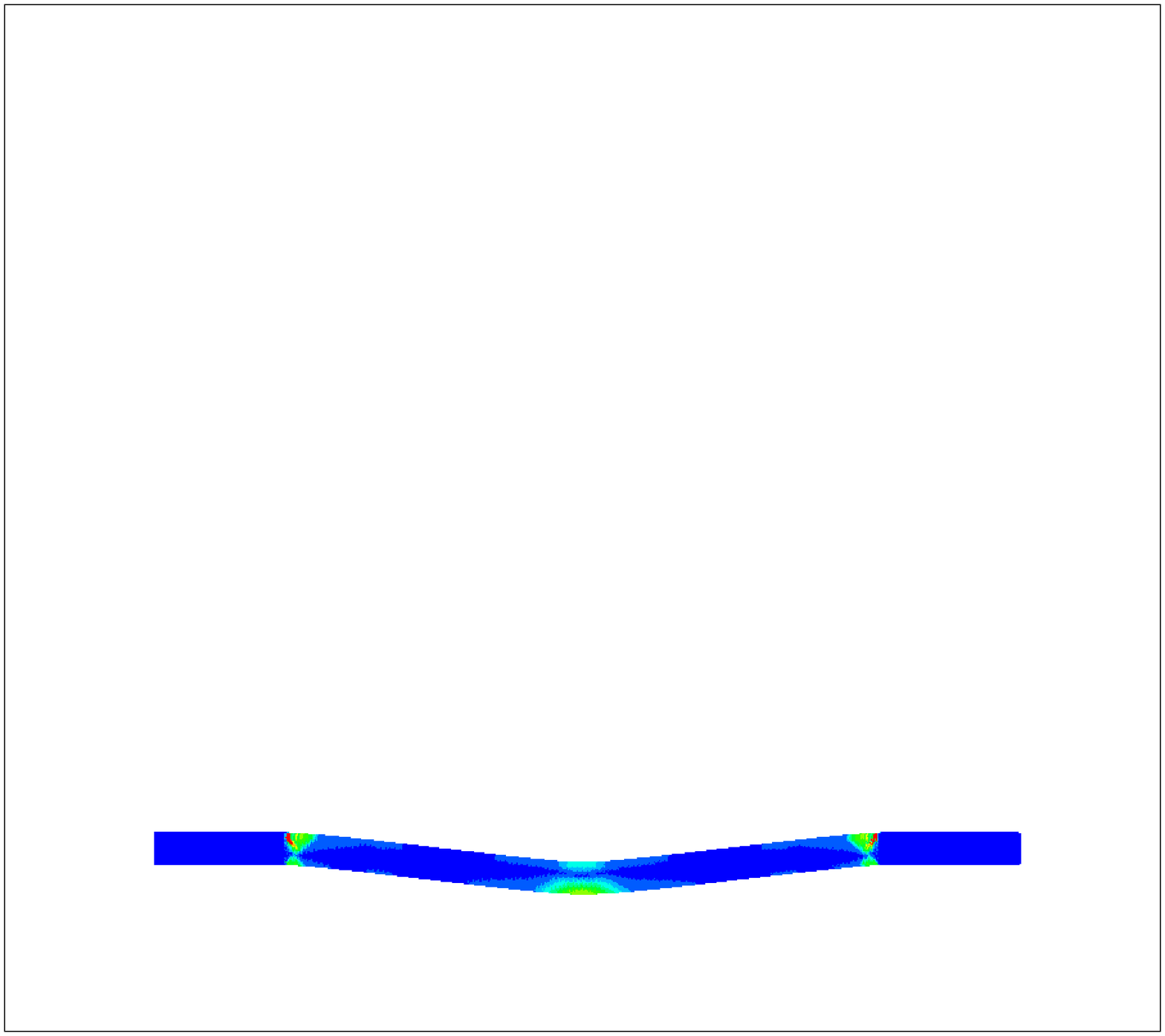}
\caption{Time = $0.5$ ms} 
\end{subfigure}
\begin{subfigure}[t]{0.4\textwidth}
\includegraphics[width=\textwidth, trim={10 50 10 300}, clip]{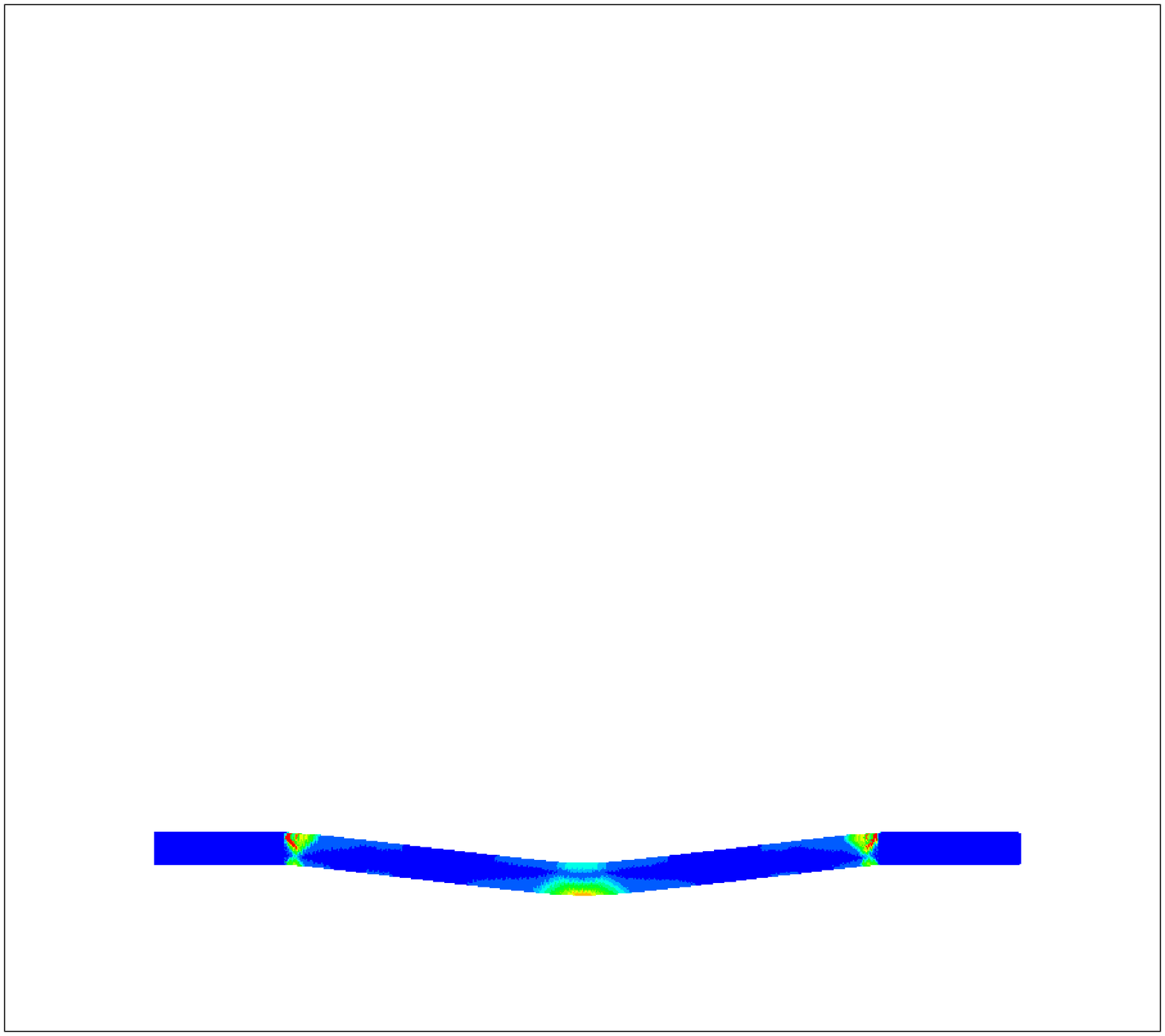}
\caption{Time = $1.0$ ms} 
\end{subfigure}
\begin{subfigure}[t]{0.4\textwidth}
\includegraphics[width=\textwidth, trim={10 50 10 300}, clip]{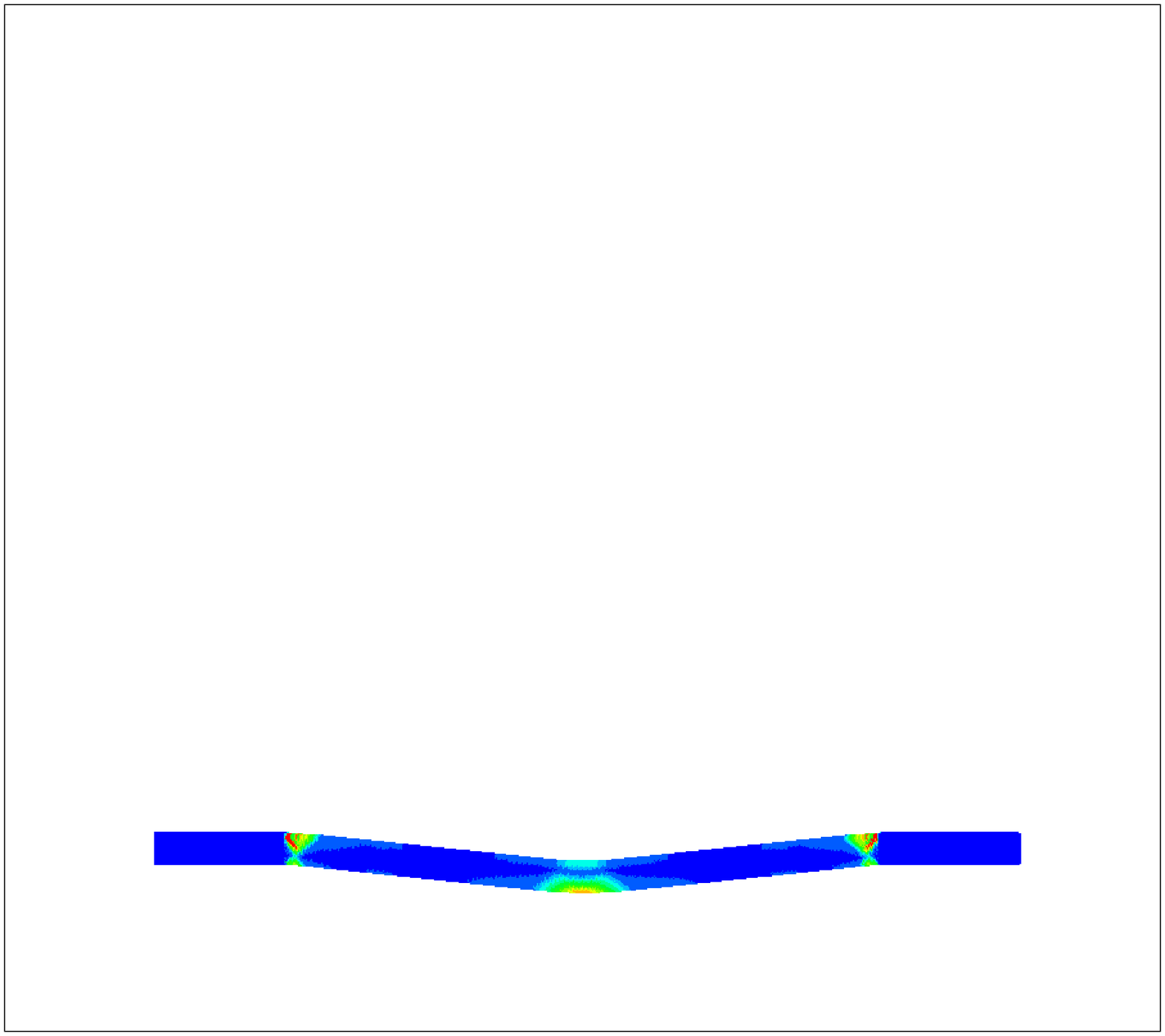}
\caption{Time = $1.5$ ms} 
\end{subfigure}
\begin{subfigure}[t]{0.5\textwidth}
\includegraphics[width=\textwidth, trim={10 300 10 200}, clip]{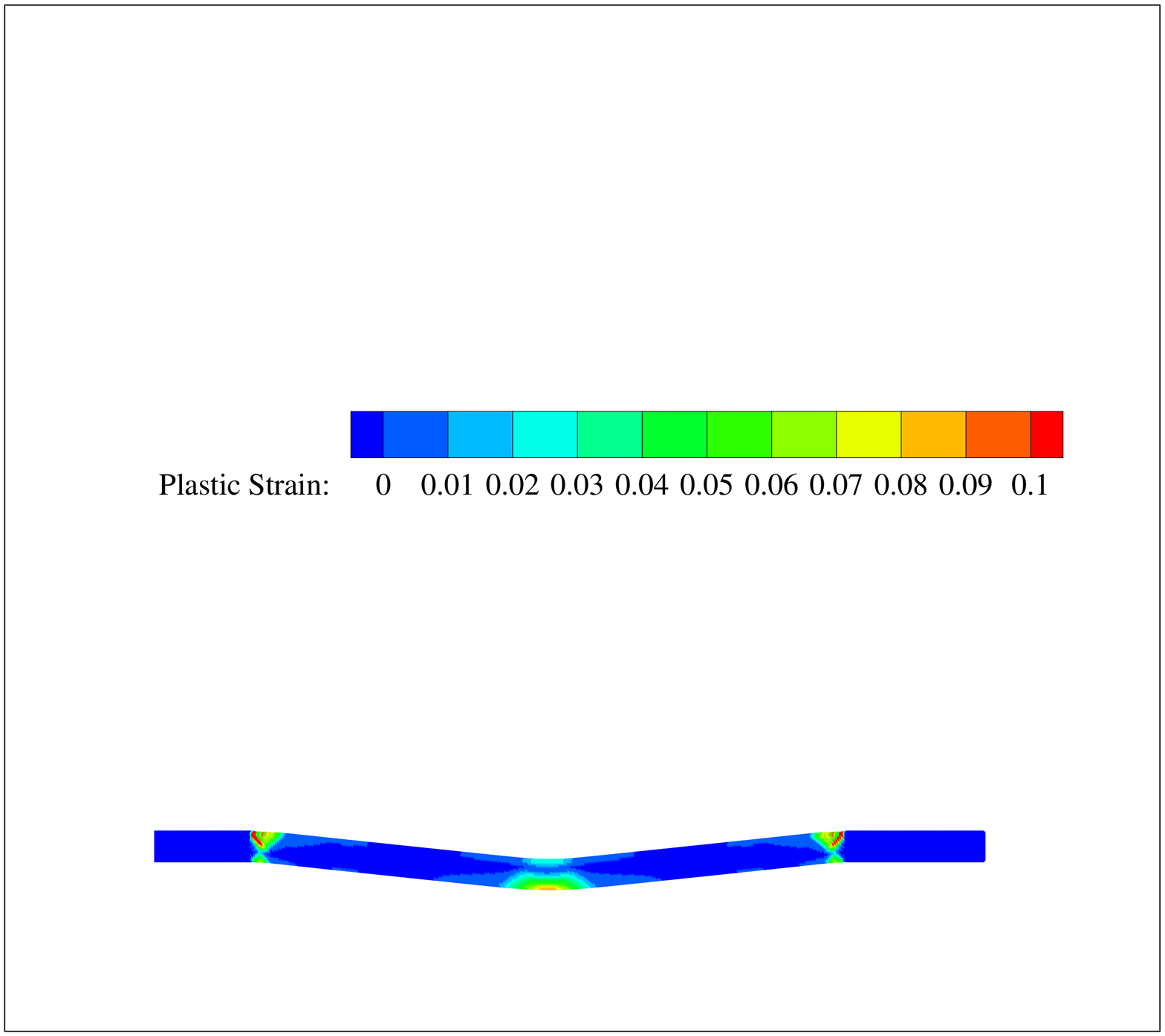} 
\end{subfigure}
\caption{Accumulation of plastic strain at different time step ($\Delta p=0.423$ mm, $v_0=20$ m/s)}\label{pl_perfect}
\end{figure}

Four simulations with different particle discretisation are performed using the presented TLSPH method. Impact velocity of 20 m/s is considered in these four simulations. The midpoint defection of the aluminium beam is compared with the values obtained from the analytical solution, as shown in Figure \ref{convergence}. The present approach demonstrates monotonous convergence. The history of the deflection in the simulation with $\Delta p = 0.423$ mm is shown in Figure~\ref{beam4}. The elastic oscillation of the beam is observed. The accumulated equivalent plastic strain at different locations for the 20 m/s impact velocity case is shown in Figure \ref{pl_perfect}. The deformation pattern and distribution of the plastic strain are well captured using the presented method. 

\begin{figure}[hbtp!]
\centering
\includegraphics[width=0.5\textwidth]{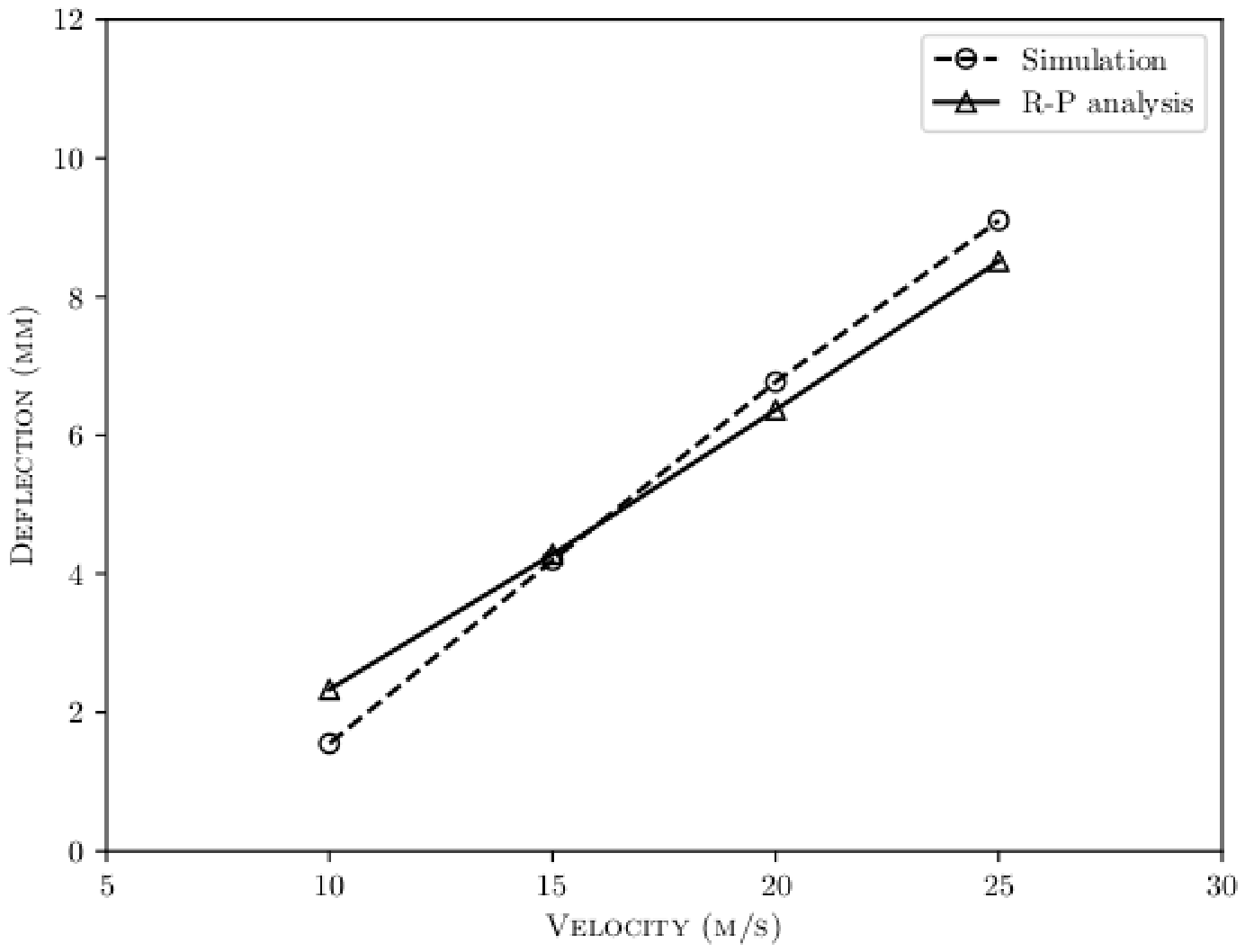}
\caption{Transverse deflection at mid-span for different velocities ($\Delta p =0.423$ mm).}\label{vel_def}
\end{figure}

To measure the effect of impact energy, another four simulations with fixed particle discretisation ($\Delta p=0.423$ mm) but varying impact velocity are carried out. The deflection at the midpoint is shown in Figure \ref{vel_def}. 

Although only immediate neighbouring particles are considered in the simulation, it is found that the numerical results are well corroborated by the analytical solution. Furthermore, good convergence behaviour is observed. This is mainly because of the kernel gradient correction which restores the first-order consistency. Thus, the truncated kernel interaction does not lead to unphysical behaviour. 

\subsection{Crack propagation in notched beam}
\begin{figure}[hbtp!]
\centering
\includegraphics[width=0.6\textwidth]{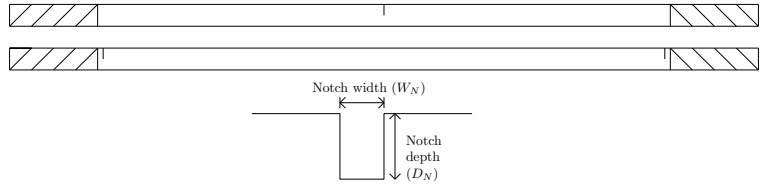}
\caption{Location and geometry of the notch for the clamped aluminium beam}\label{notch_p}
\end{figure}

\begin{figure}[hbtp!]
\centering
\includegraphics[width=0.6\textwidth]{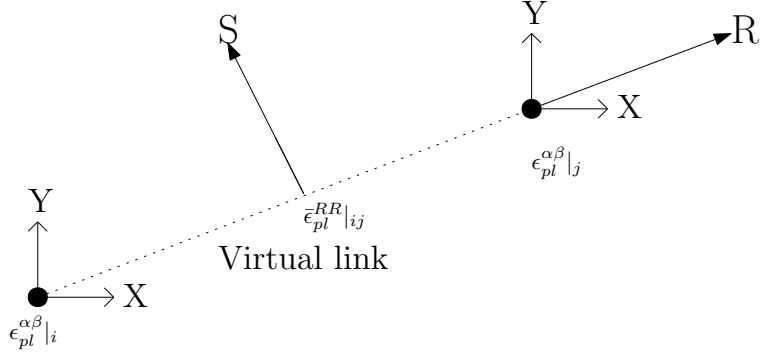}
\caption{Effective plastic strain in the virtual link}\label{vir_bond}
\end{figure}

Chen and Yu \cite{chen2004experimental} analysed the crack propagation and failure of beams with initial notches of different sizes and locations. In this section, several representative beams are selected and modelled using the present approach. The capabilities of the scheme to predict the crack propagation, plastic strain accumulation and failure are tested. The details of the notch are shown in Figure \ref{notch_p}. Three different notch dimensions are used - Type I ($W_N=0.8~mm$; $D_N=2.12~mm$), Type II ($W_N=1.5~mm$; $D_N=2.12~mm$) and Type III ($W_N=0.8~mm$; $D_N=1.59~mm$). The material parameters are the same as those used in Section~\ref{sec_fullbeam} given in Table~\ref{rp_p1_t1}. The damage in the virtual links is calculated based on the accumulated plastic strain, which is obtained by transferring the plastic strain from the particles to the links, as illustrated in Figure \ref{vir_bond}. Firstly, the plastic strain is rotated to the local coordinate systme $R$-$S$ ($R$ and $S$ are the axes that are parallel and perpendicular to the considered virtual link, respectively) \cite{chakraborty2015prognosis}

\begin{equation}
\epsilon_{pl}^{RR}|_i = \bm{T}_R \bm{\epsilon}_{pl}|_i \bm{T}_R^{\mathrm{T}} = \frac{x_{ij}^2 \epsilon_{pl}^{xx}|_i + y_{ij}^2 \epsilon_{pl}^{yy}|_i + 2 x_{ij} y_{ij} \epsilon_{pl}^{xy}|_i}{x_{ij}^2 + y_{ij}^2}
\end{equation}
where $\bm T_R=[-x_{ij}/(x_{ij}^2+y_{ij}^2),~-y_{ij}/(x_{ij}^2+y_{ij}^2]$. Then the value of plastic strain at the virtual link $i$-$j$ ($\bar{\epsilon}_{pl}^{RR}|_{ij}$) in the $R$ direction is calculated as the average value of particle $i$ and $j$ (equation \ref{avg_pl})

\begin{equation}\label{avg_pl}
   \bar{\epsilon}_{pl}^{RR}|_{ij} = \frac{\rho_i C_i \epsilon_{pl}^{RR}|_j + \rho_j C_j \epsilon_{pl}^{RR}|_i}{\rho_i C_i + \rho_j C_j}
\end{equation} 
 
The relation between the damage index $D$ and the plastic strain in the virtual link $\bar{\epsilon}_{pl}^{RR}|_{ij}$ is given in equation \ref{dam_pl}, where $(\epsilon_{pl})_{\mathrm{max}}$ is a material parameter taken as 0.17 in this work \cite{sen2019analytical}. The employed relation indicates a sudden cracking; however, more sophisticated cracking criterion can be used in the current framework. 

\begin{equation}\label{dam_pl}
    D=\begin{cases}
    1,          & \epsilon_{pl} \ge (\epsilon_{pl})_{\mathrm{max}}\\
    0,          &  \mathrm{otherwise}
\end{cases}\\
\end{equation}

\begin{figure}[hbtp!]
\centering
\begin{subfigure}[t]{0.49\textwidth}    
\includegraphics[width=\textwidth]{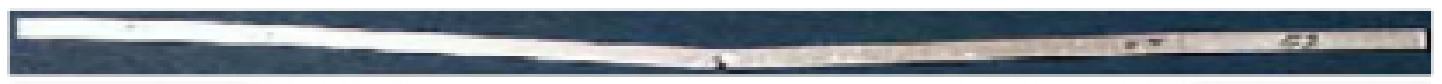}
\caption{14.2 m/s (Experimental)}\label{14_2e}
\end{subfigure}
\begin{subfigure}[t]{0.49\textwidth}
\includegraphics[width=\textwidth, trim={80 230 70 230}, clip]{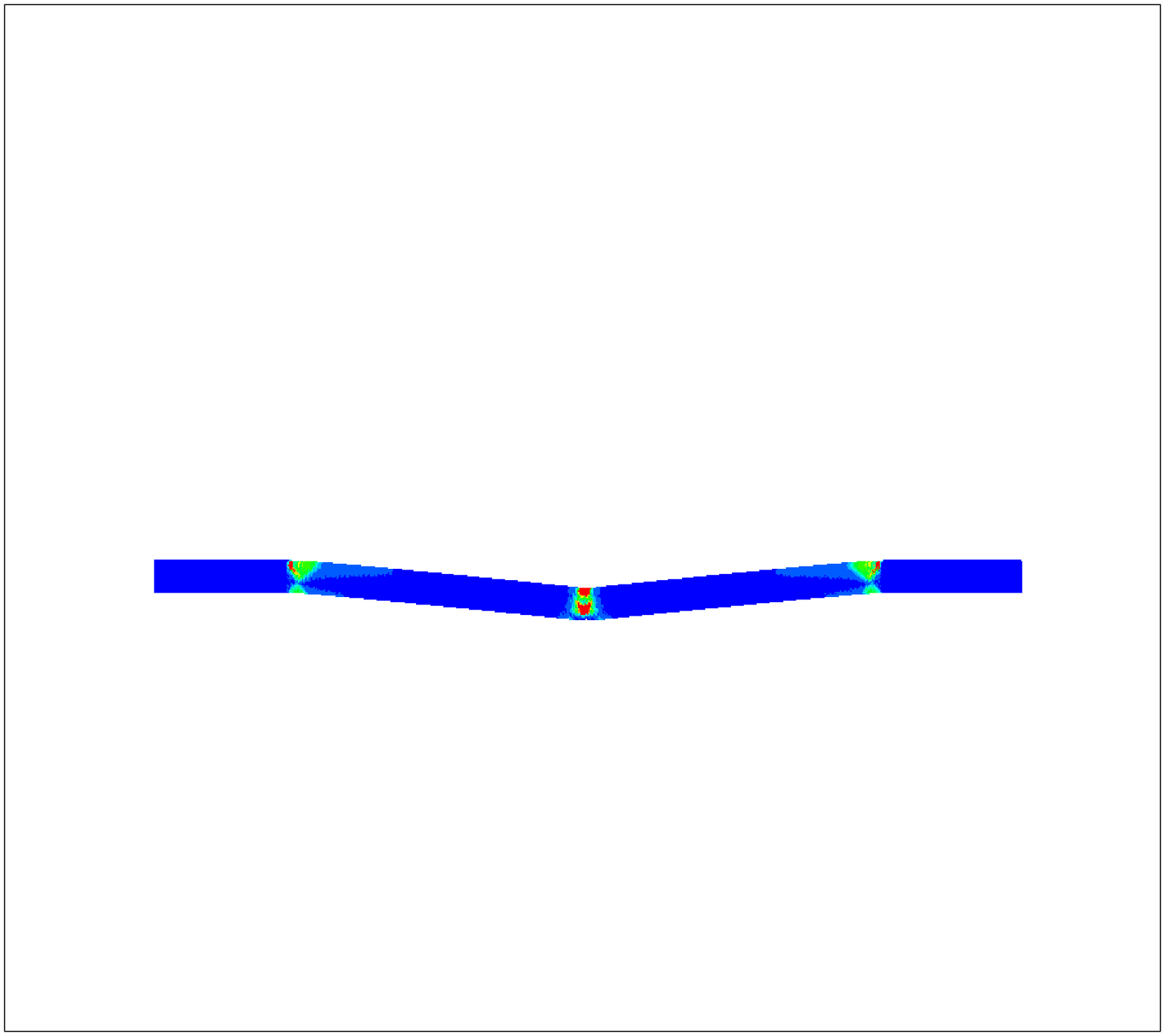}
\caption{14.2 m/s (Present simulation)}\label{14_2n}
\end{subfigure}
\begin{subfigure}[t]{0.49\textwidth}
\includegraphics[width=\textwidth]{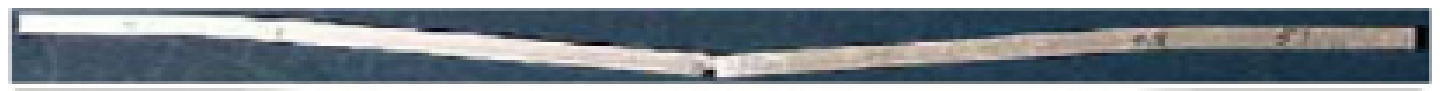}
\caption{18.2 m/s (Experimental)}\label{18_2e}
\end{subfigure}
\begin{subfigure}[t]{0.49\textwidth}
\includegraphics[width=\textwidth, trim={80 220 70 230}, clip]{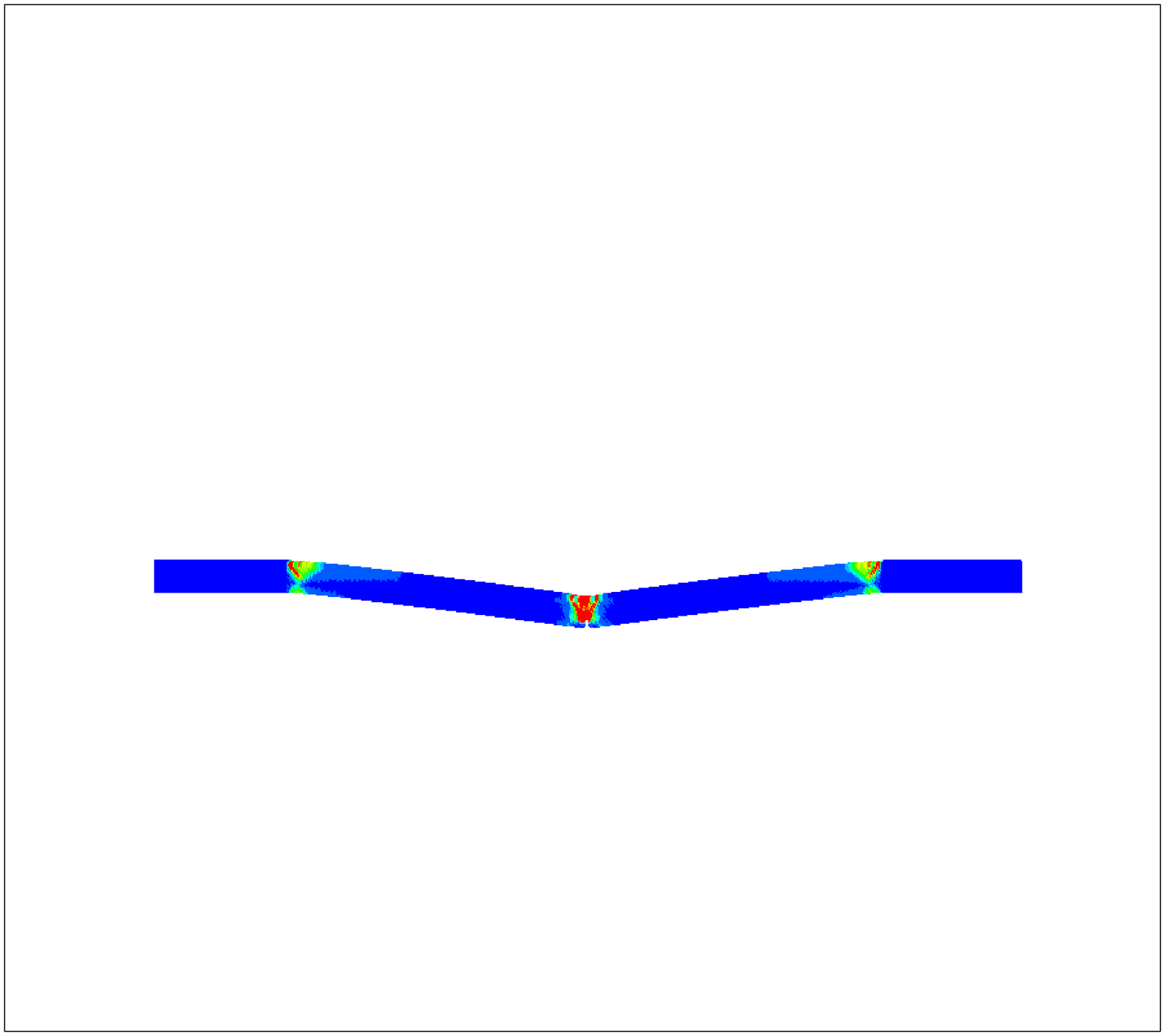}
\caption{18.2 m/s (Present simulation)}\label{18_2n}
\end{subfigure}
\begin{subfigure}[t]{0.49\textwidth}    
\includegraphics[width=\textwidth]{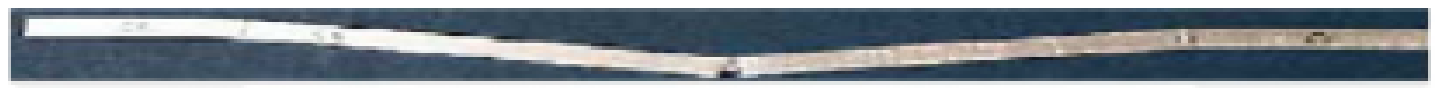}
\caption{19.2 m/s (Experimental)}\label{19_2e}
\end{subfigure}
\begin{subfigure}[t]{0.49\textwidth}
\includegraphics[width=\textwidth, trim={90 220 50 230}, clip]{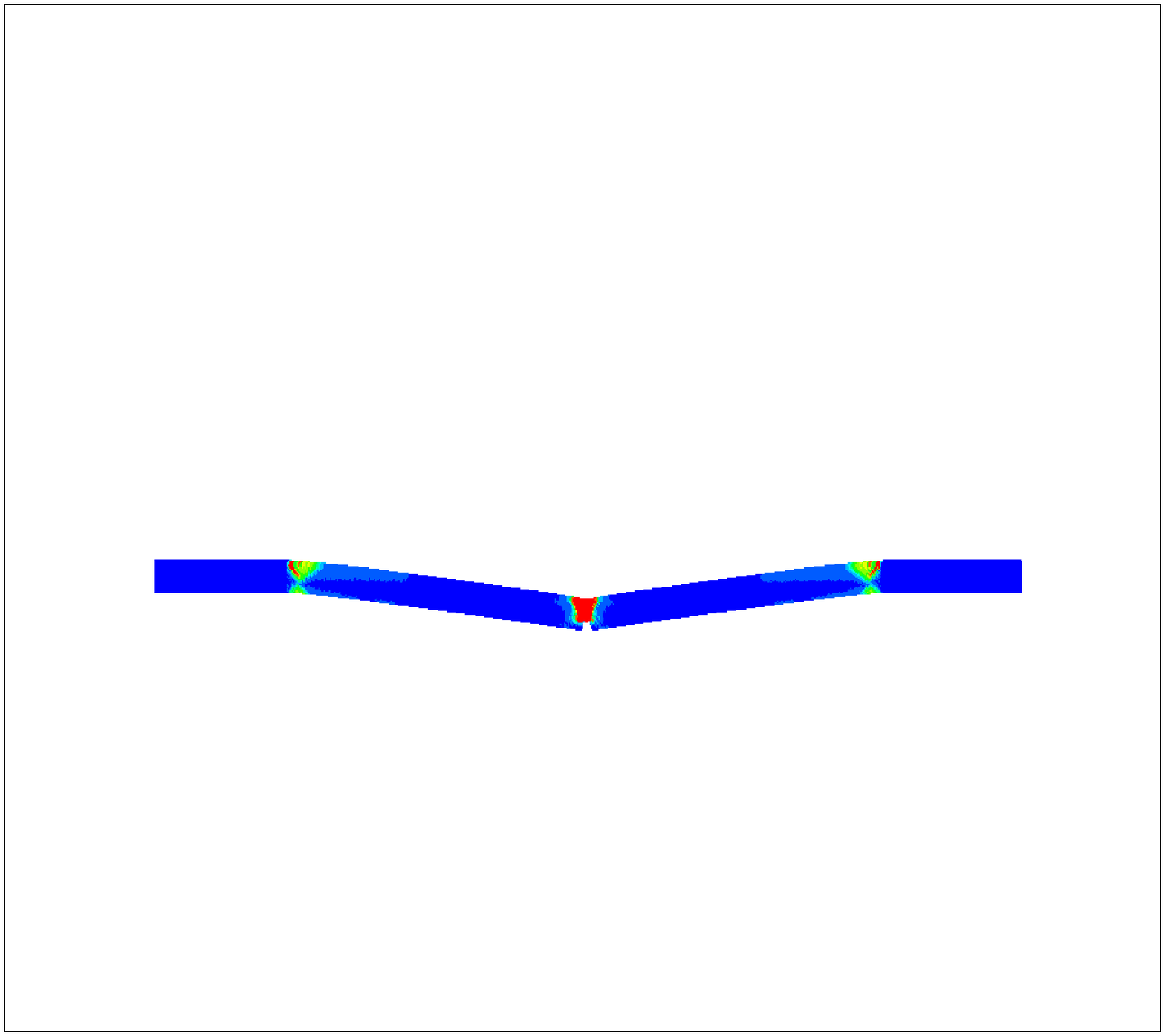}
\caption{19.2 m/s (Present simulation)}\label{19_2n}
\end{subfigure}
\begin{subfigure}[t]{0.49\textwidth}
\includegraphics[width=\textwidth]{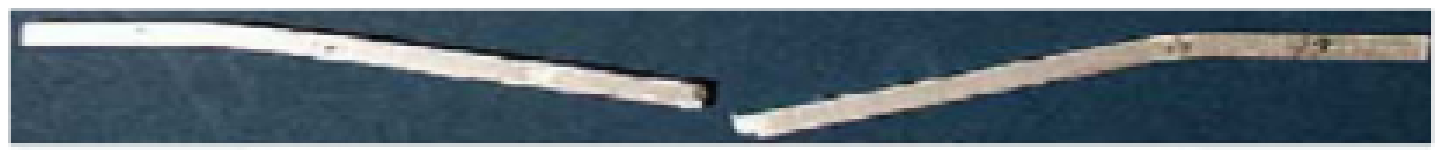}
\caption{27.1 m/s (Experimental)}\label{27_1e}
\end{subfigure}
\begin{subfigure}[t]{0.45\textwidth}
\includegraphics[width=\textwidth, trim={90 200 80 230}, clip]{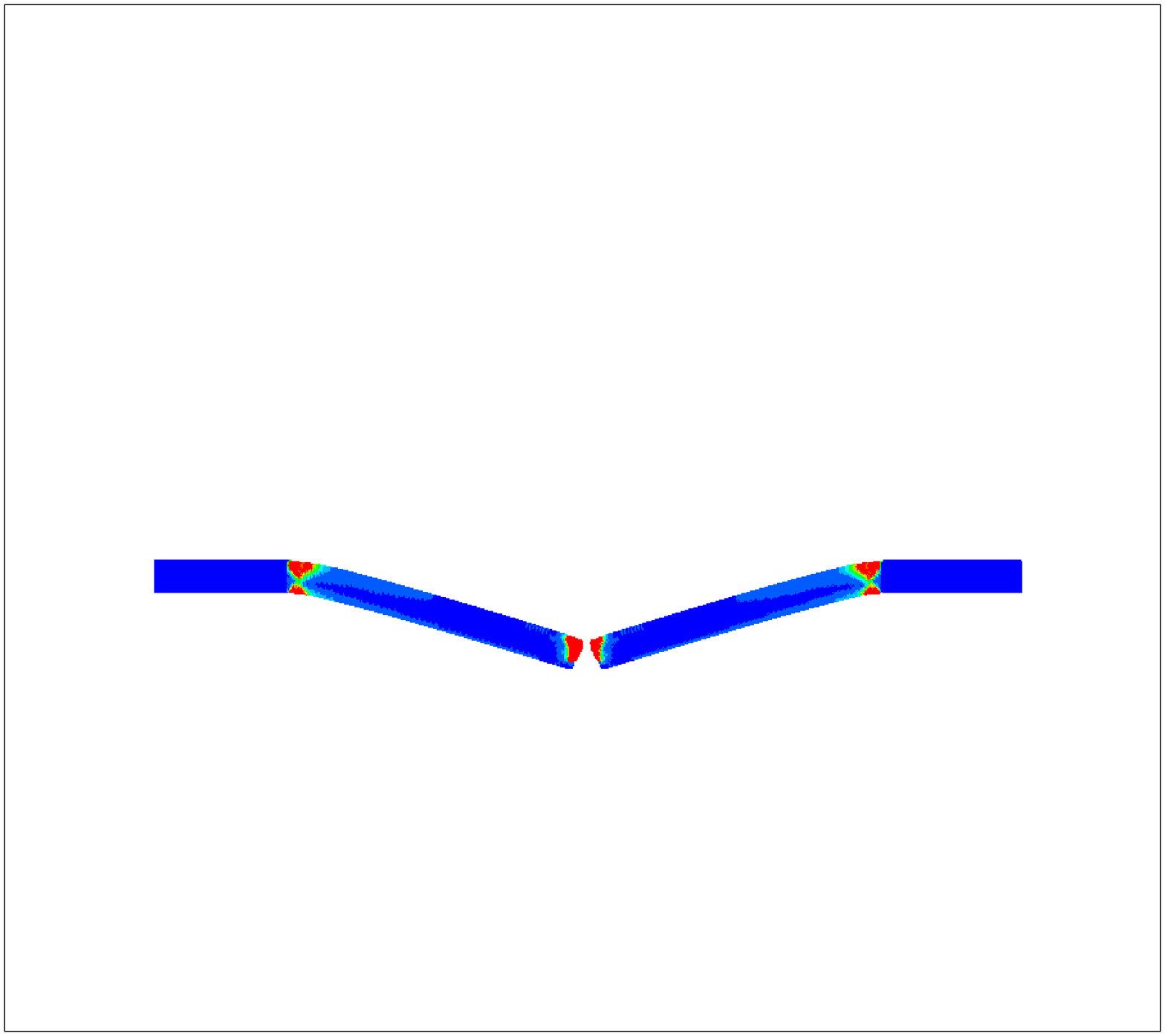}
\caption{27.1 m/s (Present simulation)}\label{27_1n}
\end{subfigure}
\begin{subfigure}[t]{0.5\textwidth}
\includegraphics[width=\textwidth, trim={10 300 10 200}, clip]{leg_pl.eps} 
\end{subfigure}
\caption{Comparison of present simulation with the experimental observation \cite{chen2004experimental}}\label{mid_crack}
\end{figure}

First, the beams with midpoint notches are modelled. The deformation and failure patterns are compared with the experimental results from \cite{chen2004experimental} in Figure \ref{mid_crack}. Large plastic deformation can be observed near the notch and the support. As the notch becomes wider and the impact velocity higher, the deflection increases with a higher plastic deformation near the notch. The cracking initiates at the tip of the notch and propagates deeper into the beam with increasing notch width and impact velocity. With impact velocity $v_0=27.1$ m/s, a complete failure is observed where the left and right parts of the beam separate completely (Figure \ref{27_1e}, \ref{27_1n}). 

All the results from the simulation with a notch at the midpoint are summarised in Table~\ref{notch-mid}. It is found that for different notch types and impact velocities, the overall agreement between the numerical and experimental results are good. The deflection, deformation pattern and cracking propagation are correctly captured using the presented method. Importantly, the present method does not use any enrichment or geometry-based treatment. The crack propagation and material separation are modelled naturally without complex formulation or implementation. 

\begin{table}[hbtp] 
\caption{Summary of the results from simulations with notches at the midpoint}
\label{notch-mid}
\centering
\begin{tabular}{lllll}
\toprule
Notch type & I & I & II & II \\
Initial velocity (m/s) & 14.2 & 18.2 & 19.2 & 27.1 \\
Experimental deflection (mm) & 7.92 & 8.62 & 10.18 & NA \\
Numerical deflection (mm) & 6.29 & 8.65 & 9.01 & NA \\
Experimental observation & LN and CI & LN and just B & LN and just B & B \\
Numerical observation & LN and CI & LN and just B & LN and just B & B \\
\bottomrule               
\end{tabular}
\\
\raggedright{\footnotesize ~~~~~~~~~~~~~~~~~~~~~}\\
\raggedright{\footnotesize ~~~~~~~~~~~~~~~~~~~~~NA: Not apply}\\
\raggedright{\footnotesize ~~~~~~~~~~~~~~~~~~~~~LN: Local necking}\\
\raggedright{\footnotesize ~~~~~~~~~~~~~~~~~~~~~CI: Crack initiated}\\
\raggedright{\footnotesize ~~~~~~~~~~~~~~~~~~~~~B:  Broken completely at notch location}
\end{table}

\begin{figure}[hbtp!]
\centering
\begin{subfigure}[t]{0.49\textwidth}    
\includegraphics[width=\textwidth]{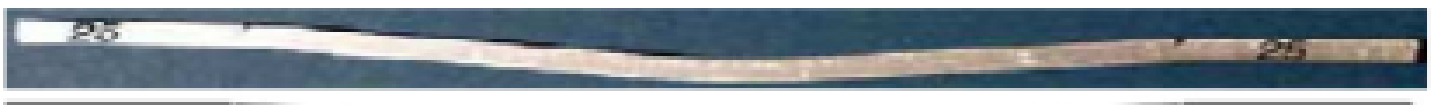}
\caption{18.5 m/s (Experimental)}\label{18_5e}
\end{subfigure}
\begin{subfigure}[t]{0.49\textwidth}
\includegraphics[width=\textwidth, trim={80 200 70 200}, clip]{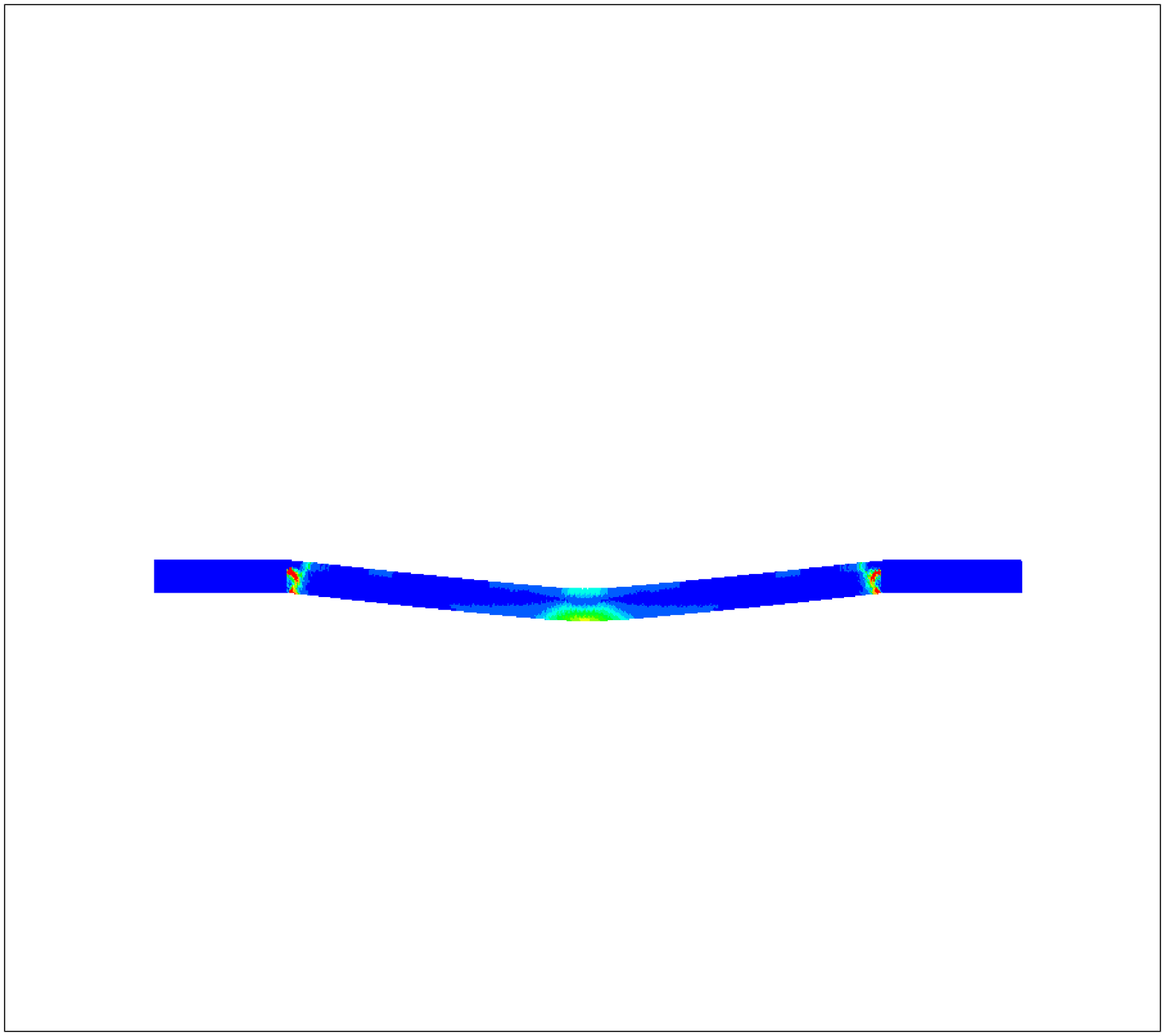}
\caption{18.5 m/s (Present simulation)}\label{18_5n}
\end{subfigure}
\begin{subfigure}[t]{0.49\textwidth}
\includegraphics[width=\textwidth]{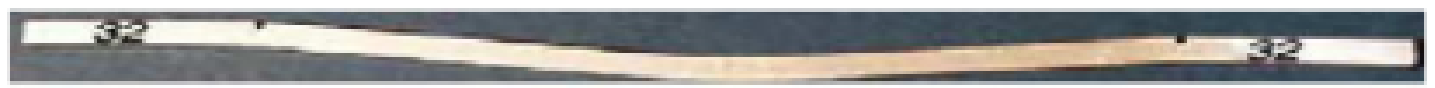}
\caption{17.7 m/s (Experimental)}\label{17_7e}
\end{subfigure}
\begin{subfigure}[t]{0.49\textwidth}
\includegraphics[width=\textwidth, trim={80 200 70 200}, clip]{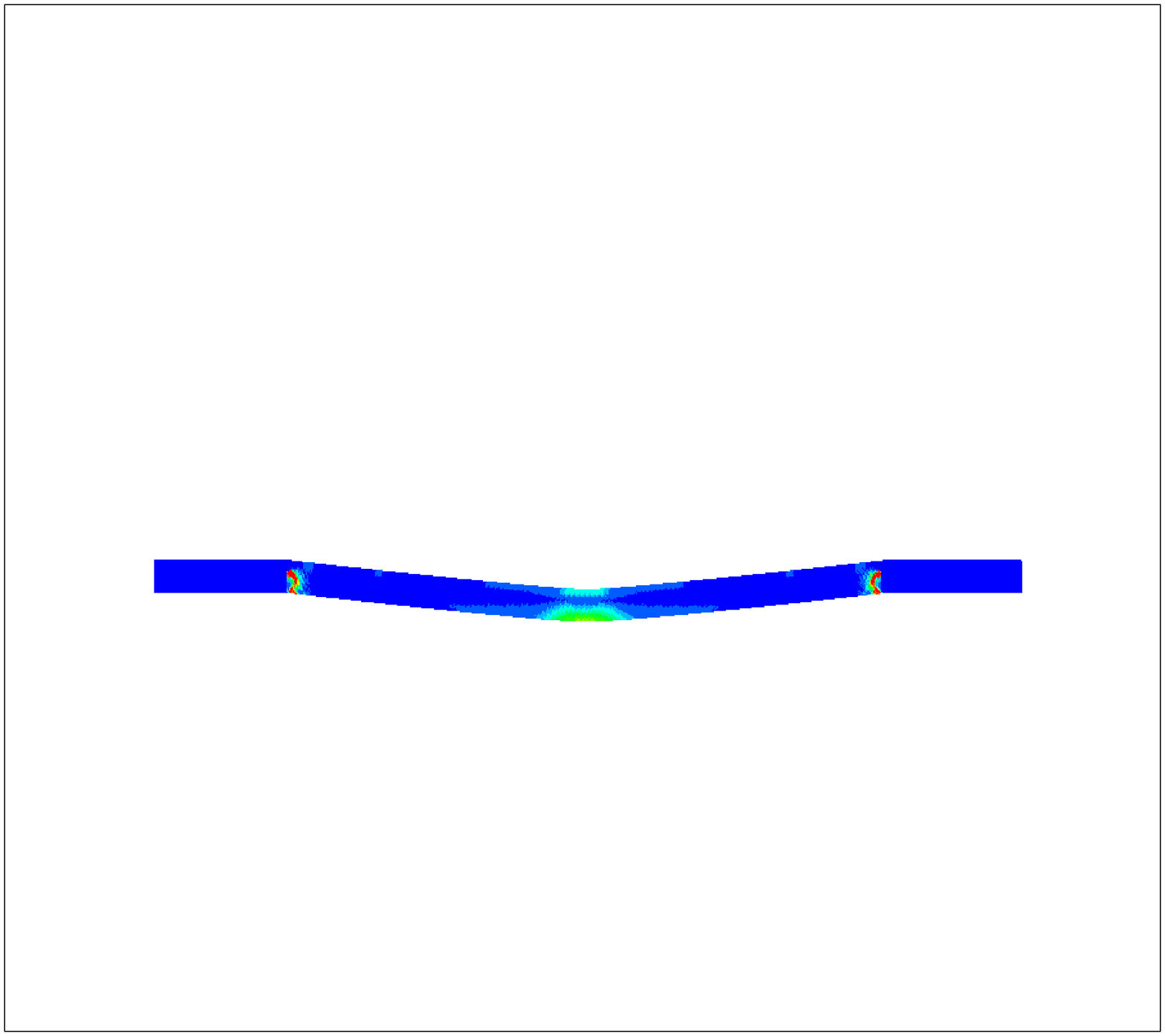}
\caption{17.7 m/s (Present simulation)}\label{17_7n}
\end{subfigure}
\begin{subfigure}[t]{0.49\textwidth}    
\includegraphics[width=\textwidth]{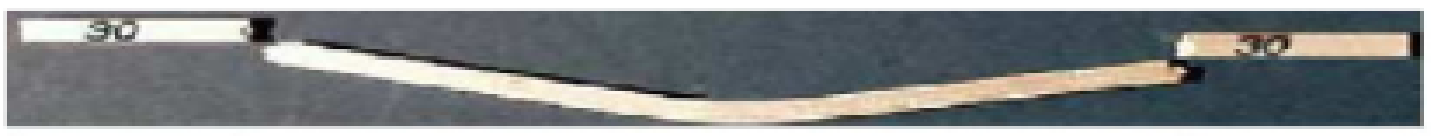}
\caption{31.6 m/s (Experimental)}\label{31_6e}
\end{subfigure}
\begin{subfigure}[t]{0.49\textwidth}
\includegraphics[width=\textwidth, trim={80 205 70 200}, clip]{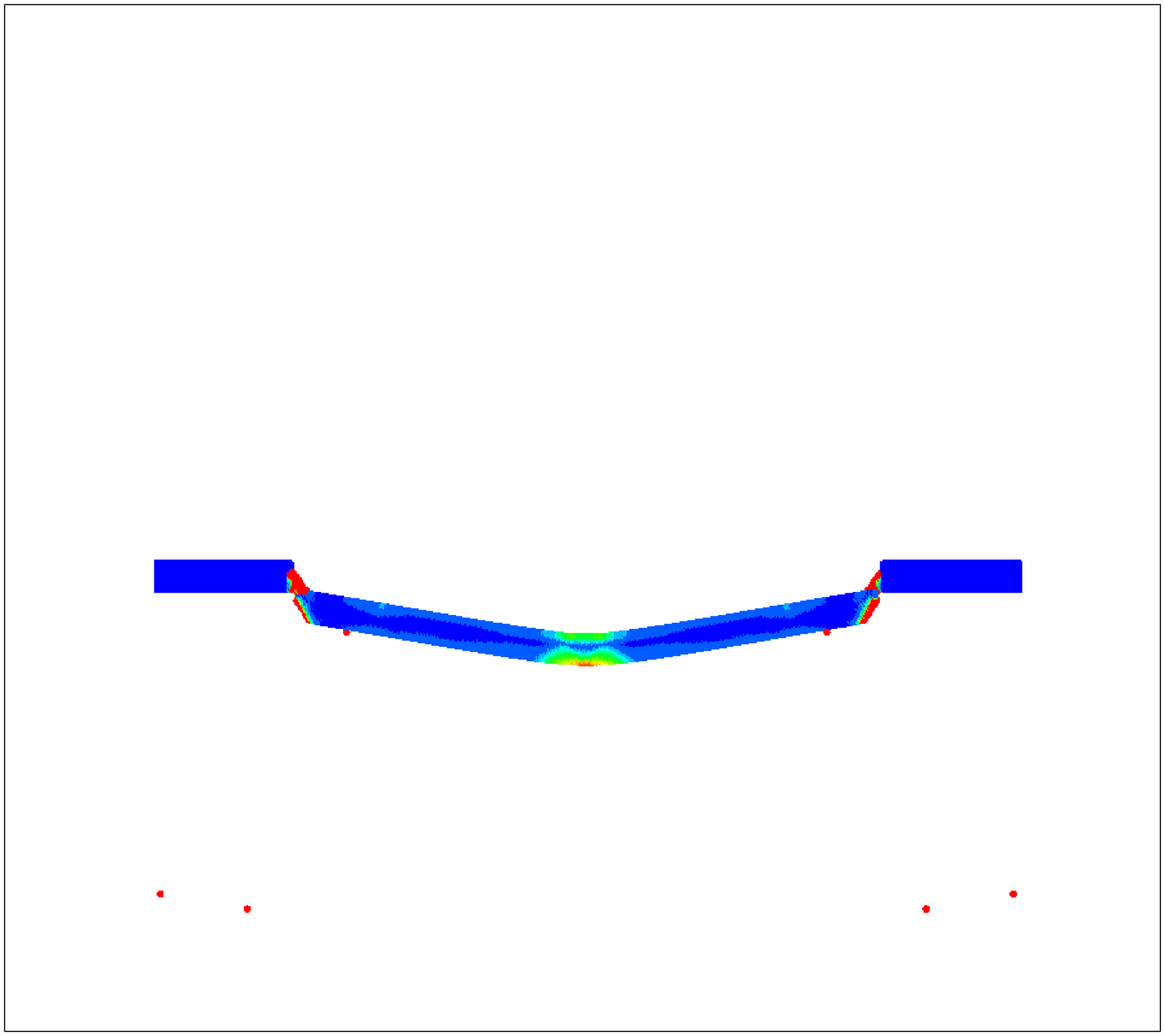}
\caption{31.6 m/s (Present simulation)}\label{31_6n}
\end{subfigure}
\begin{subfigure}[t]{0.7\textwidth}
\includegraphics[width=\textwidth, trim={10 300 10 200}, clip]{leg_pl.eps} 
\end{subfigure}
\caption{Comparison of present simulation with the experimental observation \cite{chen2004experimental}}\label{support_crack}
\end{figure}

Next, the beams with notches at the two supports are considered. Three simulations with type III and I notch and different velocities are performed. The deformation modes are compared with the experimental observation in Figure \ref{support_crack}. The contour of effective plastic strains is also given. Table~\ref{notch-support} summaries the numerical results from all the simulations. The simulation with 31.6 m/s impact velocity shows complete failure near the supports, while other beams in the other two simulations undergo large plastic deformation and initial cracking. These numerical results are consistent with the experimental observations. Therefore, it can be observed that the present formulation performs well in predicting the permanent deflection, crack initiation, plastic strain accumulation, crack propagation and failure.

\begin{table}[hbtp] 
\caption{Summary of the results from simulations with notches at the supports}
\label{notch-support}
\centering
\begin{tabular}{llll}
\toprule
Notch type & III & I & I \\
Initial velocity (m/s) & 18.5 & 17.7 & 31.6 \\
Experimental deflection (mm) & 7.55 & 7.07 & NA \\
Numerical deflection (mm) & 6.72 & 6.70 & NA \\
Experimental observation & PD, LN and just CI & Small PD, LN and CI & B \\
Numerical observation & PD, LN and just CI & Small PD, LN and CI & B \\
\bottomrule               
\end{tabular}
\\
\raggedright{\footnotesize ~~~~~~~~~~~~~~~~~~~~~}\\
\raggedright{\footnotesize ~~~~~~~~~~~~~~NA: Not apply}\\
\raggedright{\footnotesize ~~~~~~~~~~~~~~PD: Plastic deformation}\\
\raggedright{\footnotesize ~~~~~~~~~~~~~~LN: Local necking}\\
\raggedright{\footnotesize ~~~~~~~~~~~~~~CI: Crack initiated}\\
\raggedright{\footnotesize ~~~~~~~~~~~~~~B:  Broken completely at notch location}
\end{table}

\subsection{Kalthoff-Winkler numerical experiments}
\begin{figure}[hbtp!]
\centering
\begin{subfigure}[t]{0.45\textwidth}    
\includegraphics[width=\textwidth]{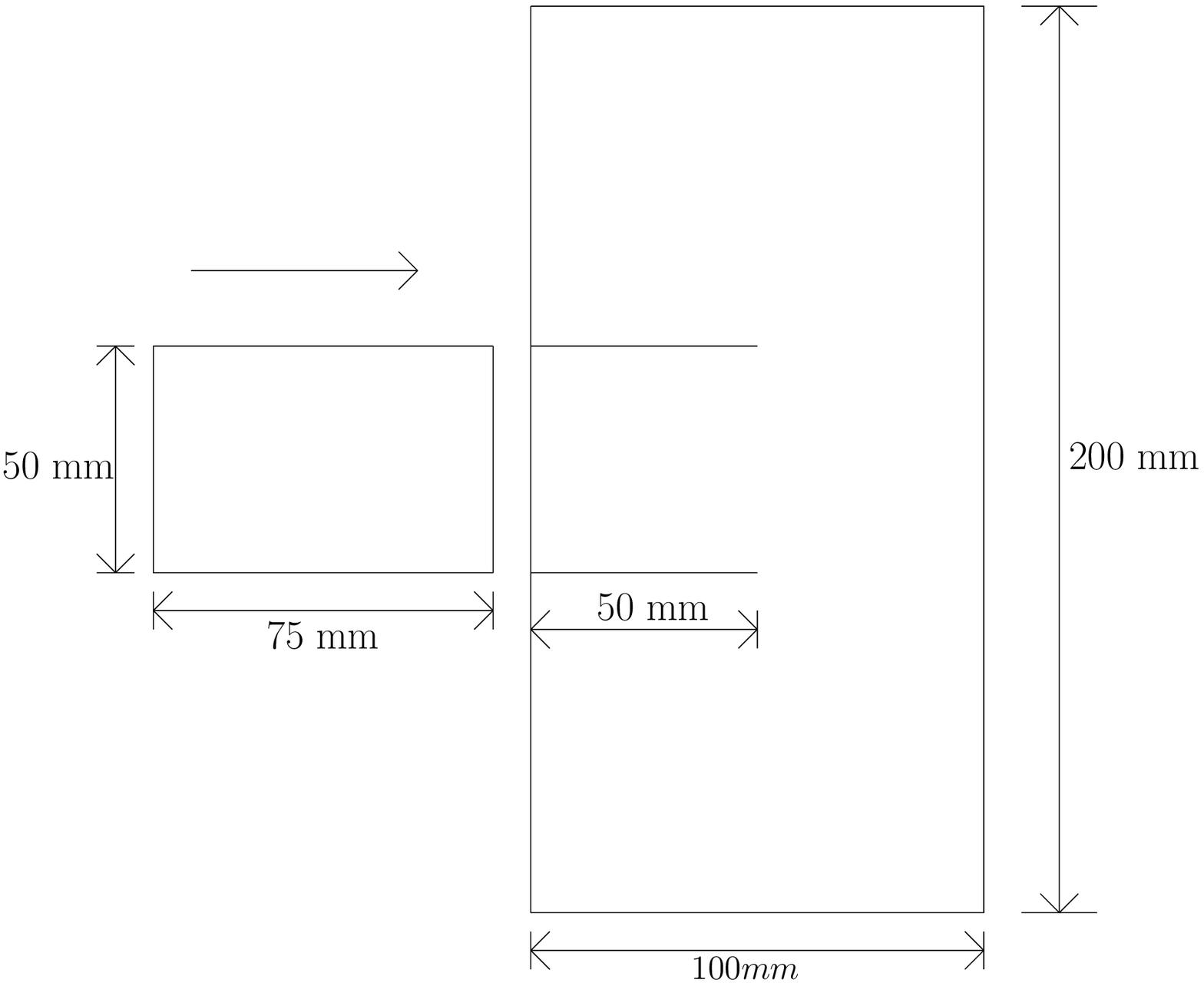}
\caption{}\label{kal1}
\end{subfigure}
\begin{subfigure}[t]{0.45\textwidth}
\includegraphics[width=\textwidth]{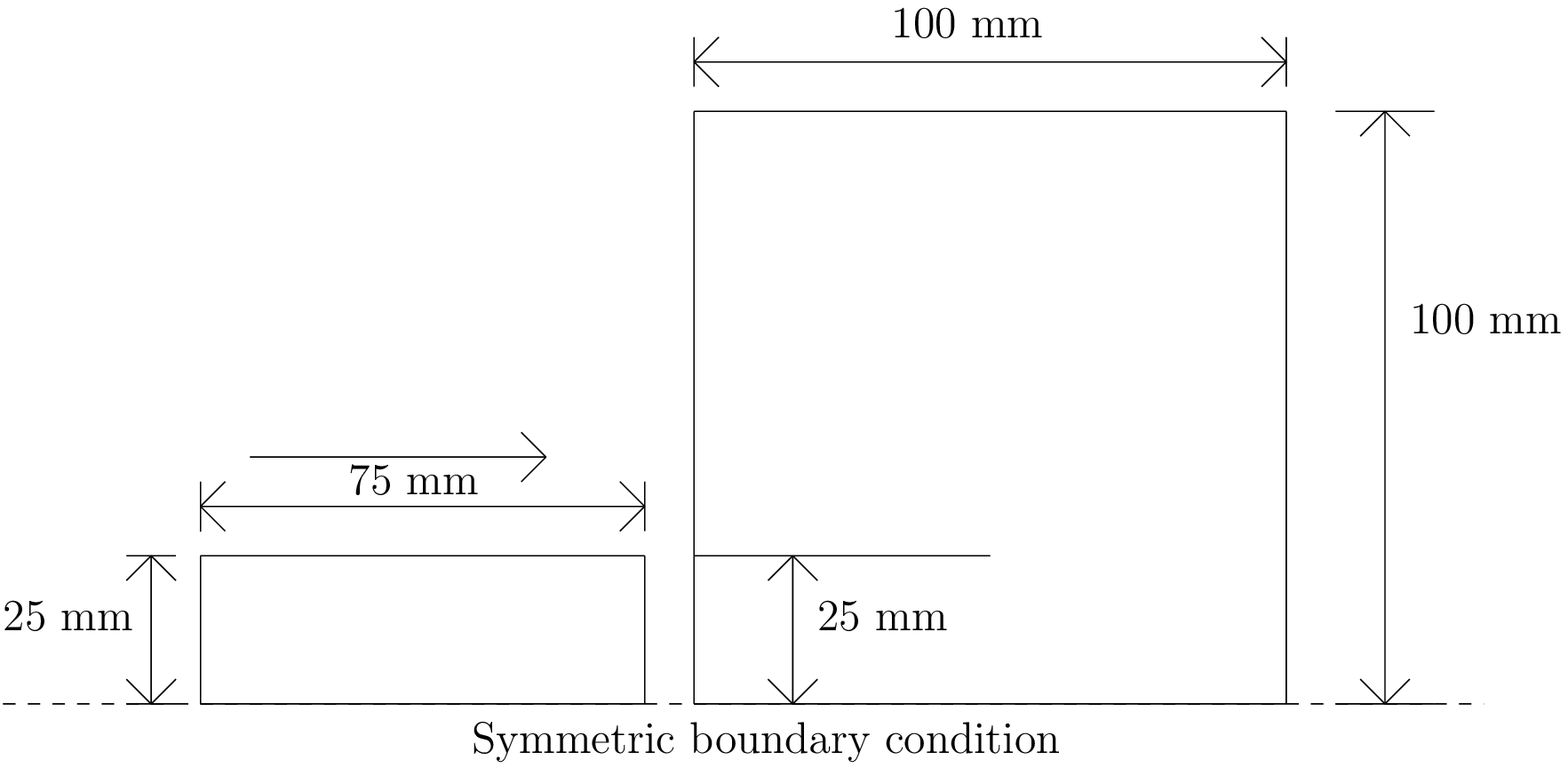}
\caption{}\label{kal2}
\end{subfigure}
\caption{Initial set-up for the Kalthoff-Winkler experiment}\label{kalthoff_setup}
\end{figure}

For the third example, the crack propagation from Kalthoff-Winkler \cite{kalthoff1988failure} is simulated. A double-notched target is subjected to an impact loading, then crack initiates at the tip of the notch and propagates at an angle of $70^\circ$. The geometry and boundary condition of the experiment are shown in Figure~\ref{kal1}. Owing to the symmetricity of the geometry, only half of the specimen is modelled. The symmetric boundary condition is employed, as shown in Figure \ref{kal2}.

The material and computational parameters are shown in Table \ref{kal_mat} and \ref{kal_sph}. The material is assumed to be elastic. The following Rankine criterion is used for the damage evolution
\begin{equation}\label{dam_el}
    D=\begin{cases}
    1,          & \dfrac{r_{ij}|_t -r_{ij}|_{t0}}{r_{ij}|_{t0}} \geq \epsilon_{max}\\
    0,          &  \mathrm{otherwise}
\end{cases}\\
\end{equation}
where $r_{ij}|_t$ and $r_{ij}|_{t0}$ are the interparticle distances between the $i$-th and $j$-th particles at the reference and the current configurations. The criterion indicates a brittle failure once the deformation between two particles is larger than the threshold. Similar cracking criterion is employed in other particle-based methods such as peridynamics \cite{silling2000reformulation} and the smoothed particle Galerkin method (SPG) \cite{wu2018stable}.

\begin{table}[h!]
\caption{Material parameters for crack propagation in the deep beam}\label{kal_mat}
\centering
\begin{tabular}{ll}
\toprule
Mechanical properties  &   Failure parameter       \\
\cmidrule{1-2}
$\rho=8000$ kg/m$^3$     & $\epsilon_{max} = 0.0044$    \\
$E=190$ GPa         & (for Rankine criterion)         \\
$\nu=0.3$     \\  
\bottomrule                                                             
\end{tabular}
\end{table}

\begin{table}[h!]
\caption{Computational parameters used for Kalthoff-Winkler}\label{kal_sph}
\centering
\begin{tabular}{cccc}
\toprule
$\Delta p$ (mm)        & $h$ (mm)            & $\beta_1$ & $\beta_2$\\
\cmidrule{1-4}
0.5               & 0.65      & 0.5 & 0.5\\
\bottomrule
\end{tabular}
\end{table}

\begin{figure}[hbtp!]
\centering
\begin{subfigure}[t]{0.4\textwidth}    
\includegraphics[width=\textwidth, trim={20 20 20 20}, clip]{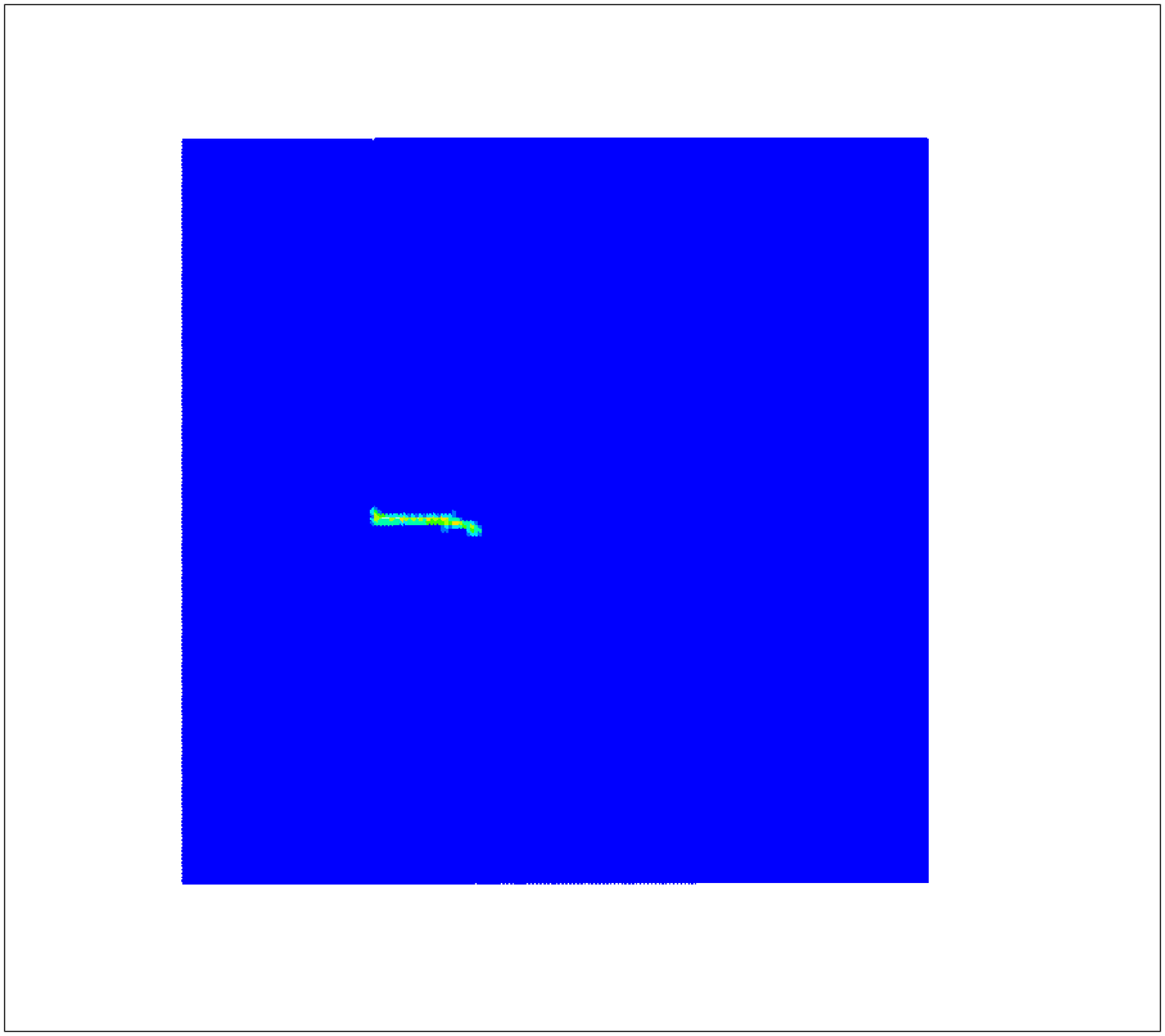}
\caption{Time = 55.0 $\mu$s}
\end{subfigure}
\begin{subfigure}[t]{0.4\textwidth}
\includegraphics[width=\textwidth, trim={20 20 20 20}, clip]{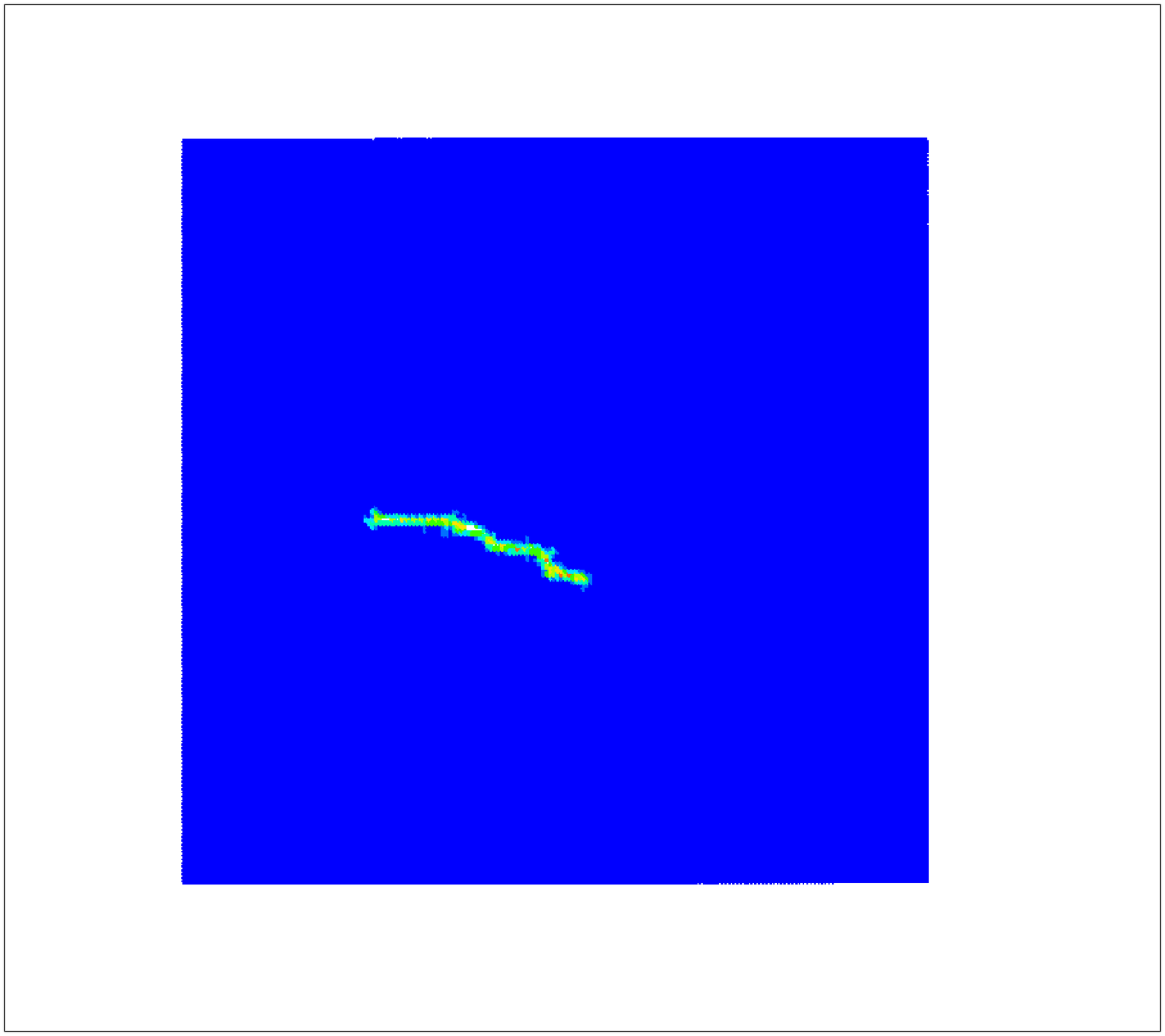}
\caption{Time = 70.0 $\mu$s}
\end{subfigure}
\begin{subfigure}[t]{0.4\textwidth}
\includegraphics[width=\textwidth, trim={20 20 20 20}, clip]{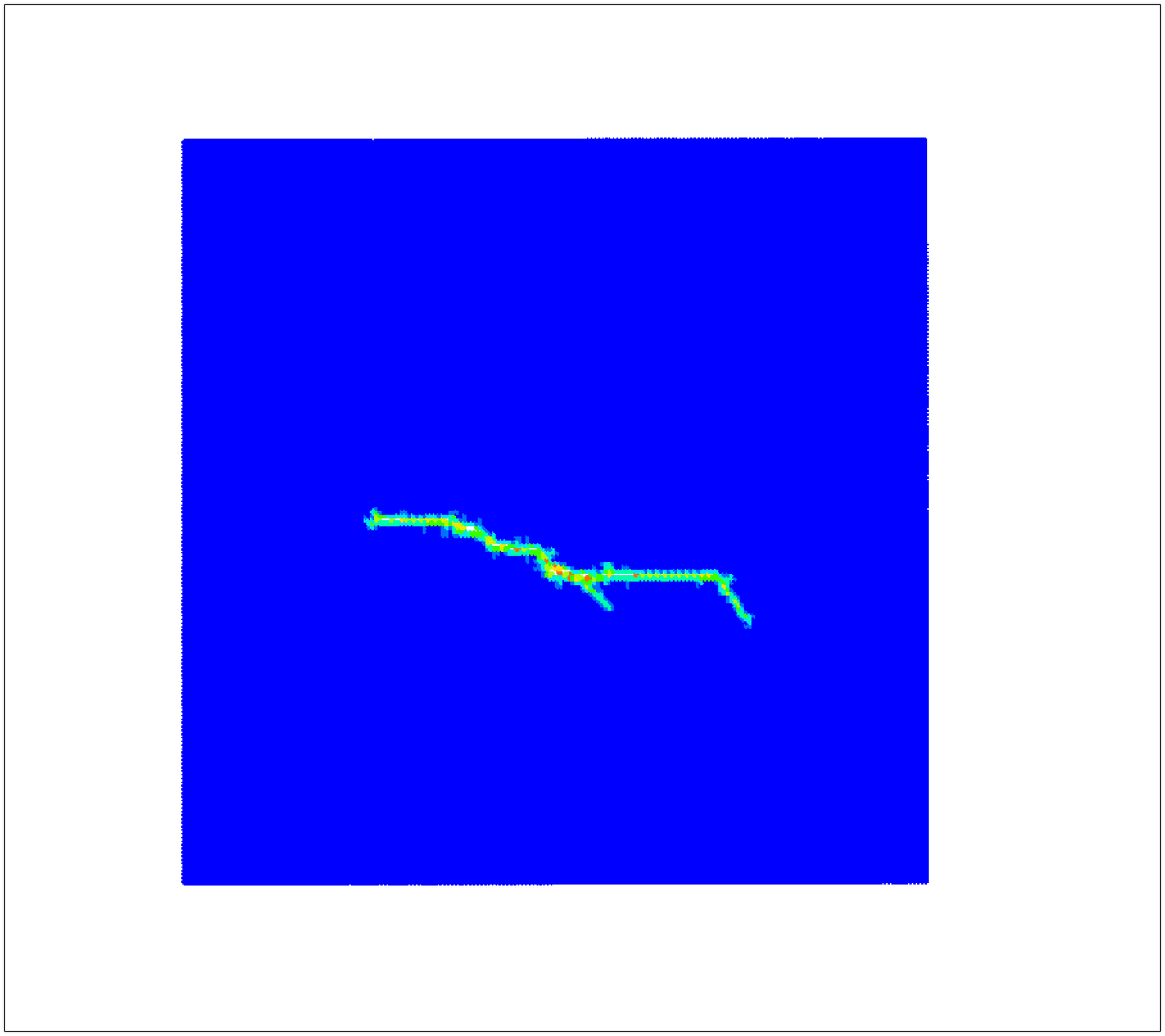}
\caption{Time = 90.0 $\mu$s}
\end{subfigure}
\begin{subfigure}[t]{0.4\textwidth}
\includegraphics[width=\textwidth, trim={20 20 20 20}, clip]{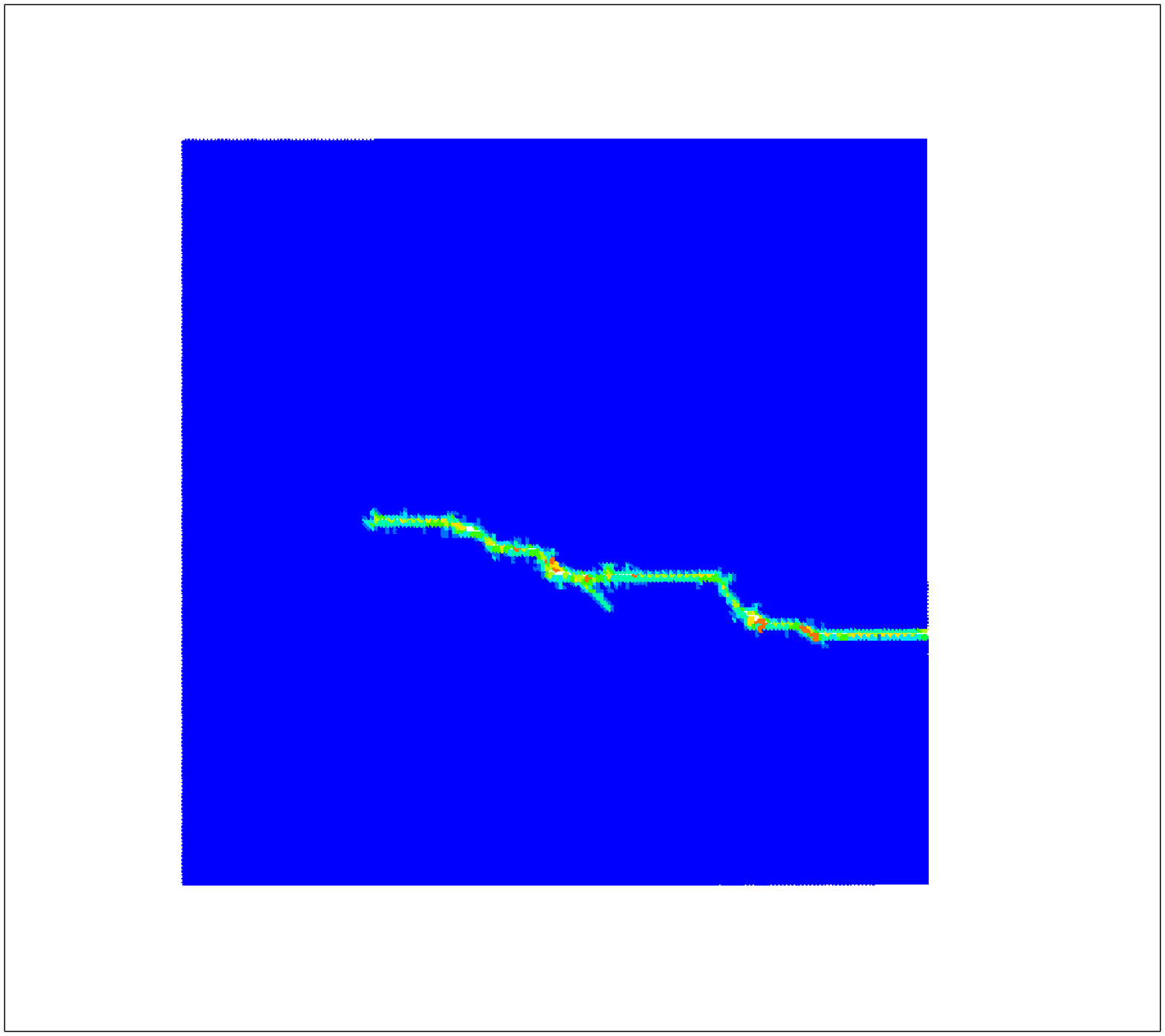}
\caption{Time = 115.0 $\mu$s}
\end{subfigure}
\begin{subfigure}[t]{0.5\textwidth}
\includegraphics[width=\textwidth, trim={25 250 50 200}, clip]{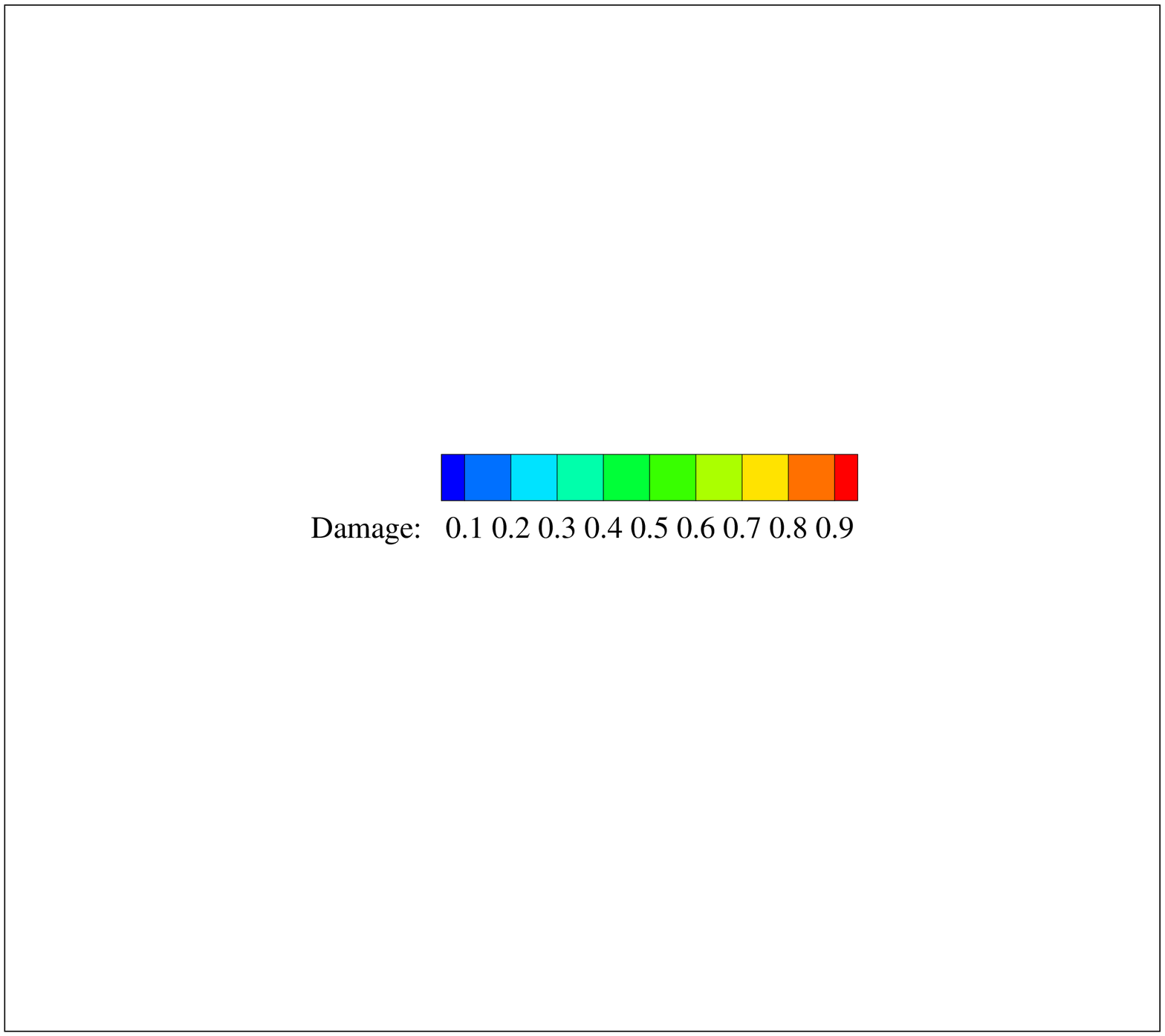} 
\end{subfigure}
\caption{Damage evolution for the Kalthoff-Winkler experiment at different time step}\label{kalthoff_crack}
\end{figure}

The dynamic crack propagation at different time step is shown in Figure \ref{kalthoff_crack}. Under the impact loading, a crack initiates at the tip of the notch and propagates to the edge of the boundary. The angle of the whole cracking path is approximately $77^\circ$, close to the cracking angle obtained in the test and other numerical simulations. The averaged damage index $D$ is also shown in Figure \ref{kalthoff_crack}, which is obtained as $D_i=(\sum_j D_{ij})/n$, where $D_{ij}$ is the damage index of the virtual link between particle $i$ and $j$, and $n$ is the number of particles in the modified support domain of particle $i$.

\begin{figure}[hbtp!]
\centering
\begin{subfigure}[t]{0.38\textwidth}    
\includegraphics[width=\textwidth, trim={0 0 0 0}, clip]{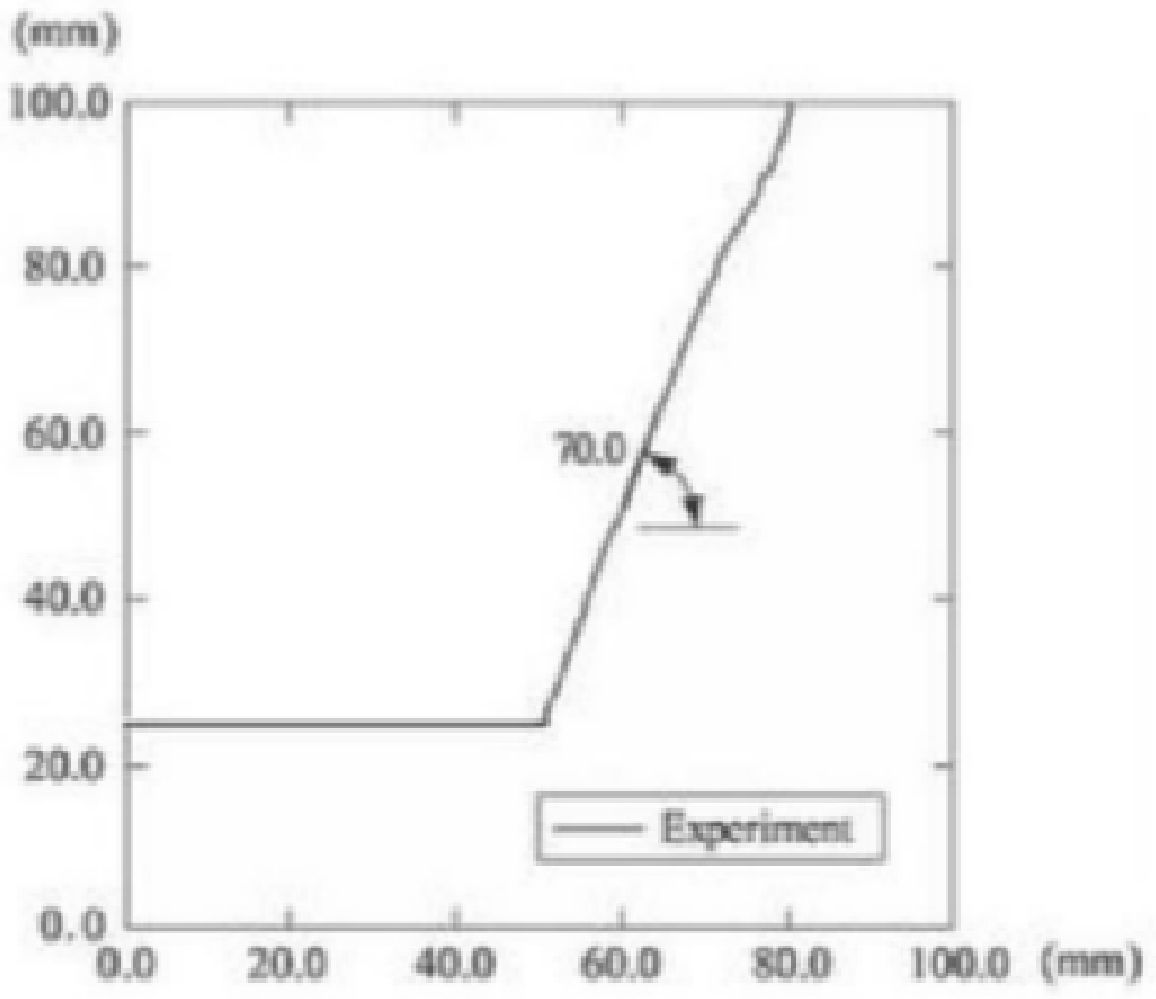}
\caption{Kalthoff-Winkler \cite{kalthoff1988failure}}\label{e1}
\end{subfigure}
\begin{subfigure}[t]{0.33\textwidth}
\includegraphics[width=\textwidth, trim={0 0 0 0}, clip]{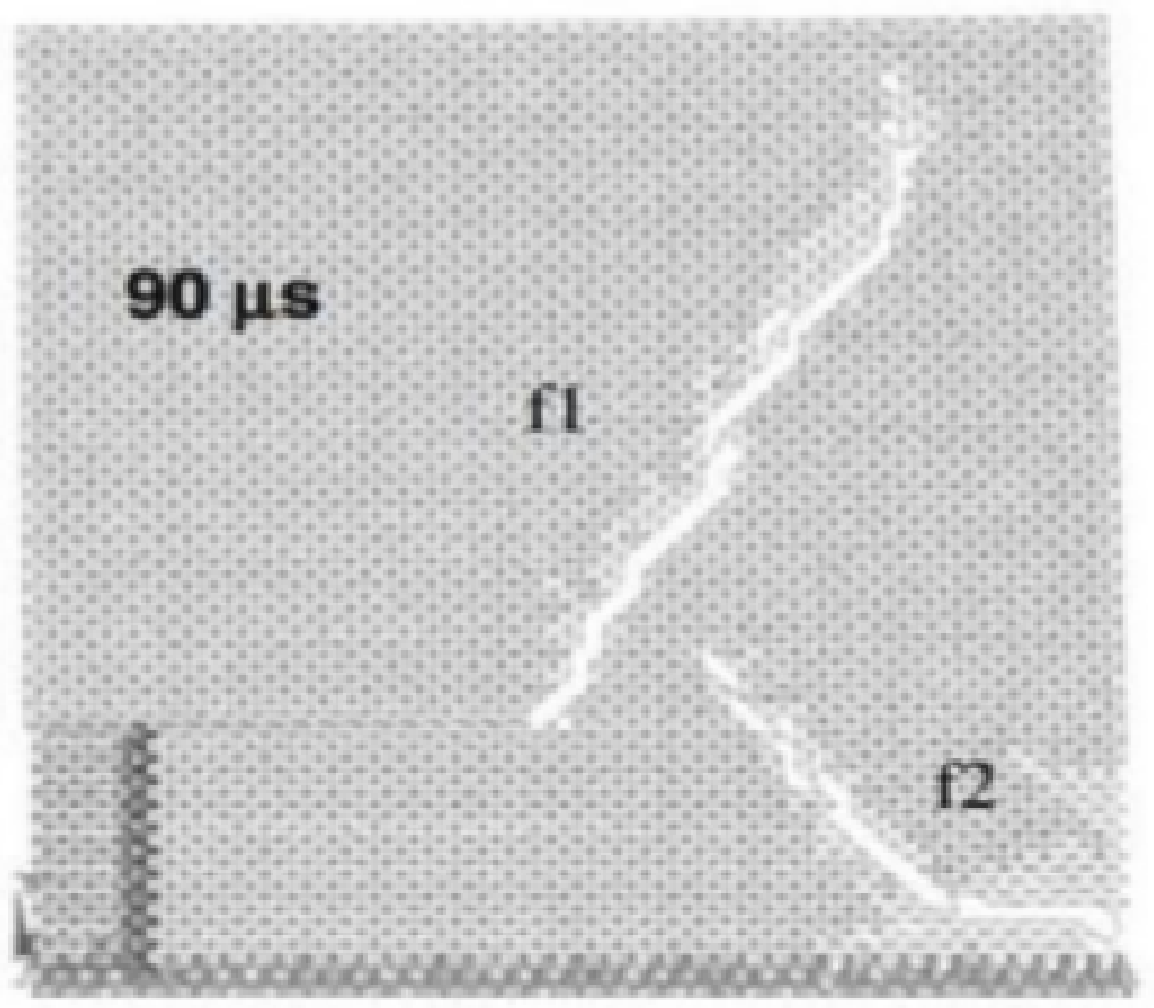}
\caption{Kosteski et al. \cite{kosteski2012crack}}\label{f1}
\end{subfigure}
\begin{subfigure}[t]{0.33\textwidth}
\includegraphics[width=\textwidth, trim={0 0 0 0}, clip]{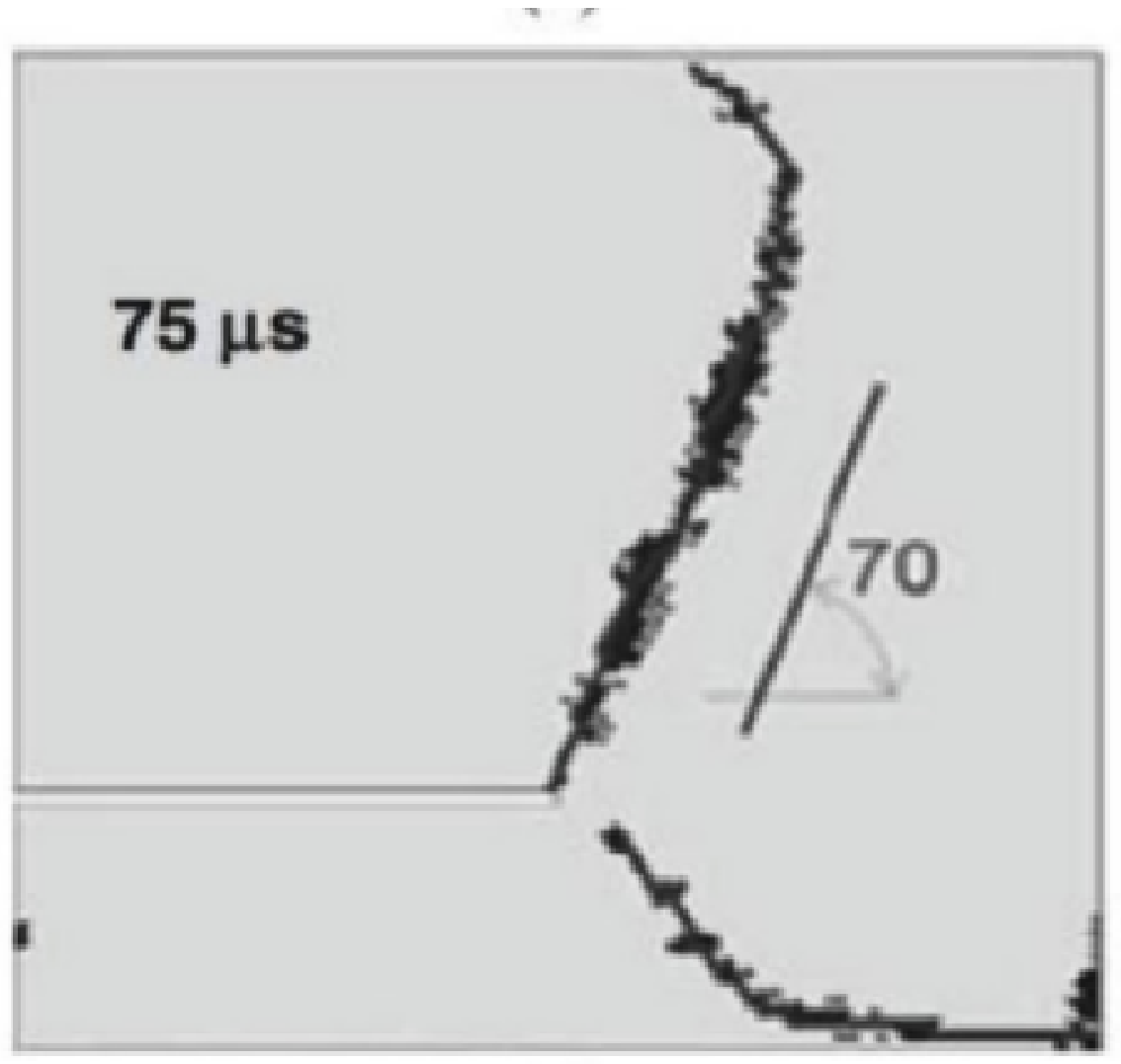}
\caption{Dipasquale et al. \cite{dipasquale2014crack}}\label{f2}
\end{subfigure}
\begin{subfigure}[t]{0.33\textwidth}
\includegraphics[width=\textwidth, trim={0 0 0 0}, clip]{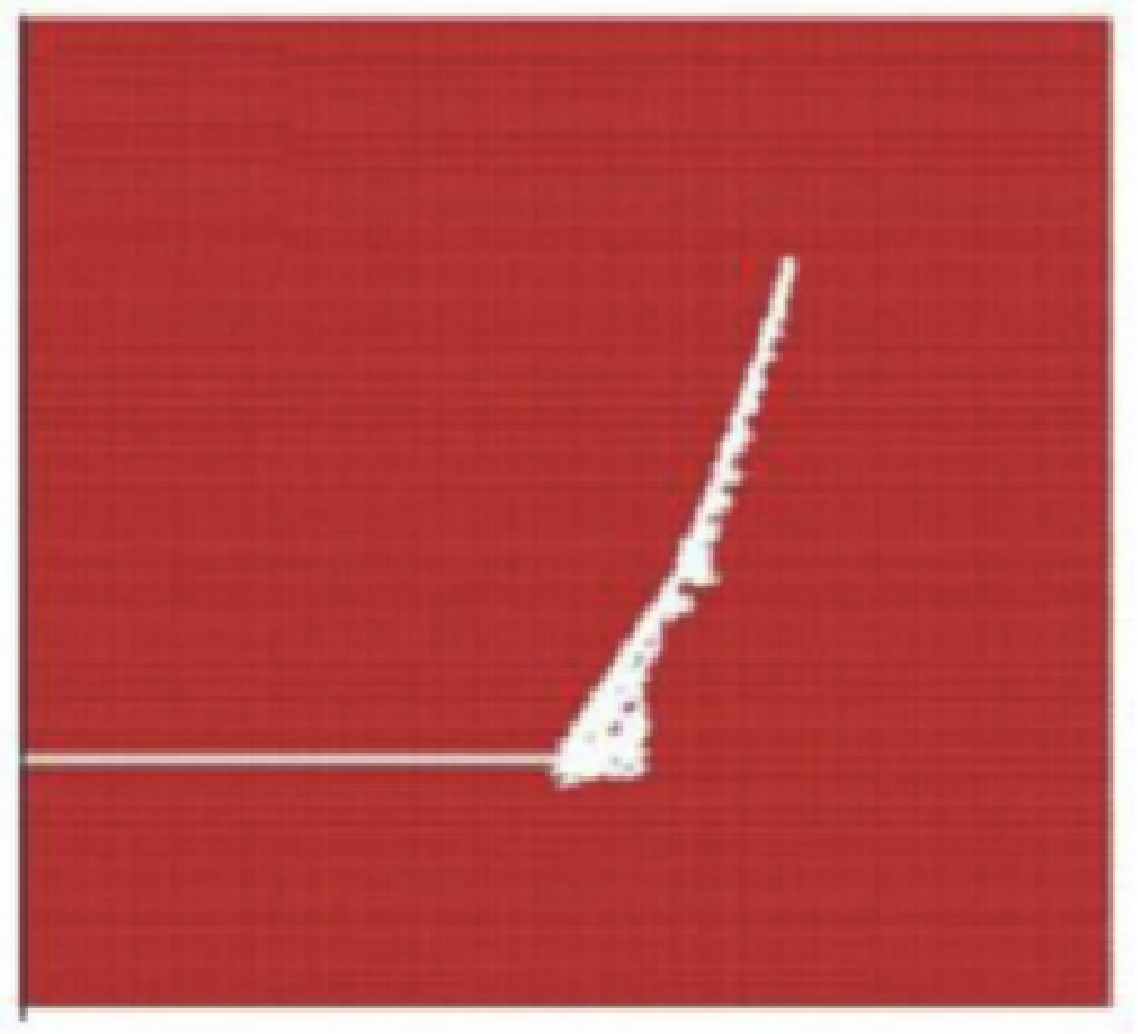}
\caption{Huespe et al. \cite{huespe2006strong}}
\end{subfigure}
\begin{subfigure}[t]{0.33\textwidth}
\includegraphics[width=\textwidth, trim={0 0 0 0}, clip]{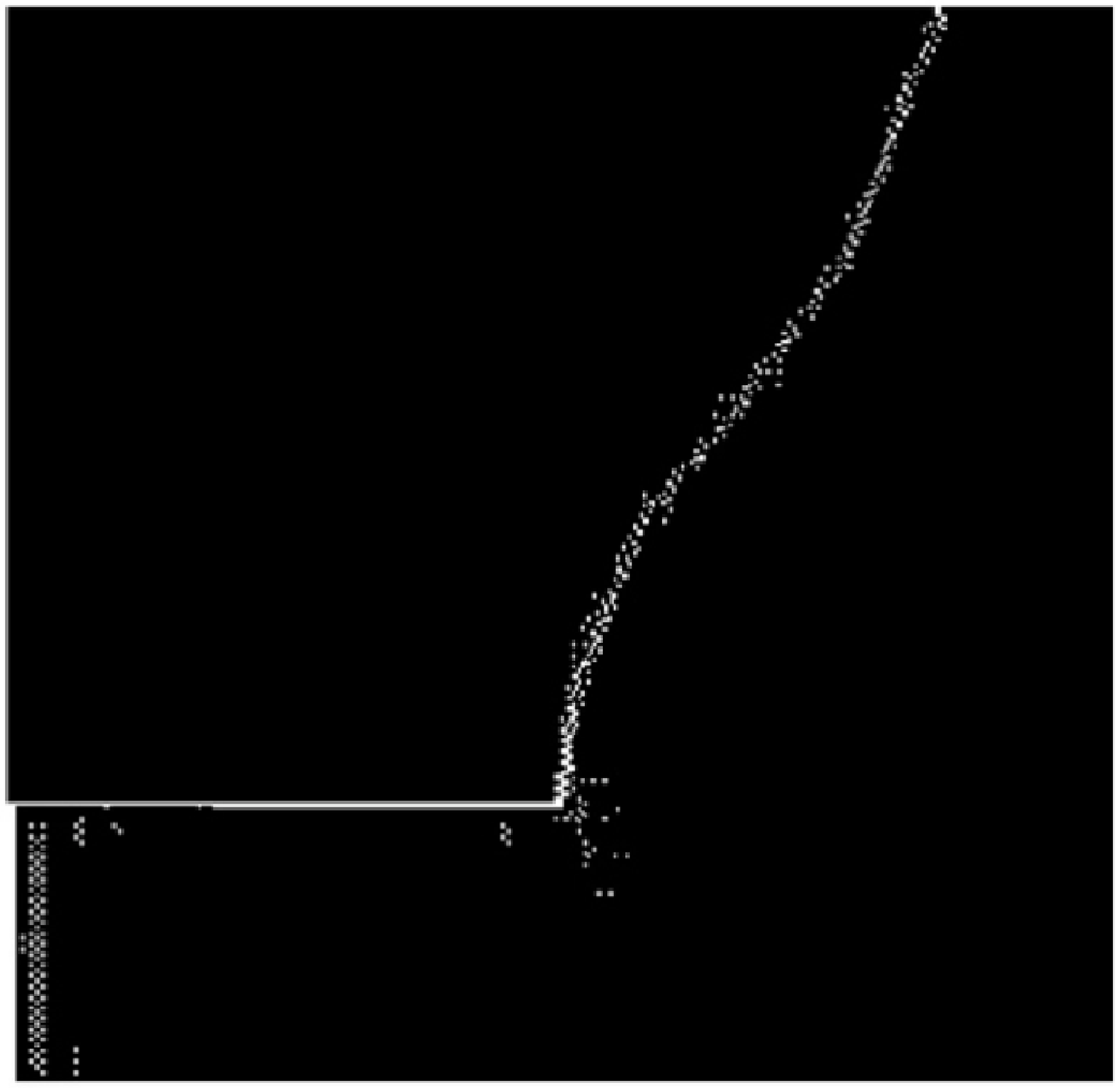} 
\caption{Chakraborty and Shaw \cite{chakraborty2013pseudo}}
\end{subfigure}
\begin{subfigure}[t]{0.36\textwidth}
\includegraphics[width=\textwidth, trim={0 0 0 0}, clip]{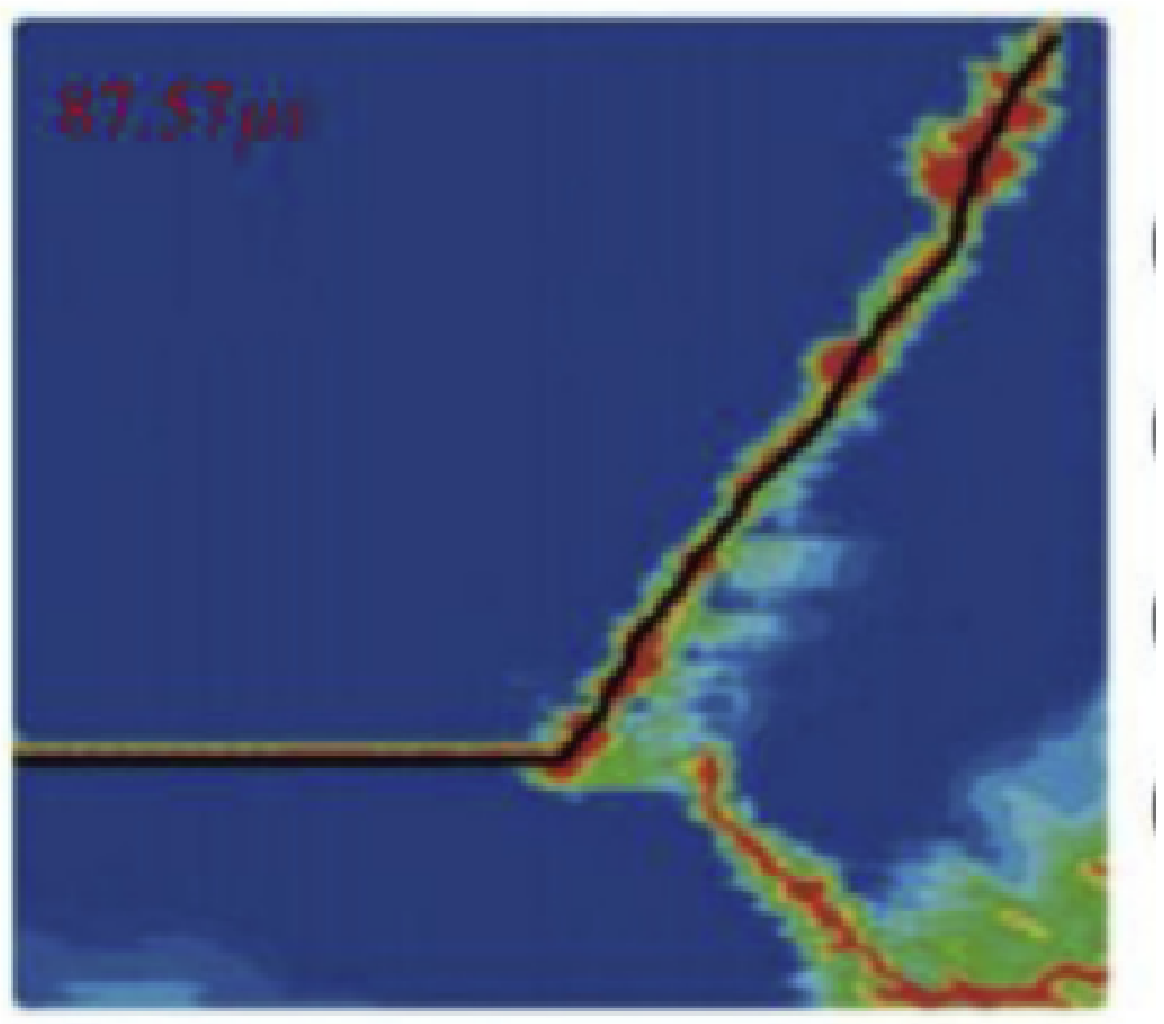} 
\caption{Belytschko et al. \cite{belytschko2003dynamic}}\label{f3}
\end{subfigure}
\begin{subfigure}[t]{0.33\textwidth}
\includegraphics[width=\textwidth, trim={0 0 0 0}, clip]{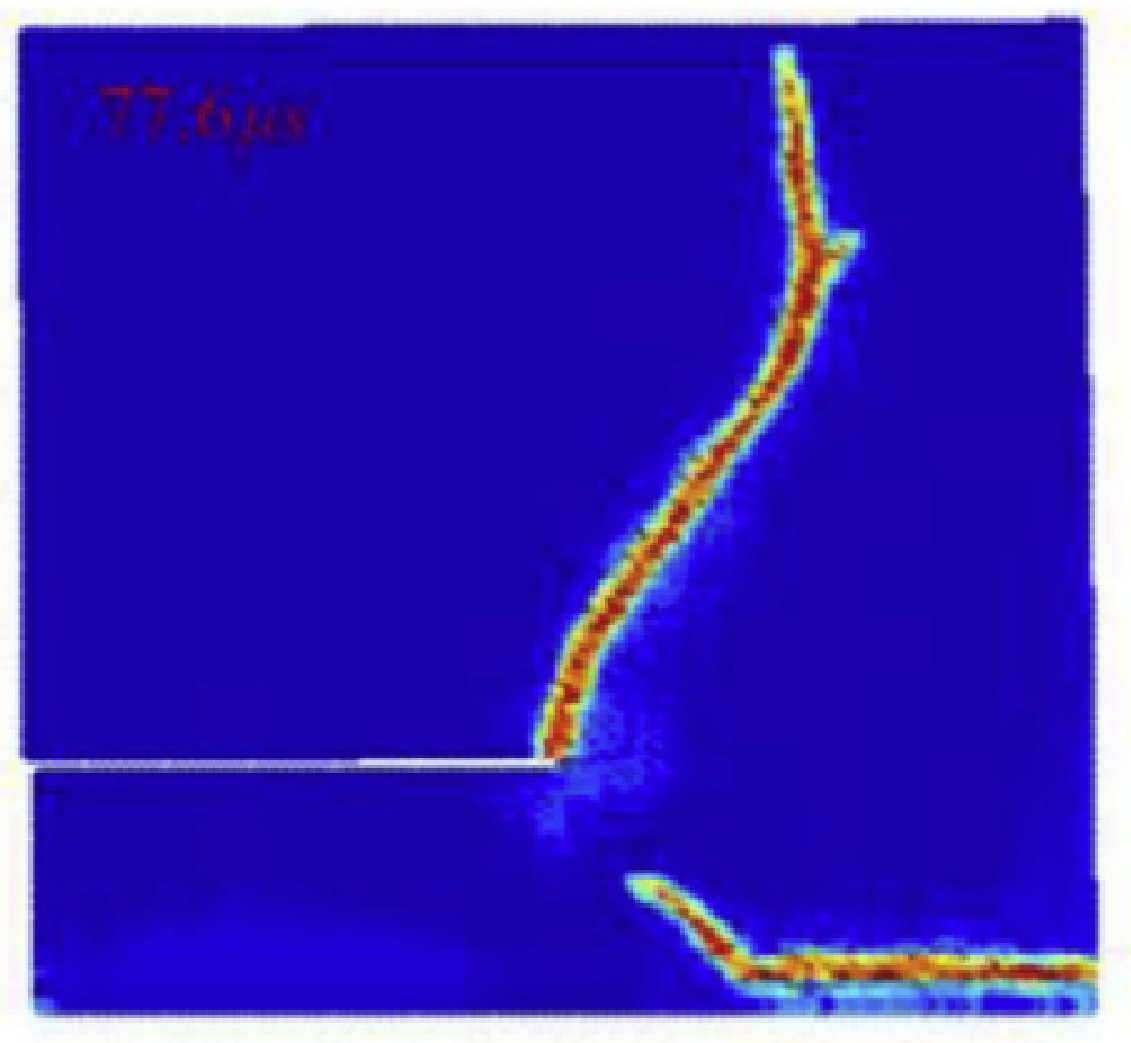} 
\caption{Zhou et al. \cite{zhou2016numerical}}\label{f4}
\end{subfigure}
\begin{subfigure}[t]{0.4\textwidth}
\includegraphics[width=\textwidth, trim={50 50 50 50}, clip]{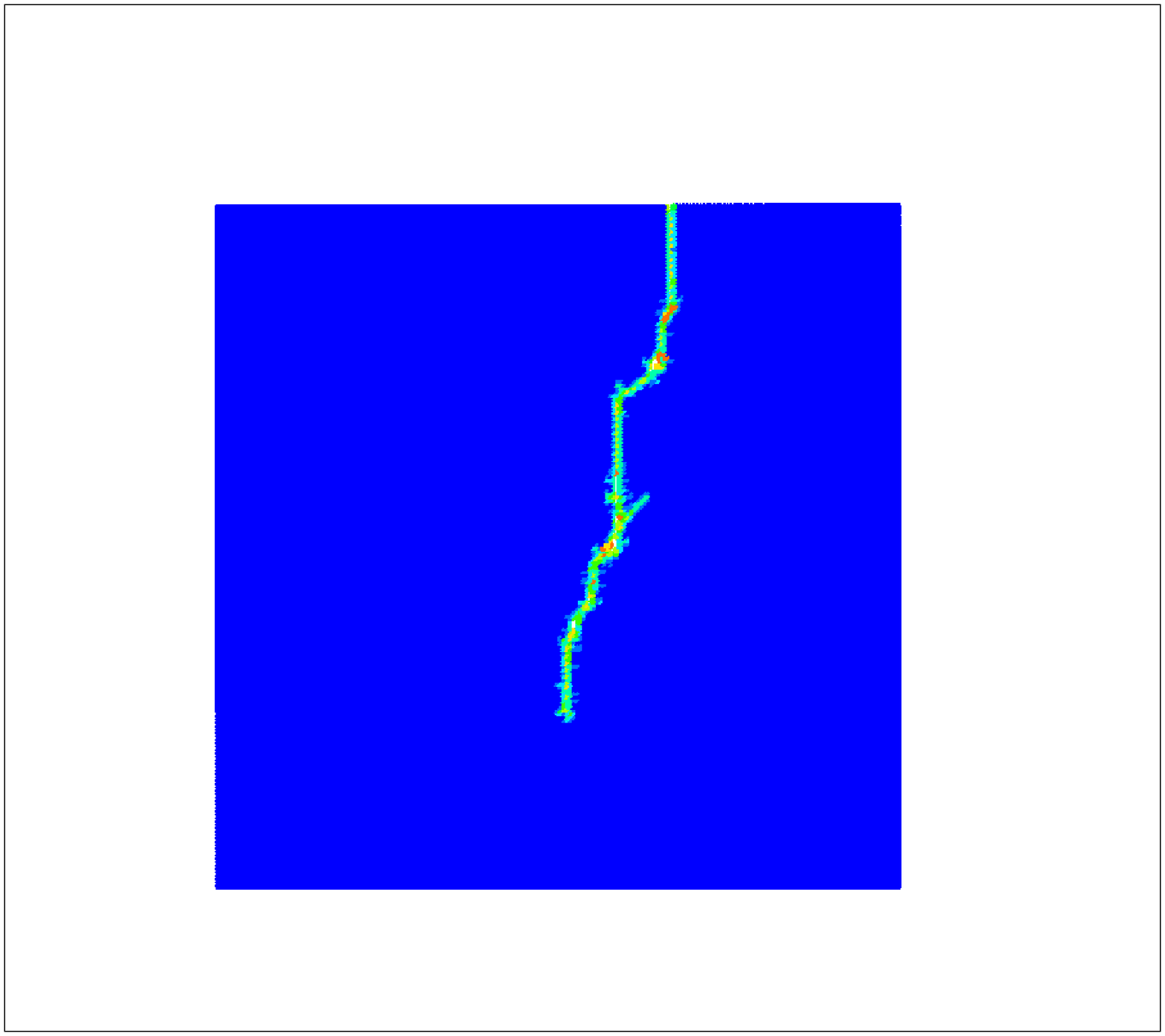} 
\caption{Present study}\label{p1}
\end{subfigure}
\caption{Comparison of fianl crack propagation paths}\label{kalthoff_compa}
\end{figure}

As this cracking problem is investigated using several other numerical methods with various cracking or damage criterion, the obtained crack path is compared with experimental and other numerical approaches in Figure \ref{kalthoff_compa}. In some of the numerical simulations a secondary crack path is also observed as shown in Figure \ref{f1}, \ref{f2}, \ref{f3}, \ref{f4}, which is not present in the experimental observation (Figure \ref{e1}). The present approach captures the crack path without any secondary crack paths (Figure \ref{p1}). Among all the numerical results, there are discrepancies regarding the cracking path and curve shape, which are mainly caused by different cracking criteria. Furthermore, some of the numerical results are obtained using a crack tracking algorithm, which we do not employ in this work. Therefore, our numerical results show a certain degree of particle distribution dependency. Nevertheless, this example demonstrates the capability of the present algorithm to capture the crack initiation and propagation at low strain rates.

\subsection{Crack propagation in deep beams}
\begin{figure}[hbtp!]
\centering
\begin{subfigure}[t]{0.45\textwidth}    
\includegraphics[width=\textwidth]{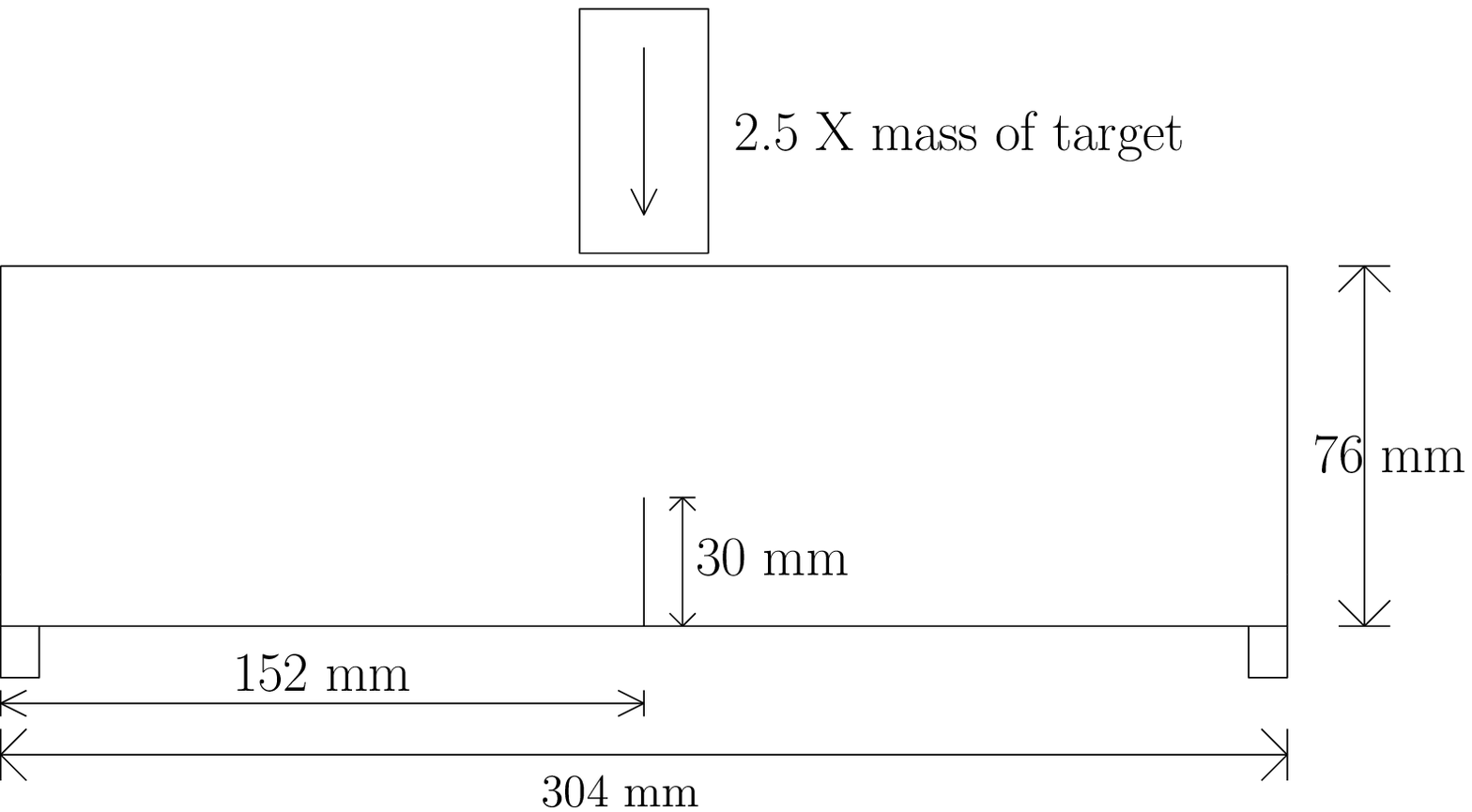}
\caption{Notch at mid-point} 
\end{subfigure}
\begin{subfigure}[t]{0.45\textwidth}
\includegraphics[width=\textwidth]{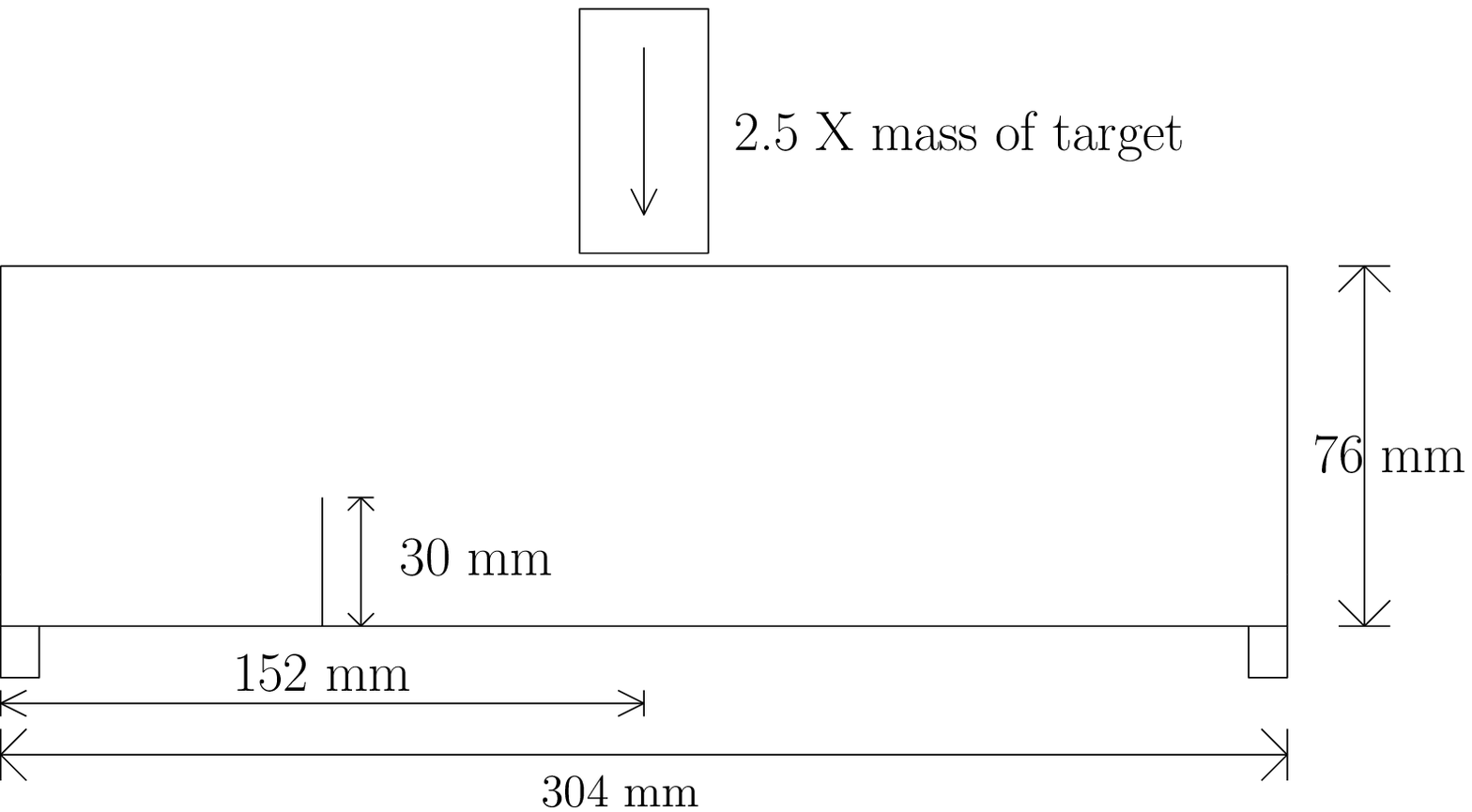}
\caption{Notch away from mid-point} 
\end{subfigure}
\caption{Set-up of the deep beam under impact with the notch at different locations}\label{deep_beam_confi}
\end{figure}

As the last example, a simply supported deep beam with the notch at different location \citep{ortiz1999finite, chakraborty2013pseudo} is chosen (Figure \ref{deep_beam_confi}) to demonstrate the capabilities of the present solution to capture the different crack propagation and failure modes. A beam of 308 mm length and 76 mm depth is considered under the impact of a projectile with 20 m/s velocity. The material and computational parameters are shown in Table \ref{deep_beam} and \ref{deep_beam_sph}. The notches are located at the midpoint and a distance of 75 mm from the left support. 

\begin{table}[h!]
\caption{Material parameters for crack propagation in deep beams}\label{deep_beam}
\centering
\begin{tabular}{ll}
\hline
Mechanical properties  &   Failure parameter       \\ \hline
$\rho=7830kg/m^3$     & $\epsilon_{max} = 0.03$    \\
$E = 200 GPa$         & (for Rankine criterion)         \\
$\nu=0.3$     \\   \hline                                                             
\end{tabular}
\end{table}

\begin{table}[h!]
\caption{Computational parameters used for deep beam}\label{deep_beam_sph}
\centering
\begin{tabular}{ccc}
\hline
Inter-Particle                 & Smoothing           & Artificial Viscosity                                \\
Spacing ($\Delta$ p)        & Length (h)            & Parameters ($\beta_1, \beta_2$)                \\ \hline
1.0 mm                        & 0.65 mm      & (2.0,2.0)                                                \\ \hline
\end{tabular}
\end{table}

\begin{figure}[hbtp!]
\centering
\begin{subfigure}[t]{0.49\textwidth}    
\includegraphics[width=\textwidth, trim={80 50 80 300}, clip]{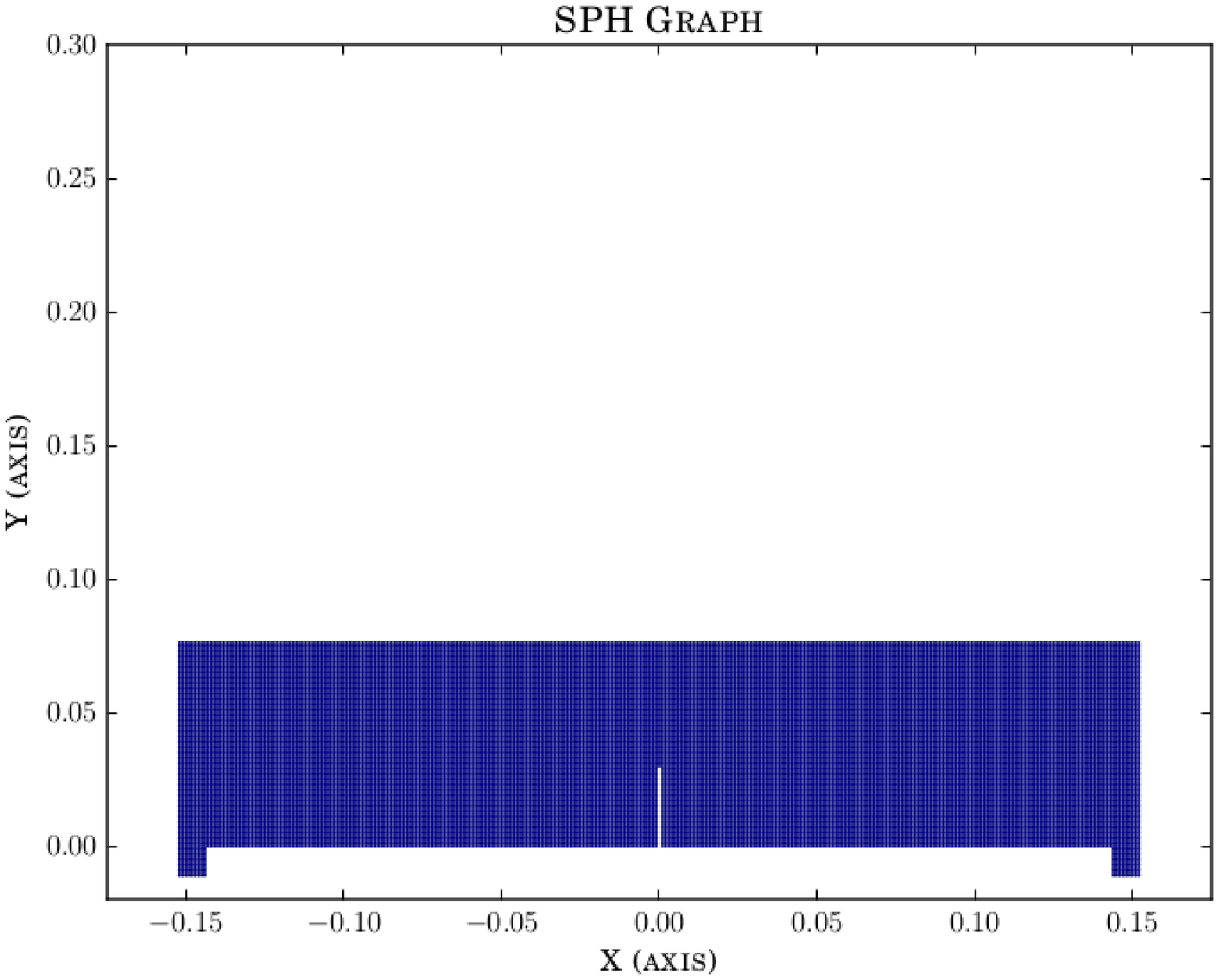}
\caption{Time = 0.0 ms} 
\end{subfigure}
\begin{subfigure}[t]{0.49\textwidth}
\includegraphics[width=\textwidth, trim={80 50 80 300}, clip]{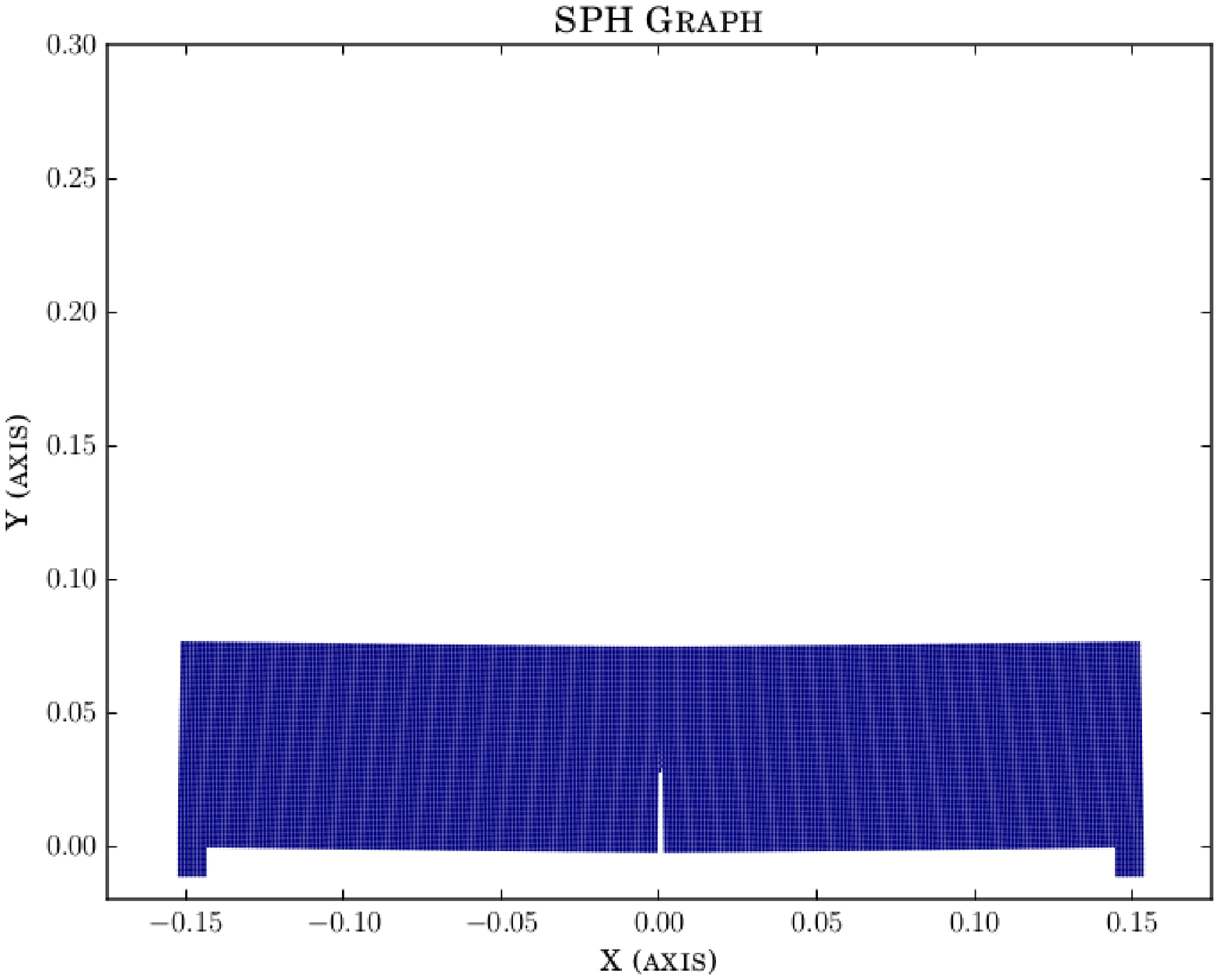}
\caption{Time = 0.5 ms} 
\end{subfigure}
\begin{subfigure}[t]{0.49\textwidth}
\includegraphics[width=\textwidth, trim={80 50 80 300}, clip]{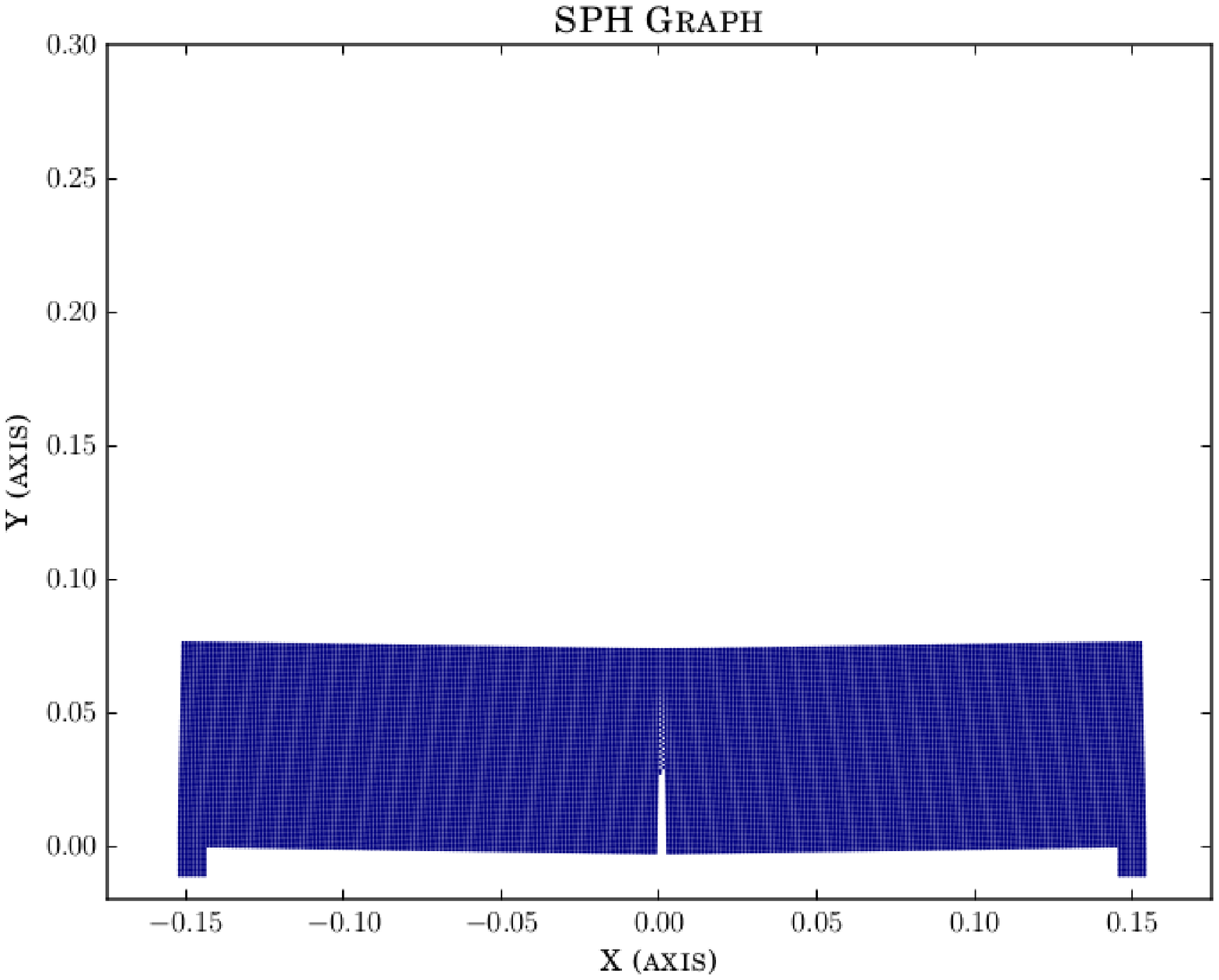}
\caption{Time = 0.6 ms} 
\end{subfigure}
\begin{subfigure}[t]{0.49\textwidth}
\includegraphics[width=\textwidth, trim={80 50 80 300}, clip]{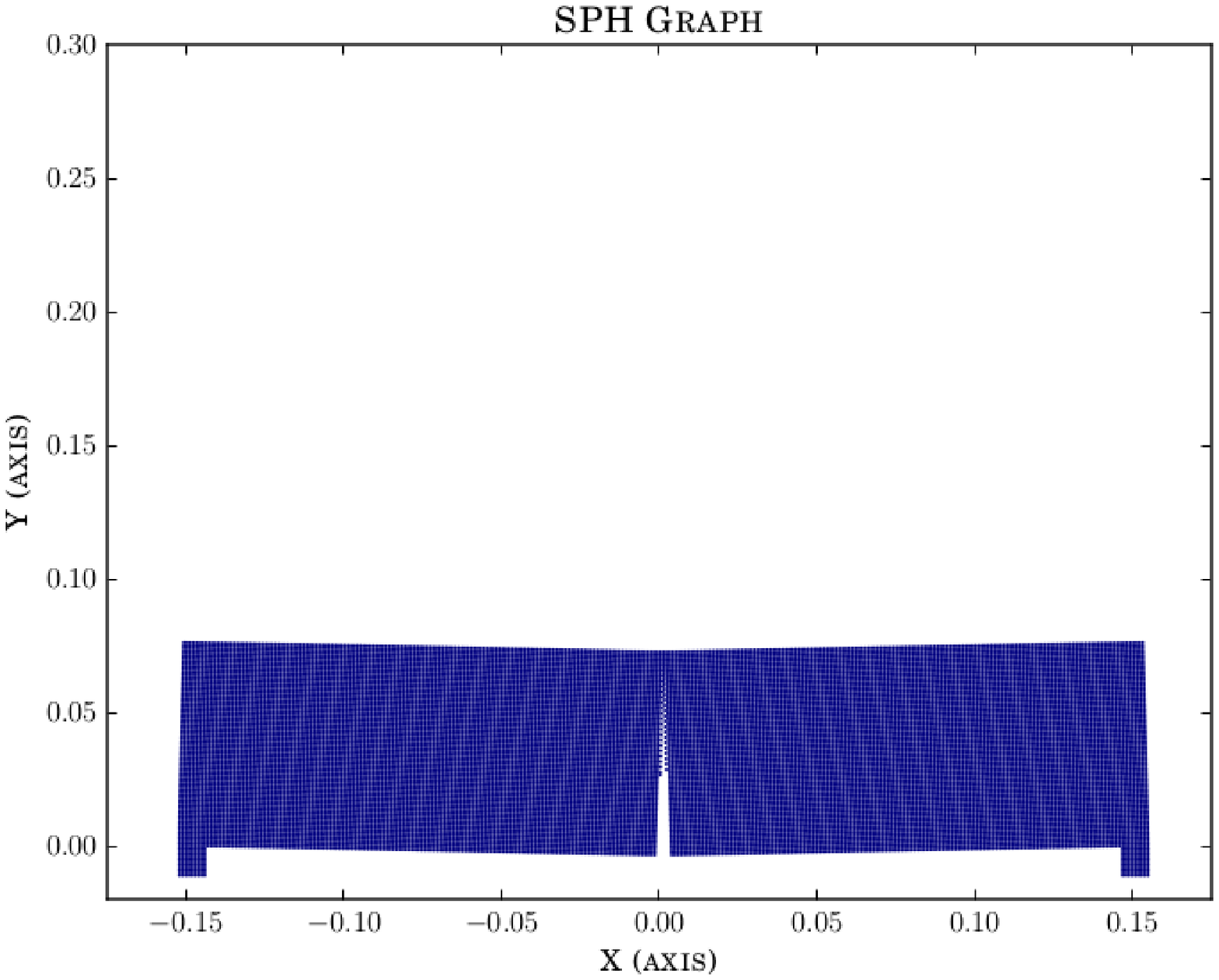}
\caption{Time = 0.7 ms} 
\end{subfigure}
\caption{Crack propagation in mode I (notch at mid-point) at different time step}\label{deep_mid_crack}
\end{figure}

\begin{figure}[hbtp!]
\centering
\begin{subfigure}[t]{0.49\textwidth}
\includegraphics[width=\textwidth, trim={0 0 0 0}, clip]{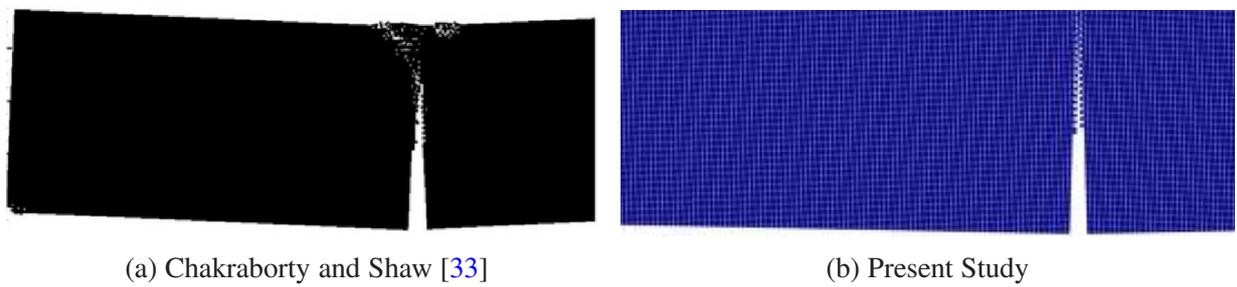}
\caption{Chakraborty and Shaw \cite{chakraborty2013pseudo}} 
\end{subfigure}
\begin{subfigure}[t]{0.49\textwidth}
\includegraphics[width=\textwidth, trim={150 60 230 300}, clip]{deep_mid4.eps}
\caption{Present Study} 
\end{subfigure}
\caption{Comparison of crack path in mode I (notch at mid-point)}\label{deep_mid_crack_comp}
\end{figure}

\begin{figure}[hbtp!]
\centering
\begin{subfigure}[t]{0.49\textwidth}    
\includegraphics[width=\textwidth, trim={80 50 80 300}, clip]{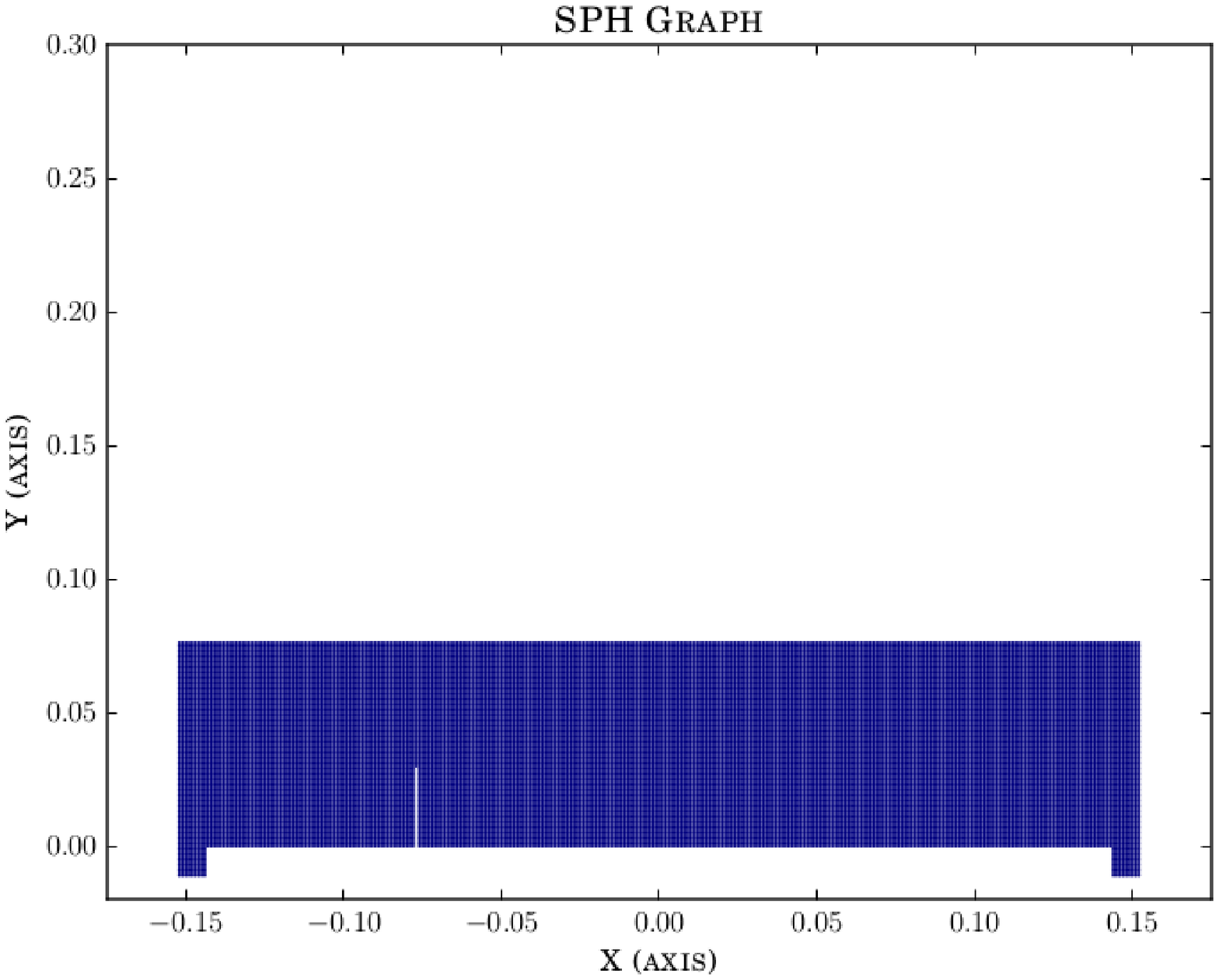}
\caption{Time = 0.0 ms} 
\end{subfigure}
\begin{subfigure}[t]{0.49\textwidth}
\includegraphics[width=\textwidth, trim={80 50 80 300}, clip]{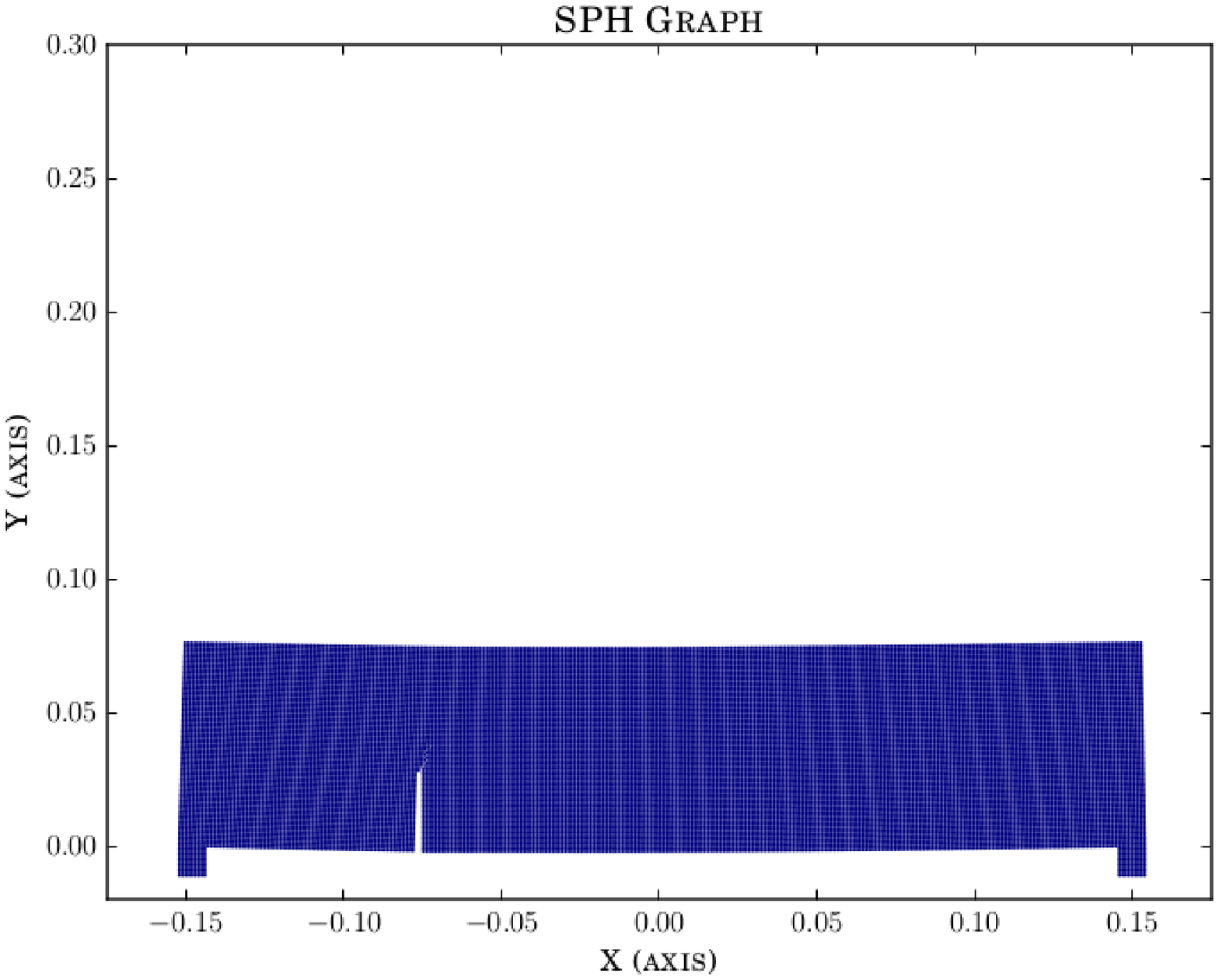}
\caption{Time = 0.75 ms} 
\end{subfigure}
\begin{subfigure}[t]{0.49\textwidth}
\includegraphics[width=\textwidth, trim={80 50 80 300}, clip]{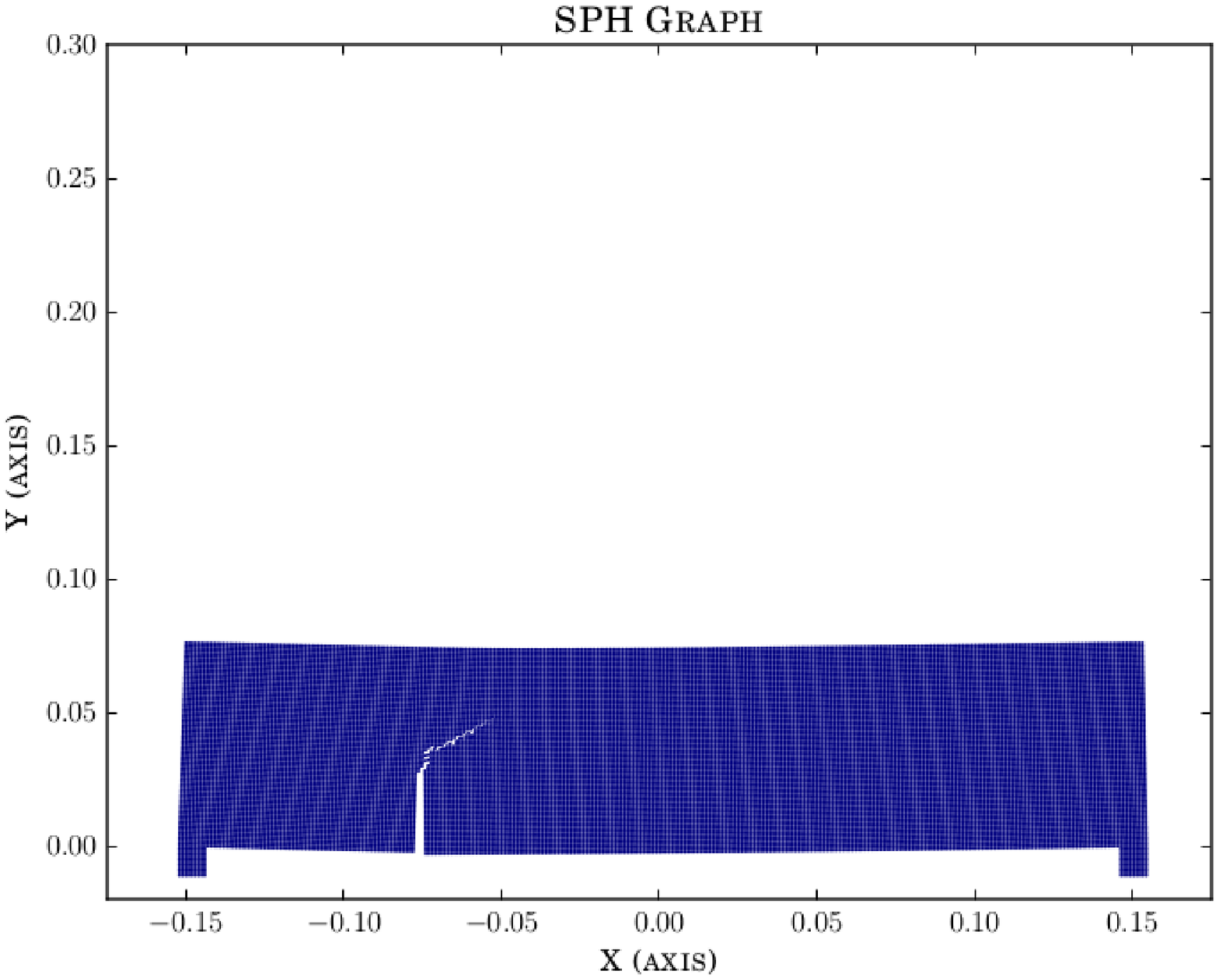}
\caption{Time = 0.6 ms} 
\end{subfigure}
\begin{subfigure}[t]{0.49\textwidth}
\includegraphics[width=\textwidth, trim={80 50 80 300}, clip]{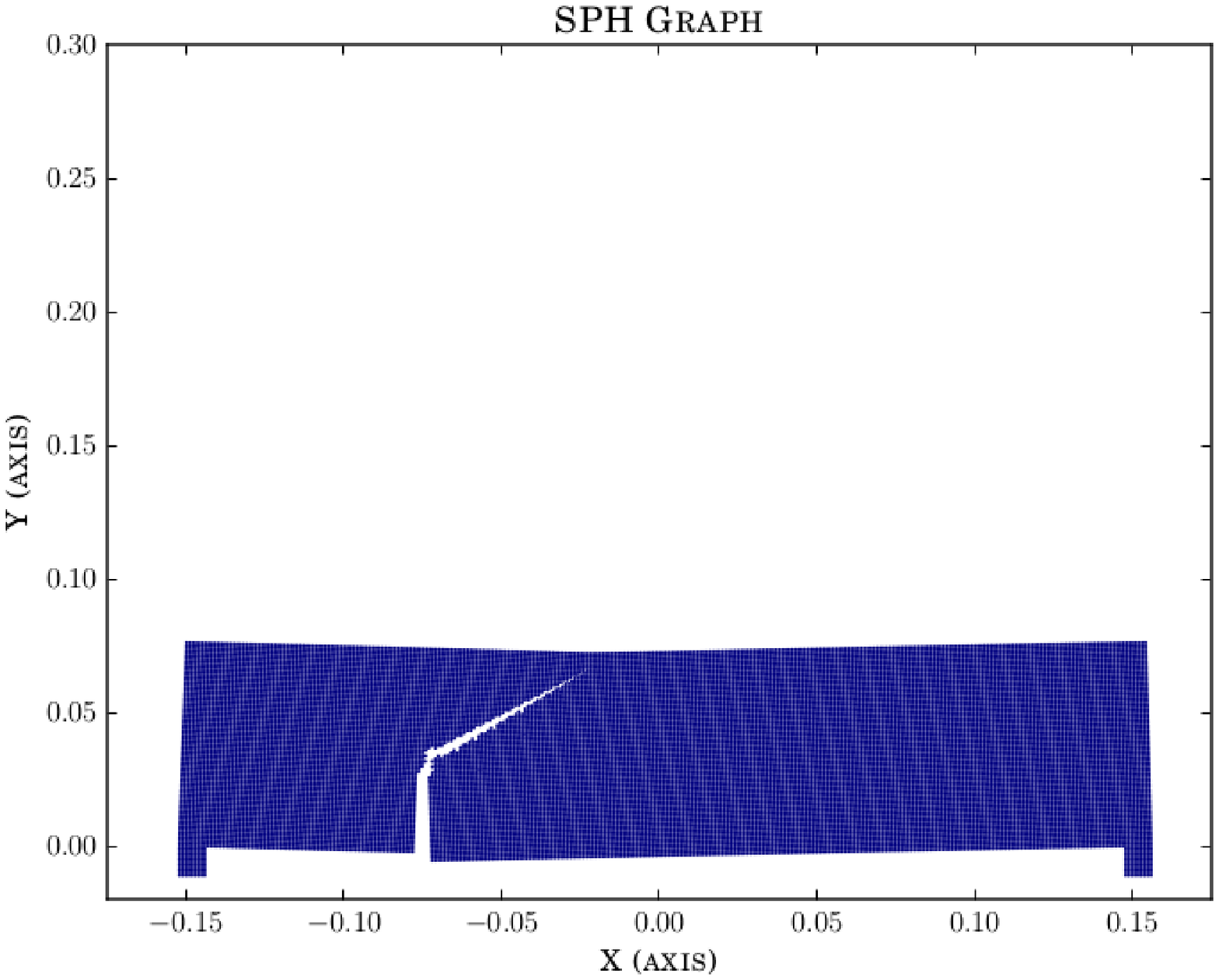}
\caption{Time = 1.0 ms} 
\end{subfigure}
\caption{Crack propagation in mixed mode (notch away from mid-point) at different time step}\label{deep_edge_crack}
\end{figure}

\begin{figure}[hbtp!]
\centering
\begin{subfigure}[t]{0.49\textwidth}
\includegraphics[width=\textwidth, trim={0 0 0 0}, clip]{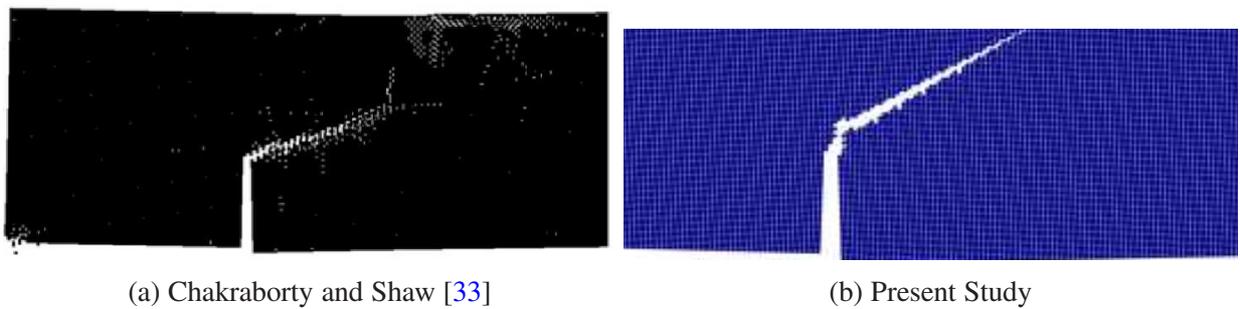}
\caption{Chakraborty and Shaw \cite{chakraborty2013pseudo}} 
\end{subfigure}
\begin{subfigure}[t]{0.49\textwidth}
\includegraphics[width=\textwidth, trim={135 60 250 300}, clip]{deep_edge4.eps}
\caption{Present Study} 
\end{subfigure}
\caption{Comparison of crack propagation in mixed mode (notch away from mid-point)}\label{deep_edge_crack_comp}
\end{figure}

As reported in \citep{ortiz1999finite, chakraborty2013pseudo}, the beam undergoes elastic bending, and as a result, the crack starts propagation from the notch tip. For the beam with the notch at the midpoint, the crack propagates vertically leading to mode I failure. The crack propagation at different time step can be observed in Figure \ref{deep_mid_crack}. The crack path captured in the present study is compared with the results from \cite{chakraborty2013pseudo} in Figure \ref{deep_mid_crack_comp}. It can be observed that the crack propagates vertically from the notch edge. On the other hand, for the beam with the notch away from the midpoint, a mixed mode type of crack propagation is found due to the combined influence of shear and transverse tension. The mixed mode crack propagation at different time step is shown in Figure \ref{deep_edge_crack}. This is found to be consistent with the observations in \cite{chakraborty2013pseudo} as shown in Figure \ref{deep_edge_crack_comp}. Hence the present observations are consistent with the results in \citep{ortiz1999finite, chakraborty2013pseudo}.

\section{Conclusion}\label{conclu}
In TLSPH, the computations are based on the reference framework; thus, it is free from tensile instability. This makes it a perfect candidate to model crack initiation, propagation and material failure. TLSPH in its original form cannot capture the forming and development of discontinuities as it is a continuum-based method. To provide a way out, a virtual link approach is proposed in the framework of TLSPH. In this approach, the particle's kernel support is modified to its immediate neighbour. A network of virtual links connects the particles. These links are called virtual as they do not provide any extra stiffness to the system but only defines the level of interaction between the connecting particles. Initially, due to the absence of any material damage, the links allows complete interaction. However, as the damage develops, the level of interaction is reduced between the connecting particles. When the connection is completely damaged, the interaction between particles stops, and the link is considered to be broken, implying the generation of a crack surface. The material constitutive models and cracking criteria guide the damage variables of the links.

The proposed method is employed to model the crack initiation, propagation and failure of notched beams under impact. The accumulation of plastic strain, initialisation of cracking and ultimate failure are well captured using the present approach. The proposed model is also able to simulate the brittle failure at low strain rates (Kalthoff-Winkler experiment) and the different modes of crack propagations in a deep beam. As the damage and failure are considered using the virtual links, the method does not need any cracking treatments such as particle deleting, splitting or visibility criterion. The initialisation and propagation are naturally modelled. The present method is observed to be stable in all the simulations.

Moreover, the crack path/surfaces can be tracked quite easily by monitoring the broken virtual links. There is no need for any explicit crack tracking algorithm. Therefore, the computational cost of the present framework is low, and the implementation is straightforward. Based on the current investigation, the method seems to be entirely natural in tracking crack paths and failure of solids. In the future, the approach will be explored for more complex brittle and ductile failure modes with arbitrary material flaws.

\section*{Acknowledgement}
This work receives funding from the European Union’s Horizon 2020 programme under grant agreement No. 778627 and the Austrian Research Promotion Agency (FFG) under the project No. 865963.

\bibliographystyle{elsarticle-num}

\end{document}